\documentclass[aps,twocolumn,reprint]{revtex4-2}

\usepackage[colorlinks=true, allcolors=blue]{hyperref}
\usepackage{graphicx}
\usepackage{bbm}
\usepackage{amsmath}
\usepackage{amssymb}
\usepackage{xcolor}
\usepackage{braket}
\usepackage{ulem}
\usepackage{MnSymbol}

\newcommand{\ISpectrum}{{\cal I}}

\renewcommand{\Re}{\operatorname{Re}}

\newcommand{\tr}{\operatorname{tr}}

\newcommand{\trlight}{\tr_{\rm light}}
\newcommand{\rholight}{\rho_{\rm light}}
\newcommand{\psilight}{\psi_{\rm light}}

\newcommand{\rmi}{\mathrm{i}}
\newcommand{\rme}{\mathrm{e}}
\newcommand{\rmd}{\mathrm{d}}

\newcommand{\opI}{\mathbbm{1}}

\newcommand{\opH}{\mathbb{H}}

\newcommand{\opD}{\mathbb{D}}
\newcommand{\opS}{\mathbb{S}}
\newcommand{\opM}{\mathbb{M}}
\newcommand{\opQ}{\mathbb{Q}}

\newcommand{\opU}{\mathbb{U}}

\newcommand{\opa}{a}

\newcommand{\wIR}{\omega_0} 
\newcommand{\aIR}{\opa_0} 
\newcommand{\nIR}{n_0} 
\newcommand{\ratioIR}{\varepsilon_0} 
\newcommand{\alphaIR}{\alpha_0} 

\newcommand{\Fxuv}{\mathbf{F}_{\rm xuv}}

\newcommand{\LCPMR}{Sorbonne Universit\'{e}, CNRS, Laboratoire de Chimie Physique – Mati\`{e}re et Rayonnement, LCPMR, 75005 Paris, France}
\newcommand{\LUND}{Department of Physics, Lund University, PO Box 118, SE-221 00 Lund, Sweden}
\newcommand{\LCF}{Universit\'{e} Paris-Saclay, CNRS, Institut d'Optique Graduate School, Laboratoire Charles Fabry, 91127 Palaiseau, France}
\newcommand{\CEA}{Universit\'{e} Paris-Saclay, CEA, LIDYL, 91191 Gif-sur-Yvette, France}

\begin{document}
\title{Quantum optical photoelectron interferometry}



\author{Jonathan Dubois} \email[Corresponding author: ]{jonathan.dubois@sorbonne-universite.fr} 
\author{Viviane Cotte}
\author{Richard Ta\"{i}eb}
\author{Camille L\'{e}v\^{e}que}
\author{J\'{e}r\'{e}mie Caillat}
\affiliation{\LCPMR}

\author{Pranshu Dave}
\author{Pascal Sali\`{e}res}
\author{David Bresteau}
\affiliation{\CEA}

\author{Charles Bourassin-Bouchet}
\affiliation{\LCF}

\author{Anne L'Huillier}
\author{David Busto}
\affiliation{\LUND} 

\begin{abstract}
We present a general theoretical framework for multiphoton processes driven by quantum light fields, establishing a direct link between photon statistics and photoelectron observables. 
Our results show that the autocorrelation and cross-correlation functions, which quantify the underlying photon statistics, are directly mapped onto the resulting photoelectron spectra. Although our framework is broadly applicable, we demonstrate specifically in the example of reconstruction of attosecond beating by interference of two-photon transitions (RABBIT) the influence of the light statistical properties. In this approach, the amplitude, contrast and phase of the oscillations of the sideband signal as a function of pump-probe delay reveal the quantum nature of light. 
We analyze these observables across several quantum configurations, including correlated infrared and harmonic modes, as well as the uncorrelated case with non-classical harmonic statistics, thereby establishing a general framework for quantum-light RABBIT spectroscopy.
We compare the analytical theory with numerical simulations for the case of classical harmonics and an infrared field in a squeezed coherent state, obtaining excellent agreement. Our results reveal how the interplay between classical and quantum correlations dictates the coherence of the photoemission process, providing a new window into the quantum-optical foundations of attosecond science.
\end{abstract}

\maketitle

\section{Introduction}

The conversion of infrared coherent light into sub-femtosecond light pulses through the generation of high-order harmonics in a gas medium~\cite{Ferray1988, Paul2001, Hentschel2001, Sansone2006} led to remarkable advances by opening the possibility of probing electron dynamics in matter at their natural timescales~\cite{Calegari2014, Nisoli2017, Merritt2021, Alexander2023}.
In recent decades, different interferometric techniques have been employed to measure the duration of these light pulses with attosecond resolution~\cite{Paul2001, Hentschel2001, Goulielmakis2004, Mairesse2005}. 
The first characterization of a train of attosecond light pulses has been achieved by extracting the relative phases between consecutive harmonics using RABBIT~\cite{Veniard1996, Maquet1998, Paul2001, Dahlstrom2013, Busto2019, Berkane2024} (reconstruction of attosecond beating by interference of two-photon transitions). 
This interferometric technique is fundamentally dependent on the \textit{coherence} of both the photoelectrons and the driving fields. Without such coherence, the interferometric signal in the spectrum would vanish along with the information it encodes about light and photoemission dynamics.

Almost all studies relied on the classical nature of light, in approaches referred to as \textit{semiclassical}, while recently questions about the quantum nature of the generated high-order harmonics have attracted particular attention~\cite{Gorlach2020, Stammer2024}.
For instance, theoretical works predict statistical properties of each harmonic that deviate from those of a coherent state~\cite{Gorlach2020, Gombkoto2021}, including bunching, anti-bunching, and entanglement between the harmonics~\cite{Gombkoto2021, Stammer2023}. 
Initially developed for interactions with gas media, high-order harmonic generation was later extended to solids~\cite{Gonoskov2024, Rivera2024}. 
The predictions on the non-classicality of the generated harmonics have been experimentally tested in this context and partially validated by the observation of bunching~\cite{Theidel2024, Theidel2025}.
Further theoretical and experimental work studied the impact of non-classical infrared light sources~\cite{Gorlach2023}, including bright squeezed vacuum~\cite{Tzur2023, Rasputnyi2024, Tzur2024, Wang2025, Li2025, Rivera2025str, Lemieux2025} (BSV), on high-order harmonic generation and above-threshold ionization~\cite{Rivera2025mic}.

\begin{figure}
    \centering
    \includegraphics[width=0.4\textwidth]{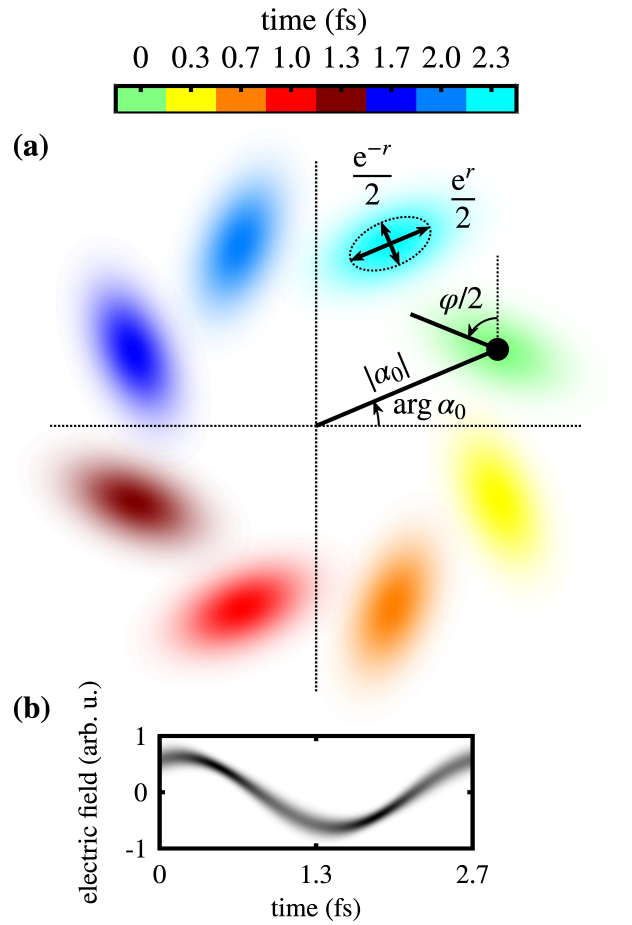}
    \caption{(a) The Wigner distribution $W(\alpha)$ of a squeezed coherent state of light [see Eq.~\eqref{eq:squeezed_coherent_state}], represented at successive instants over one optical cycle of approximately 2.7 fs (laser wavelength of 800 nm), as a function of the real and imaginary parts of $\alpha$.
    It displays the geometry of the coherent part characterized by $\alpha_0$ and the squeezing amplitude $\zeta = r \rme^{\rmi \varphi}$. The coherent-state limit corresponds to $\zeta=0$. 
    We use $10$ photons. The coherent part is $|\alphaIR| = 3.15$ and $\arg \alphaIR = 0.4$ rad. The squeezing ratio is $r \sim 0.37$ with a phase $\varphi = 2.37$ rad.
    (b) Corresponding distribution of the electric field $F_{\alpha} (t) = \bra{\alpha} F(t) \ket{\alpha}$, with $F(t)$ the electric field operator in the interaction picture and a coherent state $\ket{\alpha}$, as a function of time. 
    Each value of $\alpha$ sampled from $W(\alpha)$ generates one classical field trajectory, and the spread of trajectories reflects the quantum fluctuations of the squeezed state around the coherent part, i.e. its expectation value.}
    \label{fig:squeezed_light}
\end{figure}

Non-classical states of light are formally defined by Glauber–Sudarshan $P$-representations that are negative or more singular than a Dirac delta function~\cite{Glauber1963}. A fundamental example is the squeezed coherent state: although its $P$-function is more singular than a Dirac delta, its Wigner distribution remains well-defined and positive throughout phase space~\cite{Schnabel2017}, making it the natural representation to visualize the amplitude and phase-space structure of the field.
Figure~\ref{fig:squeezed_light}a depicts the time evolution of such a distribution, while Fig.~\ref{fig:squeezed_light}b illustrates how these states can be treated as an ensemble of ``classical'' electric fields with varying amplitudes and phases, offering a landscape of photon statistics and quantum correlations far richer than those of standard coherent states.
Currently, the most accessible realization of this regime is the BSV~\cite{Iskhakov2012, Manceau2019}, which features a mean photon number high enough to trigger multiphoton emission~\cite{Heimerl2024, Heimerl2025}. The BSV is a purely squeezed state that remains centered at the origin of the complex phase space and would be obtained by setting $\alphaIR = 0$ in Fig.~\ref{fig:squeezed_light}. Beyond their fundamental interest, these states are instrumental in suppressing quantum noise to enhance measurement precision, a capability that has led to landmark applications in high-precision metrology, most notably in gravitational wave detection~\cite{Aasi2013}.

Although photon statistics does not influence one-photon absorption~\cite{Mandel1995}, it significantly impact both the strong-field above-threshold ionization spectrum~\cite{Rivera2025mic} and low-order multiphoton processes~\cite{Dahan2021}. 
In this article, we show that interferometric photoemission measurements are strongly affected by the quantum properties of the infrared and its harmonics, which makes them ideal candidates for probing the role of the quantum properties of light.
We show that the amplitude, contrast and phase of the oscillations in a RABBIT spectrogram encode the quantum statistics of the emitted and absorbed photons of the infrared and harmonic lights.

The paper is organized as follows: In Section~\ref{sec:Hformalism}, we establish a theoretical framework to analyze intensity spectra in attosecond interferometric measurements involving quantum light. We describe the photoelectron spectrum using the density matrix of the complete system and develop a treatment based on time-dependent perturbation theory (TDPT).
Appendix~\ref{sec:derivation_H} provides a first-principles derivation of the Hamiltonian, starting from its most general form and reducing it to the one central to our framework via the length gauge in the dipole approximation.
In Section~\ref{sec:RABBIT}, the TDPT is applied to the RABBIT scheme.
We recover the standard form of the RABBIT signal, here expressed as a combination of \textit{semiclassical} transition matrix elements~\cite{Veniard1996, Maquet1998, Klunder2011, Dahlstrom2013, Busto2019, Berkane2024} coupled with high-order moments of quantum light fields. The formal expressions for these matrix elements are provided in Appendix~\ref{sec:formal_expressions}.

Finally, in Section~\ref{sec:application}, we explore several applications using a series of illustrative quantum states.
We first analyze the RABBIT signal in terms of normalized correlation functions between infrared and harmonic pulses (Section~\ref{sec:quantum_correlations}). We then examine the specific case of uncorrelated infrared and harmonic fields, explicitly deriving the RABBIT contrast and phase as functions of the normalized first-order cross-correlation of the harmonics (Section~\ref{sec:uncorrelated_light}).
Finally, assuming classical harmonics, we investigate the influence of infrared squeezing on the interferometric signal, ranging from squeezed coherent states to the BSV regime, and validate our perturbative results against full numerical solutions of the time-dependent Schr\"{o}dinger equation (Section~\ref{sec:classicalXUV}). 
Further details on the Wigner distributions of these states and the specifics of the numerical simulations are available in the Appendix~\ref{sec:TDSE}. 
The conclusions are presented in Section~\ref{sec:conclusions}.
Atomic units are used unless otherwise stated.

\section{Multiphoton processes induced by quantum light \label{sec:Hformalism}}
\subsection{Intensity spectrum from the density matrix}
We describe the combined state of light and matter by a density matrix $\rho$. 
The light field can take any state, from classical to fully quantum.
Photoemission measurements~\cite{Bourassin2020, Berkane2025} consist in measuring the energy of the photoelectron and can be formally described as a partial measurement in the photoelectron energy space. The spectrum $\ISpectrum (E)$ is obtained as
\begin{equation}
\label{eq:ISpectrum}
    \ISpectrum (E) = \lim_{t \to \infty}  \bra{E} \trlight \left\lbrace \rho (t) \right\rbrace \ket{E} ,
\end{equation}
where $\trlight$ denotes the partial trace over the light's degrees of freedom and $\ket{E}$ is the continuum state of the photoelectron of energy $E$. We assume that photoemission occurs in a single channel (not specified in the notations for the sake of readability).
We consider an arbitrary initial state of light $\rholight (-\infty)$ and an electron in the ground state of an atom $\ket{\rm g}$ such that the initial state is separable: $\rho (- \infty) = \ket{\rm g}\bra{\rm g} \otimes \rholight (-\infty)$.
In general, the state is no longer separable under time evolution.
The density matrix of the total system evolves in time from its initial state to the detector following the von Neumann equation
\begin{equation}
\label{eq:main_vonNeumann}
    \dot{\rho} = - \rmi [ \opH (t) , \rho ] ,
\end{equation}
with the Hamilton operator~\cite{Gorlach2020}
\begin{equation}
\label{eq:main_Hamiltonian}
    \opH (t) = \opH_{A} +\opH_{\rm int}(t) ,
\end{equation}
where $\opH_A$ is the field-free Hamilton operator and $\opH_{\rm int}(t)$ describes the interaction with the quantum field. 
The latter is treated in the interaction picture, which explicitly introduces time dependence into the operator and effectively eliminates the free-field harmonic oscillator terms. This allows us to recover the standard semiclassical picture of a dipole interacting with a time-dependent classical field~\cite{Veniard1996, Maquet1998, Klunder2011, Dahlstrom2013, Busto2019, Berkane2024}. 
Appendix~\ref{sec:derivation_H} details this derivation from the standard form of the quantum optical Hamilton operator~\cite{Gorlach2020, Lange2024}.
In the dipole approximation and in length gauge,
\begin{equation}
\label{eq:HFint}
    \opH_{\rm int}(t) = \mathbf{D} \cdot \mathbf{F}(t) ,
\end{equation}
where $\mathbf{D}$ is the dipole moment and 
\begin{equation}
\label{eq:main_Ft}
    \mathbf{F}(t) = \sum_{\mathbf{k}\sigma} f_{\bf k} \left( \mathbf{e}_{\sigma} \opa_{\mathbf{k}\sigma} \rme^{- \rmi \omega_{\mathbf{k}}t} + \mathbf{e}_{\sigma}^{\star} \opa_{\mathbf{k}\sigma}^{\dagger}  \rme^{\rmi \omega_{\mathbf{k}}t} \right) ,
\end{equation}
the quantum optical electric field.
The annihilation (creation) operator of the mode labeled by the wave vector $\mathbf{k}$ and the polarization $\mathbf{e}_{\sigma}$ is $\opa_{\mathbf{k}\sigma}$ ($\opa_{\mathbf{k}\sigma}^{\dagger}$), such that $[\opa_{\mathbf{k} \sigma} , \opa^{\dagger}_{\mathbf{k}^{\prime}\sigma^{\prime}}] = \delta_{\mathbf{k}\mathbf{k}^{\prime}} \delta_{\sigma \sigma^{\prime}}$. The frequency $\omega_{\bf k}$ obeys the dispersion relation $\omega_{\mathbf{k}} = c |\mathbf{k}|$. The vacuum electric field amplitude is 
\begin{equation}
\label{eq:gk}
    f_{\bf k} = \sqrt{\dfrac{\omega_{\bf k}}{ 2\epsilon_0V}} ,    
\end{equation}
where $\epsilon_0$ is the permittivity and $V$ the quantization volume.

\subsection{Time-dependent perturbation theory for density matrices}
We introduce the unitary transformation $\opU_A (t) = \exp ( -\rmi \opH_A t )$.
The equation of motion~\eqref{eq:main_vonNeumann} becomes 
\begin{equation}
\label{eq:TDSE_I}
    \dot{\rho}_I = - \rmi [ \opH_I (t) , \rho_I ] ,
\end{equation}
where $\rho_I (t) = \opU_A^{\dagger} (t) \: \rho (t) \: \opU_A (t)$ and $\opH_I (t) = \opU_A^{\dagger} (t) \opH_{\rm int} (t) \opU_A (t)$. 
The TDPT consists of expanding the density matrix as a series of terms corresponding to different perturbation orders,
\begin{equation}
\label{eq:rhoI_TDPT}
    \rho_I (t) = \sum_{n} \rho_I^{(n)} (t) .
\end{equation}
To maintain consistency with the established \textit{semiclassical} framework~\cite{Veniard1996, Maquet1998, Klunder2011, Dahlstrom2013, Busto2019, Berkane2024}, we evaluate the density-matrix coefficients through a time-dependent perturbative expansion of the evolution operator $\opU (t,t^{\prime})$.
This approach, detailed in~\cite{Messiah1962}, is summarized here for self-consistency. 
The time evolution of $\opU (t,t^{\prime})$ is described by
\begin{equation}
\label{eq:rho_time_evolution_U}
    \dot{\opU} (t,t^{\prime}) = - \rmi \opH_I (t) \opU (t,t^{\prime}) ,
\end{equation}
with the condition $\opU (t ,t ) = \opI$. The density matrix evolves as $\rho_I (t) = \opU (t,t^{\prime}) \rho_I (t^{\prime}) \opU^{\dagger} (t,t^{\prime})$.
Introducing the perturbation expansion
\begin{equation}
    \opU (t,t^{\prime}) = \sum_n \opU^{(n)} (t,t^{\prime}) ,
\end{equation}
and substituting into~\eqref{eq:rho_time_evolution_U} leads to the recurrence relation
\begin{equation}
\label{eq:Un1_Un_TDPT}
    \opU^{(n+1)} (t,t^{\prime}) = - \rmi \int_{t^{\prime}}^{t} \opH_I (t^{\prime\prime}) \opU^{(n)} (t^{\prime\prime} , t^{\prime}) \; \rmd t^{\prime\prime} ,
\end{equation}
with the zeroth-order term $\opU^{(0)} (t,t^{\prime}) = \opI$.
The $n$th order coefficient of the density matrix in~\eqref{eq:rhoI_TDPT} can be expanded as
\begin{equation}
\label{eq:rhoI_U_TDPT}
    \rho_I^{(n)} (t) = \sum_{m=0}^n \opU^{(m)} (t,t^{\prime}) \rho_I (t^{\prime}) \left[ \opU^{(n-m)} (t,t^{\prime}) \right]^{\dagger} .
\end{equation}
The intensity spectrum~\eqref{eq:ISpectrum} can then be written as the sum of contributions of order $n$
\begin{equation}
\label{eq:ISpectrum_TDPT}
    \ISpectrum (E) = \sum_n  \ISpectrum^{(n)} (E) ,
\end{equation}
where each term is defined as
\begin{equation}
\label{eq:ISpectrum_n}
    \ISpectrum^{(n)} (E) = \lim_{t \to \infty}  \bra{E} \trlight \left\lbrace   \rho^{(n)}_I (t) \right\rbrace \ket{E}  .
\end{equation}

Note that Eq.~\eqref{eq:rhoI_U_TDPT} together with~\eqref{eq:Un1_Un_TDPT} are equivalent--after integrating by parts and rearranging the terms--to the usual recurrence relation 
\begin{equation}
\label{eq:VonNeumann_TDPT}
    \dot{\rho}_I^{(n+1)} (t) = - \rmi [\opH_I(t) , \rho_I^{(n)} (t)] ,
\end{equation}
expressed in TDPT~\cite{Scully_Zubairy_1997}. 
The key advantage of Eq.~\eqref{eq:rhoI_U_TDPT} is that the time-evolution operators $\opU^{(n)}$ are identical in both the wave function and density-matrix formalisms. 
This symmetry allows us to directly export established concepts from \textit{semiclassical} attosecond science, such as multiphoton transition amplitudes~\cite{Veniard1996, Maquet1998}, into the quantum-optical regime, as shown in the following section for the case of one- and two-photon processes.

In the case of pure states, the perturbative structure of the density matrix has a natural interpretation. The density matrix can be written as $\rho_I (t) = \ket{\psi_I (t)} \bra{\psi_I (t)}$ where $\ket{\psi_I (t)} = \sum_n \ket{\psi_I^{(n)} (t)}$ and $\ket{\psi_I^{(n)} (t)} = \opU^{(n)} (t,t^{\prime}) \ket{\psi_I (t^{\prime})}$.
The element of the $n$th-order density matrix $\rho_I^{(n)}$ contains contributions from all pairs $( \ket{\psi_I^{(m)}(t)}, \bra{\psi_I^{(n-m)}(t)})$. In particular, for even $n$, the diagonal elements (populations) are determined by $\braket{\psi_I^{(n/2)}(t)| \psi_I^{(n/2)}(t)}$, while the off-diagonal elements (coherences) depend on the product of coefficients describing different perturbative orders, i.e. $m$-photon and $(n-m)$-photon processes. The density-matrix formalism thus accounts for the coherence between different $n$-photon processes~\cite{Laurell2022, Laurell2025}, that would be less visible in a purely wave-function description.

The $n$th-order evolution operator $\opU^{(n)}$, and consequently the $n$th-order density matrix $\rho_I^{(n)}$, involve a product of $n$ annihilation and creation operators ($\opa_{\mathbf{k}\sigma}$ and $\opa_{\mathbf{k}\sigma}^{\dagger}$). 
This represents a net exchange of $n$ photons with the field consistent with the standard interpretations of the TDPT. 
As a primary result of this work, we have shown that $\ISpectrum^{(n)}$ is expressed as a linear combination of various $n$th moments of the distribution, or equivalently, $n$-point correlation functions of the electromagnetic field. Consequently, the photoelectron intensity spectrum produced by multiphoton processes encodes the underlying quantum nature and photon statistics of the field.
Using interferometric photoemission measurements where the temporal delay between different field modes is varied, one can experimentally extract the amplitude and phase of these high-order $n$-point correlation functions. We demonstrate this capability in the following sections by applying our formalism to the RABBIT technique.

\section{Interferometric photoemission: RABBIT scheme \label{sec:RABBIT}}

\begin{figure}
    \centering
    \includegraphics[width=0.45\textwidth]{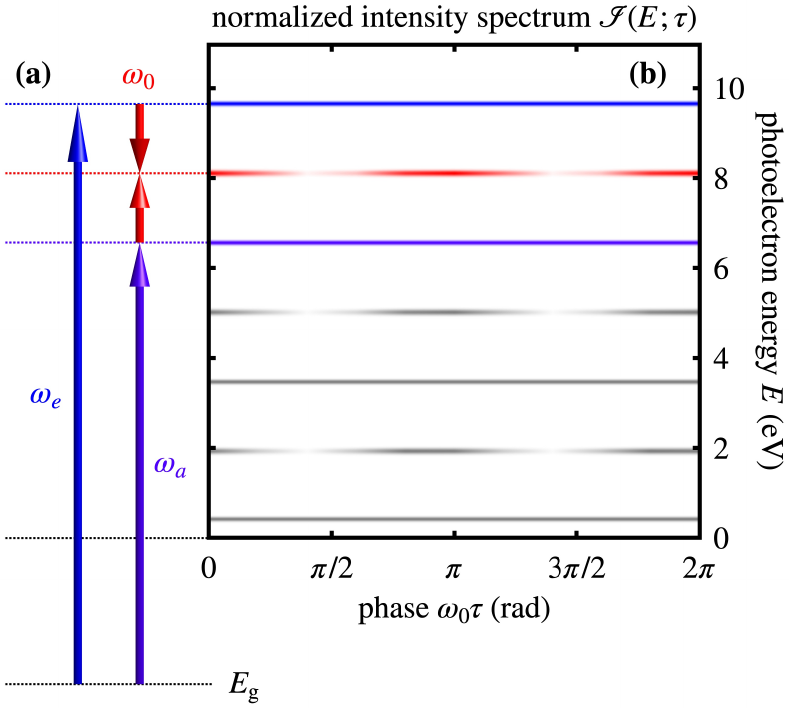}
    \caption{(a) Illustration of the RABBIT scheme in the regime of quantum light. Two types of quantum paths contribute to the photoelectron spectrum: one-photon absorption, giving rise to main bands (e.g. at frequencies $\omega_e$ and $\omega_a$, shown in blue and purple, respectively), and two-photon transitions involving an additional photon of frequency $\wIR$, giving rise to sidebands (e.g. the one shown in red between the two highlighted harmonic bands). 
    (b) Photoelectron intensity spectrum obtained from the solution of the TDSE (see Appendix~\ref{sec:TDSE} for details) as a function of $\tau$ and the energy of the photoelectron at the detector $E$, with $\tau$ the delay between the harmonic and infrared pulses.
    Other harmonics and sidebands are shown in gray. 
    The shading of the colors is proportional to the intensity spectrum. Each harmonic peak and sideband is normalized with respect to its maximum value.}
    \label{fig:RABBIT_illustration}
\end{figure}

The RABBIT technique involves generating a photoelectron wavepacket from a target atom or molecule through the simultaneous interaction with an attosecond pulse train and an infrared laser field. As illustrated in Fig.~\ref{fig:RABBIT_illustration}, the final energy state of the photoelectron in a sideband can be reached via two distinct quantum pathways that interfere at the detector. Specifically, the sideband signal is formed by the interference between the absorption of a harmonic photon (frequency $\omega_e$) and the stimulated emission of an infrared photon ($\wIR$), and the absorption of a photon from the preceding harmonic ($\omega_a$) and the absorption of an infrared photon
\begin{equation}
     \omega_e - \omega_a = 2 \wIR .
\end{equation}
The resulting intensity spectrum consists of main bands, shown in purple and blue in Fig.~\ref{fig:RABBIT_illustration}b, and the sideband shown in red. By introducing a temporal delay $\tau$ between the harmonic and infrared pulses, the relative phase between these two pathways is modulated, resulting in a periodic oscillation of the sideband intensity.
The phase of these oscillations encodes critical information, including temporal information on the field~\cite{Veniard1996, Paul2001}, as well as intrinsic atomic and molecular quantities, such as the Wigner time delay of the photoelectron~\cite{Klunder2011, Caillat2011, Beaulieu2017, Autuori2022}.
Ultimately, the success of RABBIT as an interferometric technique is fundamentally based on the coherence of the photoelectron wavepacket and, crucially, on the coherence of the driving fields. Without this ingredient, the interference contrast would vanish and the sideband oscillations would disappear.

In the following we concentrate on the three modes outlined in color in Fig.~\ref{fig:RABBIT_illustration}, leading to the photoelectron spectra in blue, purple and red.
This choice is made without loss of generality as the framework can be straightforwardly extended to an arbitrary number of harmonic orders.
Accordingly, the electric field in Eq.~\eqref{eq:main_Ft} is composed of these three modes. Each mode $\beta = \lbrace 0,a,e \rbrace$ is characterized by its polarization $\mathbf{e}_{\beta}$, vacuum electric field amplitude $f_{\beta}$ and a pair of annihilation and creation operators $( \opa_{\beta} ,\opa_{\beta}^{\dagger} )$. 
The electric field~\eqref{eq:main_Ft} reads explicitly as a function of the delay $\tau$ as
\begin{eqnarray}
\label{eq:Ft_RABBIT}
    && \mathbf{F}(t;\tau) = \sum_{\beta = a,e} f_{\beta} \left( \mathbf{e}_{\beta} \opa_{\beta} \rme^{- \rmi \omega_{\beta}t} + \mathbf{e}_{\beta}^{\star} \opa_{\beta}^{\dagger}  \rme^{\rmi \omega_{\beta}t} \right) \nonumber \\
    && \qquad\qquad + f_{0} \left[ \mathbf{e}_{0} \aIR \rme^{- \rmi \wIR (t-\tau)} + \mathbf{e}_{0}^{\star} \aIR^{\dagger}  \rme^{\rmi \wIR (t-\tau)} \right] . \label{eq:Ft_ae0}
\end{eqnarray}
In this regime, the intensity spectrum~\eqref{eq:ISpectrum_TDPT} has two main contributions such that
\begin{equation}
    \ISpectrum (E) \approx \ISpectrum^{(2)} (E) + \ISpectrum^{(4)} (E) .
\end{equation}
It depends on the photon statistics of the infrared and harmonic light pulses through $\rholight (-\infty)$.
Restricting our analysis to one- and two-photon transitions, the term $n=2$ describes the main bands and the term $n=4$ describes the sideband.
Thereafter, we analyze $\ISpectrum^{(2)} (E)$ and $\ISpectrum^{(4)} (E)$ in terms of the expectation values over the light's degrees of freedom. We use as an expression of the expectation value of an operator $O$
\begin{equation}
    \braket{O} = \trlight \left\lbrace \rholight (-\infty) O \right\rbrace .
\end{equation}

\subsection{Intensity spectrum}
\subsubsection{Main bands}

The intensity spectrum of the main bands is obtained from $\rho^{(2)} (t)$ in Eq.~\eqref{eq:ISpectrum_n} by retaining only one-photon processes.
Furthermore, only $\opU^{(1)}$ contributes to $\bra{E} \rho^{(2)}\ket{E}$, since $\bra{E} \opU^{(0)} \ket{\rm g} = \braket{E |\rm g}$ is equal to zero. The first-order transition amplitude thus reduces to
\begin{equation}
\label{eq:U1_Eg}
    \lim_{\substack{t \to \infty \\ t^{\prime} \to -\infty}} \bra{E} \opU^{(1)} (t,t^{\prime}) \ket{\rm g} = M_e^{(1)} (E) \; f_e \: \opa_e + M_a^{(1)} (E) \; f_a \: \opa_a .
\end{equation}
Note that here we keep only the non-vanishing terms corresponding to the energy of the harmonic bands.
Here $M_{\beta}^{(1)} (E)$ is the \textit{semiclassical} one-photon transition matrix element~\cite{Veniard1996, Maquet1998}, explicitly given in Appendix~\ref{sec:SemiclassicalM1M2}.
The intensity spectrum reads 
\begin{equation}
\label{eq:I2_spectrum}
    \ISpectrum^{(2)} (E) = \sum_{\beta = a,e}  |M_{\beta}^{(1)} (E)|^2 \: f_{\beta}^2  \braket{\opa_{\beta}^{\dagger} \opa_{\beta}} .
\end{equation}
As expected, the photoelectron signal due to absorption of harmonic mode $\beta$ is proportional to the mean number of photons $n_{\beta} = \braket{\opa_{\beta}^{\dagger} \opa_{\beta}}$. In the classical limit, it directly maps to the intensity of the electric field. Because both frameworks describe the main bands via this first-order correlation, their predicted intensities are identical, rendering them indistinguishable regardless of whether a classical or quantum-optical treatment is employed.

\subsubsection{Sidebands: RABBIT oscillations}

The intensity of the sidebands is obtained by considering two-photon processes described by the second-order transition amplitude
\begin{eqnarray}
    && \lim_{\substack{t \to \infty \\ t^{\prime} \to - \infty}} \bra{E}\opU^{(2)} (t,t^{\prime}) \ket{\rm g} = M_e^{(2)} (E) \; f_e f_0 \: \opa_e \aIR^{\dagger} \rme^{-\rmi \wIR \tau} \nonumber \\
    && \qquad \qquad + M_a^{(2)} (E) \; f_a f_0 \: \opa_a \aIR \rme^{\rmi \wIR \tau} , \label{eq:U2_Eg}
\end{eqnarray}
with $M^{(2)}_{\beta} (E)$ the \textit{semiclassical} two-photon transition matrix element associated with the paths $a$ and $e$~\cite{Veniard1996, Maquet1998, Klunder2011, Dahlstrom2013, Busto2019, Berkane2024}, explicitly given in Appendix~\ref{sec:SemiclassicalM1M2}.
We find that the intensity spectrum~\eqref{eq:ISpectrum_n} takes a similar form as in the case of classical light
\begin{equation}
\label{eq:I4}
    \ISpectrum^{(4)}(E ; \tau ) = A (E) \Big\lbrace 1 + C (E) \cos \left[ 2\wIR \tau + \theta (E) \right] \Big\rbrace ,
\end{equation}
with the so-called RABBIT amplitude $A$, contrast $C$ and phase $\theta$.
However, here the coefficients do depend on the quantum characteristics of the light, as the amplitude reads
\begin{subequations}
\label{eq:I4_all}
\begin{eqnarray}
    && A = |M_e^{(2)} (E)|^2 \: f_e^2 f_0^2 \braket{\opa_e^{\dagger}\aIR \aIR^{\dagger} \opa_e} \nonumber \\
    && \qquad \qquad +  |M_a^{(2)} (E)|^2 \: f_a^2 f_0^2  \braket{\opa_a^{\dagger} \aIR^{\dagger}\aIR \opa_a}  , \label{eq:A2}
\end{eqnarray}
the contrast is given by
\begin{equation}
\label{eq:C2}
    C =  \dfrac{2 \: \big| \xi^{(2)} (E) \big| \times \big|  \braket{\opa_e^{\dagger} \opa_a \: \aIR^2} \big|}{\big| \xi^{(2)} (E) \big|^2 \braket{\opa_e^{\dagger} \aIR  \aIR^{\dagger} \opa_e} + \braket{\opa_a^{\dagger} \aIR^{\dagger}\aIR \opa_a}} ,
\end{equation}
and the phase expressed as
\begin{equation}
\label{eq:theta2}
    \theta = - \arg \left[ \xi^{(2)} (E) \right] + \arg \braket{\opa_e^{\dagger}\opa_a \:\aIR^2}  ,
\end{equation}
where the complex-valued coefficient
\begin{equation}
\label{eq:xi2}
    \xi^{(2)} (E) = \dfrac{f_e \:  M_e^{(2)} (E)}{f_a \: M_a^{(2)} (E)} ,
\end{equation}
\end{subequations}
accounts for the ratio of the two-photon transition matrix elements.
Equation~\eqref{eq:theta2} allows us to recover the atomic phase $\arg \xi^{(2)} (E)$ that characterizes the intrinsic response of the atomic target. Both its modulus and phase can be determined numerically through standard second-order perturbation theory~\cite{Paul2001, Muller2002}.
Beyond the atomic response, the RABBIT signal encapsulates the joint photon statistics of the infrared light and the generated harmonics. 
Specifically, its structure and visibility are determined by the following fourth-order correlation functions: $\braket{\opa_e^{\dagger} \aIR \aIR^{\dagger} \opa_e}$, $\braket{\opa_a^{\dagger} \aIR^{\dagger} \aIR \opa_a }$, and $\braket{\opa_e^{\dagger} \opa_a \aIR^2}$. 
Notably, the numerator of the signal contrast is strictly bounded by the terms in the denominator via the Cauchy-Schwarz inequality such that
\begin{equation}
\label{eq:Cauchy_Contrast}
    |\braket{\opa_e^{\dagger} \opa_a \aIR^2}|^2 \leq  \braket{\opa_e^{\dagger} \aIR \aIR^{\dagger} \opa_e} \braket{\opa_a^{\dagger} \aIR^{\dagger} \aIR \opa_a} .
\end{equation}
This inequality implies that the maximum possible contrast is reached only if the equality holds, as shown in Appendix~\ref{sec:Cauchy_Schwartz_derivation}. Consequently, depending on the specific quantum correlations of the light fields, the contrast can significantly diminish or even vanish, even if the individual modes are non-zero.
While traditional RABBIT theory treats the laser and harmonic fields as classical waves with a well-defined phase, our results demonstrate that such hypothesis is not necessary for observing attosecond interference.

\subsection{Coherence and the nature of RABBIT}
\subsubsection{The quantum condition for RABBIT interference}
The most general condition for observing a RABBIT interferometric signal, defined by the oscillation of the photoelectron yield as a function of delay $\tau$ [see Eqs.~\eqref{eq:I4_all}], is the non-vanishing of the four-point correlation function
\begin{equation}
\label{eq:RABBIT_condition}
    \braket{\opa_e^{\dagger}\opa_a \:\aIR^2} \neq 0 .
\end{equation}
This \textit{quantum condition} constitutes the fundamental physical requirement for attosecond beating, and is the central result of our framework.
Under this condition, the phase that can be extracted from the RABBIT signal $\theta = \Delta \phi_A + \Delta \phi$ [Eq.~\eqref{eq:I4}] contains the atomic contribution $\Delta \phi_A = - \arg \xi^{(2)}$ and a field phase given by the argument of the full correlator
\begin{equation}
\label{eq:Delta_phi_quantum}
    \Delta \phi = \arg \braket{\opa_e^{\dagger} \opa_a \aIR^2} ,
\end{equation}
which is the most general expression for the field phase extracted by RABBIT.

Condition~\eqref{eq:RABBIT_condition} and Eq.~\eqref{eq:Delta_phi_quantum} are fully general: they apply regardless of the quantum state of the incoming field and holds for any field statistics, classical or non-classical. In particular, experimental imperfections such as shot-to-shot phase fluctuations and timing jitter are naturally incorporated by treating $\rholight (-\infty)$ as a statistical mixture, reproducing the well-known degradation of fringe contrast encoded in Eqs.~\eqref{eq:C2} and~\eqref{eq:theta2}.
Beyond these technical imperfections, the examples explored in Section~\ref{sec:application} illustrate the full generality of this condition: the RABBIT signal may vanish even when $\braket{\opa_{\beta}} \neq 0$, if the modes are specifically anti-correlated, while conversely a robust signal can persist even when all mean fields vanish ($\braket{\opa_{\beta}} = 0$), as in a BSV source. This establishes that RABBIT fundamentally probes \textit{multi-mode correlations}, not first-order coherences.

\subsubsection{The classical limit as a special case}
Our quantum formalism recovers the standard semiclassical RABBIT results~\cite{Veniard1996, Maquet1998, Klunder2011, Dahlstrom2013, Busto2019, Berkane2024} in the appropriate limit. Specifically, when the incoming electromagnetic field is in a product of coherent states
\begin{equation}
\label{eq:psi_classical}
    \ket{\psilight (-\infty)} = \ket{\alpha_0}_0 \otimes \ket{\alpha_e}_e \otimes \ket{\alpha_a}_a ,
\end{equation}
where $\ket{\alpha_{\beta}}_{\beta}$ denotes the coherent state of mode $\beta$ with well-defined amplitude and carrier-envelope phase, Eqs.~\eqref{eq:I4_all} reduce to the standard semiclassical expressions in the large-photon-number limit ($|\alpha_{\beta}|^2 \gg 1$). We stress that these two conditions -- a product of coherent states \textit{and} a large photon number -- must hold simultaneously to recover the semiclassical limit.

In this limit, the field reduces to its \textit{coherent part}
\begin{equation}
\label{eq:Fcoh}
    \mathbf{F}_{\rm coh} (t ; \tau) = \braket{\mathbf{F}(t ; \tau)} ,
\end{equation}
and all higher-order correlations are fully determined by the first-order expectation $\braket{\opa_{\beta}}$ of each mode. The quantum condition~\eqref{eq:RABBIT_condition} then factorizes as $\braket{\opa_{e}^{\dagger}}\braket{\opa_{a}}\braket{\aIR}^{2}$, reducing to the classical condition
\begin{equation}
    \label{eq:RABBIT_classical_condition}
    \braket{\opa_{\beta}} \neq 0 \quad (\beta = \lbrace 0 , a, e \rbrace ),
\end{equation}
and Eq.~\eqref{eq:Delta_phi_quantum} reduces to the conventional relative harmonic phase~\cite{Paul2001}
\begin{equation}
    \label{eq:RABBIT_classical_condition}
    \Delta\phi = \arg\braket{\opa_a} - \arg\braket{\opa_e} + 2 \arg \braket{\aIR} .
\end{equation}
It is important to note, however, that this classical picture is based on an assumption that is rarely made explicit: even within a \textit{single} realization, each mode is assumed to carry a well-defined amplitude and phase [Eq.~\eqref{eq:psi_classical}], so that all cross-correlations factorize. It concerns the intrinsic quantum nature of the field within each shot. It is precisely this assumption that the quantum treatment calls into question~\cite{Zurek2003, Breuer2007}.

\subsubsection{Attosecond timing}
The distinction between the general phase~\eqref{eq:Delta_phi_quantum} actually extracted by RABBIT and its classical counterpart~\eqref{eq:RABBIT_classical_condition} has direct consequences for the physical interpretation of attosecond timing measurements, such as the landmark measurement of the 250 as extreme-ultraviolet pulse duration~\cite{Paul2001}. The timing of an attosecond pulse is encoded in the phase relations across the full harmonic comb -- it is the coherent superposition of many frequency components that produces a localized pulse in time. 
To focus on this timing specifically, we consider the case where the infrared field $\mathbf{F}_{0} (t;\tau)$ and the total harmonic field $\Fxuv(t)$, defined by $\mathbf{F} (t;\tau) = \mathbf{F}_{0} (t;\tau) + \Fxuv (t)$, are initially uncorrelated. Under this assumption, the correlator for two successive harmonics in Eq.~\eqref{eq:Delta_phi_quantum} reduces to
\begin{equation}
    \Delta\phi = \arg\braket{\opa_e^{\dagger}\opa_a} + \arg \braket{\aIR^2} ,
\end{equation}
where $\braket{\opa_e^{\dagger}\opa_a}$ is one inter-mode correlator between two successive harmonics of the full two-point correlation function $\braket{\Fxuv^{(+)}(t) \cdot \Fxuv^{(-)}(t^{\prime})}$, with $\Fxuv^{(+)}(t)$ [resp. $\Fxuv^{(-)}(t)$] denoting the part of $\Fxuv(t)$ containing only annihilation (resp. creation) operators~\cite{Glauber1963}.
We stress that a single RABBIT measurement, probing only one pair of successive harmonics, gives access to only two-successive harmonic correlators $\braket{\opa_e^{\dagger}\opa_a}$, and therefore provides partial information on $\braket{\Fxuv^{(+)}(t)} \cdot \braket{\Fxuv^{(-)}(t^{\prime})}$, whose full reconstruction requires all inter-mode correlators across the harmonic spectrum, i.e. the complete quantum coherency matrix~\cite{Fabre2020}.
Only in the classical limit, where $\braket{\Fxuv^{(+)}(t) \cdot \Fxuv^{(-)}(t^{\prime})}$ factorizes as $\braket{\Fxuv^{(+)}(t)} \cdot \braket{\Fxuv^{(-)}(t^{\prime})}$ and the measured phase reduces to Eq.~\eqref{eq:RABBIT_classical_condition}, does each correlator $\braket{\opa_e^{\dagger}\opa_a}$ reduce to $\braket{\opa_e^{\dagger}} \braket{\opa_a}$, and does RABBIT give direct access to the timing of the attosecond pulse itself.
However, this interpretation relies on the assumption that the harmonic field is classical, an assumption that is increasingly challenged by recent experimental and theoretical evidence on the quantum nature of high-order harmonic generation~\cite{Gorlach2020, Stammer2024, Gombkoto2021, Stammer2023, Gonoskov2024, Rivera2024, Theidel2024, Theidel2025}. In the general quantum case, it is the temporal structure of $\braket{\Fxuv^{(+)}(t) \cdot \Fxuv^{(-)}(t^{\prime})}$ that is probed -- a subtle but fundamentally different quantity, which can for instance remain sharply localized in time even when $\braket{\Fxuv(t)} = \boldsymbol{0}$, and whose beating reflects the interference of two-photon transition amplitudes mediated by quantum field fluctuations rather than classical waves.

Crucially, this reinterpretation does not invalidate existing experimental protocols, whether for measuring atomic and molecular photoemission delays~\cite{Beaulieu2017, Isinger2017} or for characterizing light pulses~\cite{Paul2001}. In the case of photoemission delay measurements, the RABBIT phase is calibrated against a known reference species, so that the field phase $\arg \braket{\opa_e^{\dagger} \opa_a \aIR^{2}}$ cancels in the differential measurement, leaving only the difference of atomic phases -- a procedure that is insensitive to the quantum state of the field by construction. 
In the case of the characterization of light pulses, a theoretically computed atomic phase is subtracted from the RABBIT phase, directly yielding $\arg \braket{\opa_e^{\dagger} \opa_a \aIR^{2}}$.
What our framework does call into question is not the measurement itself, but its physical interpretation, as discussed above.
Our framework thus complements rather than challenges existing attosecond measurement protocols, and extends their interpretation to non-classical field states, as explored in the following section.

\section{RABBIT signal for different incoming light field states \label{sec:application}}

In this section, we apply the previously developed quantum framework to several characteristic states of light. By evaluating the RABBIT signal [Eq.~\eqref{eq:I4_all}] for different statistical field distributions, we illustrate how specific field correlations, ranging from purely classical mixtures to highly non-classical states, dictate the resulting interferometric amplitude, contrast, and phase. To facilitate this analysis, we introduce a set of normalized correlation functions that serve as a quantitative bridge between the field's quantum statistics and the observed photoelectron beating.

\subsection{Correlated infrared and harmonic fields \label{sec:quantum_correlations}}

We first note that the expressions for the amplitude [see Eq.~\eqref{eq:A2}] and contrast [see Eq.~\eqref{eq:C2}] involve the cross-correlation function
\begin{equation}
\label{eq:g2_beta0}
    g^{(2)}_{i j} = \dfrac{\braket{\opa_i^{\dagger} \opa_j^{\dagger} \opa_j \opa_i}}{\braket{\opa_i^{\dagger}\opa_i} \braket{\opa_j^{\dagger}\opa_j}} ,
\end{equation}
between modes $i$ and $j$. 
This quantity is a standard metric in quantum optics for characterizing the statistical dependence and coherence properties of electromagnetic fields~\cite{Glauber1963}.
Therefore, the signal amplitude, including the denominator of the contrast and the bound defined in Eq.~\eqref{eq:Cauchy_Contrast}, can be directly related, not only to the number of photons of both infrared $\nIR = \braket{\aIR^{\dagger} \aIR}$ and harmonics $n_{\beta}$ modes, but also to the cross-correlated terms $g^{(2)}_{\beta 0}$, for $\beta = \lbrace a,e \rbrace$. 
The case where $g^{(2)}_{\beta 0} = 1$ leads to uncorrelated infrared and harmonic modes and is detailed in the next section, while here we focus on the other cases involving non-trivial correlations.

\subsubsection{Inter-mode bunching}
The condition $g^{(2)}_{\beta 0} > 1$ denotes ``inter-mode bunching'', a regime where the infrared and harmonic modes exhibit strong mutual correlations. 
Classically, this occurs when the intensity of the harmonic $\beta$ tracks the fluctuations of the infrared driving field, a scenario intrinsic to high-order harmonic generation, where higher infrared intensities yield a greater probability of harmonic emission in an optimal regime~\cite{Wahlstrom1993, Weissenbilder2022}. In the quantum regime, however, these correlations can reach significantly higher values, potentially amplifying the signal far beyond the levels expected for uncorrelated light. A quintessential example of this state is the two-mode squeezed vacuum~\cite{Ruo2015}. Similar correlations may be inherently relevant to high-order harmonic generation since the process involves the simultaneous annihilation of infrared photons to produce a single high-energy harmonic photon. Therefore, it could establish a robust statistical link between the driving field and the resulting radiation.

\subsubsection{Inter-mode anti-bunching}
The condition $g^{(2)}_{\beta 0} < 1$ denotes ``inter-mode anti-bunching'', a purely quantum correlation regime. In this case, the joint probability of detecting a photon in the harmonic mode and another in the infrared mode simultaneously is reduced compared to the statistically independent case. 
In the limiting scenario where $g^{(2)}_{\beta 0} = 0$ for $\beta= \lbrace a, e \rbrace$, the harmonic and infrared modes become perfectly anti-correlated. 
This implies that the detection of a photon in one mode precludes the simultaneous detection of a photon in the other, a feature that has no classical counterpart.
Interestingly, the RABBIT amplitude [Eq.~\eqref{eq:A2}] does not vanish due to vacuum-mediated transitions but reduces to
\begin{equation}
    A = |M_e^{2} (E)|^2 f_e^2 f_0^2 \braket{\opa_e^{\dagger} \opa_e} ,
\end{equation}
which depend only on the harmonic $e$ photon number and is completely independent of the infrared photon number $\nIR$.
The asymmetry between the quantum paths involving absorption (a) and stimulated emission (e) is due to the fact that stimulated emission can be induced by vacuum fluctuations, while absorption cannot.
In this regime, the influence of the infrared field is restricted solely to the \textit{semiclassical} two-photon transition matrix elements, effectively rendering the infrared field a ``spectator'' to the interferometric process.
Remarkably, this also places no constraint on the interferometric contrast $C$, which may vanish ($C=0$) or remain finite ($C\neq 0$) depending on the relative phase and amplitude of the contributing pathways.

\subsubsection{Characterization of non-classicality}
Turning to the formal characterization of non-classicality, a primary indicator of quantum correlations between the infrared and harmonic modes is the violation of the classical inequality~\cite{Glauber1963} $g^{(2)}_{\beta 0} \leq [g^{(2)}_{00}g^{(2)}_{\beta\beta} ]^{1/2}$. However, it should be noted that the converse is not necessarily true; the presence of quantum states or entanglement does not strictly guaranty a violation of this bound.
As shown in the following equation, a violation becomes physically accessible only in the regime of low mean photon numbers, $n_{\beta}$ or $\nIR$, since the cross-correlation is bounded by the quantum Cauchy-Schwarz inequality
\begin{equation}
\label{eq:CSg2}
g^{(2)}_{\beta 0} \leq  \left( g^{(2)}_{\beta\beta} + \dfrac{1}{\braket{\opa_{\beta}^{\dagger} \opa_{\beta}}}\right)^{1/2} \left( g^{(2)}_{00} + \dfrac{1}{\braket{\aIR^{\dagger} \aIR}} \right)^{1/2}  ,
\end{equation}
resulting from commutation relations applied to~\eqref{eq:g2_beta0}.
The terms $1/\braket{\opa_{\beta}^{\dagger} \opa_{\beta}}$ and $1/\braket{\aIR^{\dagger} \aIR}$ represent the quantum correction that allows the cross-correlation to exceed the classical limits. This provides a strong motivation for experimental investigations of RABBIT in low-photon regimes at high repetition rates, as achievable with cavity-enhanced high-order harmonic generation~\cite{Heinrich2021}, where non-classical signatures could be most clearly resolved.

\subsection{Uncorrelated infrared and harmonic fields \label{sec:uncorrelated_light}}

We now consider the case where the infrared field state $\rho_0 (-\infty)$ and the harmonic field state $\rho_{ea} (-\infty)$ including two frequency modes, $\omega_e$ and $\omega_a$, are initially separated in the overall field state
\begin{equation}
\label{eq:rho_rho0_rhoea}
    \rholight (-\infty) = \rho_0 (-\infty) \otimes \rho_{ea} (-\infty) .
\end{equation}
In this case, the harmonic and infrared lights are initially uncorrelated, which formally translates into
\begin{subequations}
    \label{eq:uncorrelated_light}
\begin{eqnarray}
    && \braket{\opa_e^{\dagger} \opa_a \: \aIR^2} = \braket{\opa_e^{\dagger} \opa_a} \braket{\aIR^2} , \\
    && 
    g^{(2)}_{\beta 0} = 1,
\end{eqnarray}
\end{subequations}
for $\beta = \lbrace a , e \rbrace$. In addition, we consider that the number of infrared photons is large, such that $\braket{\aIR \aIR^{\dagger}} \approx \braket{\aIR^{\dagger} \aIR}$ which is in general the case in RABBIT experiments. 
Note that a large number of photons does not necessarily imply a classical limit since both the auto-correlations in $\rho_{0}$ or the cross-correlations in $\rho_{ea}$ can be non-classical.

\subsubsection{Reduced expressions}

The amplitude~\eqref{eq:A2}, contrast~\eqref{eq:C2} and phase~\eqref{eq:theta2} thus reduce to
\begin{subequations}
\begin{eqnarray}
    && A \approx f_0^2 \braket{\aIR^{\dagger}\aIR} \nonumber \\ 
    && \qquad \times  \left\lbrace   |M_e^{(2)} (E)|^2 \: f_e^2 \braket{\opa_e^{\dagger}\opa_e} + |M_a^{(2)} (E)|^2  \: f_a^2 \braket{\opa_a^{\dagger}\opa_a} \right\rbrace , \nonumber \\ \label{eq:A2_uncorrelated}
\end{eqnarray}
\begin{equation}
\label{eq:C2_uncorrelated}
    C \approx  \dfrac{2 \big| \xi^{(2)} (E) \big| \mu}{ \left\lbrace \big| \xi^{(2)} (E) \big| \mu \right\rbrace^2 + 1} \times \dfrac{\big|  \braket{\opa_e^{\dagger} \opa_a} \big|}{\sqrt{\braket{\opa_e^{\dagger}\opa_e} \braket{\opa_a^{\dagger}\opa_a}}} \times \dfrac{| \braket{\aIR^2} |}{\braket{\aIR^{\dagger} \aIR}} ,
\end{equation}
and 
\begin{equation}
\label{eq:theta2_uncorrelated}
    \theta =  - \arg \left[ \xi^{(2)} (E) \right]  + \arg \braket{\opa_e^{\dagger}\opa_a}  + \arg  \braket{\aIR^2}  .
\end{equation}
\end{subequations}
We have introduced in Eq.~\eqref{eq:C2_uncorrelated} the ratio between the square root of the number of photons of the two harmonics
\begin{equation}
\label{eq:mu}
    \mu = \sqrt{\dfrac{\braket{\opa_e^{\dagger}\opa_e}}{\braket{\opa_a^{\dagger}\opa_a}}},
\end{equation}
that can be fully determined from the intensity of the main bands $\ISpectrum^{(2)}(E)$ [see Eq.~\eqref{eq:I2_spectrum}]. 
As a consequence of separability~\eqref{eq:rho_rho0_rhoea}, we observe that the contrast in Eq.~\eqref{eq:C2_uncorrelated} and the phase in Eq.~\eqref{eq:theta2_uncorrelated} separate into three contributions. 
The last two terms depend only on the properties of light (fundamental and harmonics) and their correlations.

The contrast and phase of the RABBIT signal encode the modulus and phase of the normalized first-order coherence function~\cite{Mandel1995}
\begin{equation}
\label{eq:g1_ea}
    g^{(1)}_{ea} = \dfrac{\braket{\opa_e^{\dagger} \opa_a}}{\sqrt{\braket{\opa_e^{\dagger}\opa_e} \braket{\opa_a^{\dagger}\opa_a}}} ,
\end{equation}
between the two harmonic modes.
The modulus of $g^{(1)}_{ea}$ ranges from 0 to 1, where a value $|g^{(1)}_{ea}|=0$ indicates a complete lack of coherence between the two harmonic modes $e$ and $a$, and a value $|g^{(1)}_{ea}|=1$ represents perfect coherence between them.
Hence, the contrast can vanish if there is no coherence between the modes $a$ and $e$, which is equivalent to saying that there is no phase relation between these modes. 

\subsubsection{Examples of harmonic fields}
In the following examples, we specify the initial state of the harmonic field before the light-matter interaction. Whenever this state is pure, the corresponding density matrix is $\rho_{ea}(-\infty) = \ket{\psi_{ea}(-\infty)}\bra{\psi_{ea}(-\infty)}$.

\vspace{0.5cm}
\paragraph{Macroscopic Bell-like states:}
Several physical scenarios result in the disappearance of first-order coherence ($g^{(1)}_{ea} = 0$), thus suppressing the RABBIT beating. 
For example, it can emerge from non-local quantum correlations, even when individual modes appear perfectly coherent. A compelling example is the entangled, Bell-like state
\begin{equation}
\label{eq:Bell_state_CS}
    \ket{\psi_{ea}  (-\infty)} = \dfrac{1}{\cal N} \big( \ket{\alpha_e}_e \otimes \ket{0}_a + \ket{0}_e \otimes \ket{\alpha_a}_a \big) ,
\end{equation}
where $\cal N$ is a normalization factor.
Here, each isolated mode possesses a well-defined coherent phase, as evidenced by the non-vanishing local expectation values $\braket{\opa_e} = \alpha_e / {\cal N}$ and $\braket{\opa_a} = \alpha_a / {\cal N}$, which map to a well-defined phase difference~\eqref{eq:RABBIT_classical_condition}.
Yet, because the state is a coherent superposition of mutually exclusive macro-realizations, photons exist either entirely in mode $e$ or entirely in mode $a$, i.e. the two harmonics are anti-correlated. Consequently, the coherence function between the modes
\begin{equation}
    g^{(1)}_{ea} = \dfrac{\alpha_e^{\star} \alpha_a}{|\alpha_e^{\star} \alpha_a|} \exp \left[ - \frac{1}{2} \left( |\alpha_e|^2 + |\alpha_a|^2 \right) \right] ,
\end{equation}
exhibits a striking dependence on photon number: in the few-photon limit ($|\alpha_e|^2$ and $|\alpha_a|^2 \ll 1$), $|g^{(1)}_{ea}| \to 1$, corresponding to maximal RABBIT contrast, while it drops to zero as the number of photons increases ($|\alpha_e|^2$ or $|\alpha_a|^2 \gg 1$) completely suppressing the interference signal. Unlike a standard product of coherent states [see Eq.~\eqref{eq:psi_classical}], which lacks inter-mode entanglement, this Bell-like state represents a purely quantum mechanism for contrast loss.

\vspace{0.5cm}
\paragraph{Two photons in one frequency mode:}
Similar suppression can occur through classical phase-averaging from independent, phase-randomized coherent sources. To isolate the microscopic roles of phase uncertainty and particle-number constraints in eliminating this interference, we consider a simplified model within a minimal basis of Fock-state $\lbrace \ket{0} , \ket{1} \rbrace$.
We first examine a state containing exactly one photon in each mode
\begin{equation}
\label{eq:2photon_ea}
    \ket{\psi_{ea}  (-\infty)} = \ket{1}_e \otimes \ket{1}_a .
\end{equation}
For this state, the first-order coherence term vanishes, $\braket{\opa_e^{\dagger} \opa_a} = 0$, leading to $g_{ea}^{(1)}=0$ and a resulting RABBIT contrast of zero. This occurs because Fock states are eigenstates of the number operator, thus possessing a definite particle count but containing no information regarding the relative phase between the two modes. Mathematically, the annihilation of a photon in mode $a$ and the creation of a photon in mode $e$, result in a state (or similar) which is orthogonal to the initial state, thereby ``erasing'' any phase relationship that would otherwise drive the interferometric beating.
Note that Eq.~\eqref{eq:2photon_ea} is fundamentally distinct from the product state in Eq.~\eqref{eq:psi_classical}. Although coherent states are superpositions of Fock states, they lack the inter-mode correlations, or entanglement, characterizing Eq.~\eqref{eq:2photon_ea}.

\vspace{0.5cm}
\paragraph{Single photon with classical correlations (incoherent):}
Similarly, the interference contrast vanishes for a statistical mixture where a single photon is randomly distributed between the two modes
\begin{equation}
    \rho_{ea} (-\infty) = \dfrac{1}{2} \big( \ket{1}_e \bra{1}_e \otimes \ket{0}_a \bra{0}_a + \ket{0}_e \bra{0}_e \otimes \ket{1}_a \bra{1}_a \big) .
\end{equation}
In this case, the lack of off-diagonal terms in the density matrix, representing a complete absence of coherence between modes, prevents the formation of the sideband interference. More generally, $g_{ea}^{(1)}=0$ whenever the density matrix is diagonal in the Fock state basis, a condition that means that the relative phase between the pathways is undefined. 
This is also the case if, more generally, the occupied state is a coherent state $\ket{\alpha}$ instead of $\ket{1}$.

\vspace{0.5cm}
\paragraph{Single photon with quantum correlations (entanglement):}
Conversely, complex quantum correlations can produce a robust RABBIT signal even in non-trivial configurations. Consider the path-entangled Fock state
\begin{equation}
\label{eq:Bell-state_01}
    \ket{\psi_{ea} (-\infty)} = \dfrac{1}{\sqrt{2}} \big( \ket{1}_e \otimes \ket{0}_a + \ket{0}_e \otimes \ket{1}_a \big) .
\end{equation}
In this configuration, the individual field expectations vanish ($\braket{\opa_e} = 0$ and $\braket{\opa_a}=0$), reflecting the ill-defined phase inherent to Fock states. Nevertheless, the state remains perfectly coherent ($g^{(1)}_{ea} = 1$), as a single photon is shared in a coherent superposition between the two harmonics.
This example serves as a critical counterexample to classical intuition: it demonstrates that measuring a constant phase between consecutive harmonics via RABBIT does not necessarily imply that the phase difference~\eqref{eq:RABBIT_classical_condition} between the underlying coherent components of the fields is fixed, nor even that a \textit{semiclassical} phase is well-defined. More generally, the `zoology' of possibilities afforded by quantum light is remarkably vast.

\vspace{0.5cm}
\paragraph{Summary:}
In sum, we have shown that \textit{inter-mode coherence} between the fields of two successive harmonics, quantified in this case by $g_{ea}^{(1)}\neq 0$, is necessary to obtain interference in the RABBIT signal. This coherence can arise from fundamentally different physical mechanisms. In the classical case, such as a product of coherent states~\eqref{eq:psi_classical}, it arises from a well-defined phase carried by each individual mode, and is macroscopic. 
In the quantum case, inter-mode coherence can emerge from entanglement alone, even in the absence of a well-defined phase for each individual mode, as demonstrated by the two few-photon examples [see Eqs.~\eqref{eq:Bell_state_CS} and~\eqref{eq:Bell-state_01}]. Interestingly, this shows that single-mode coherence is not required -- it is the inter-mode coherence, whether classical or quantum, that governs the RABBIT contrast.

This distinction may play a significant role in the few-photon regime when comparing harmonics generated from free-electron lasers~\cite{Berrah2025} and high-order harmonic generation~\cite{Ferray1988}, which arise from inherently different mechanisms: free-electron lasers generate different frequency modes independently, whereas high-order harmonic generation is an intrinsically coherent quantum process.

Finally, the last term in Eq.~\eqref{eq:C2_uncorrelated} depends exclusively on the infrared mode.  
Specifically, the RABBIT oscillation vanishes if $\braket{\aIR^2}=0$, a condition met by thermal chaotic light, phase-diffused states, or Fock states. In these cases, the lack of intrinsic phase coherence in the infrared field destroys the interferometric signal. In contrast, $\braket{\aIR^2}$ is non-zero for a broader class of states, most notably squeezed coherent states, which includes both the BSV and standard coherent states. This motivates a closer examination of how the specific statistics of these states influence the visibility of the sideband.

\subsection{Infrared in a squeezed coherent state and classical harmonics \label{sec:classicalXUV}}

To isolate the infrared contributions, we consider the regime in which the generated harmonics are effectively classical and perfectly coherent. In addition to the general conditions for uncorrelated light in Eq.~\eqref{eq:uncorrelated_light}, we assume the harmonics possess a well-defined amplitude and perfect mutual coherence, such that
\begin{subequations}
\begin{eqnarray}
    && F_{\beta}^2 = 4  f_{\beta}^2\braket{\opa_{\beta}^{\dagger}\opa_{\beta}} , \\
    && g_{ea}^{(1)} = 1 ,
\end{eqnarray}
\end{subequations}
for $\beta=\lbrace a , e \rbrace$. 

\subsubsection{Reduced expressions}
Under these assumptions, the field ratio reduces to $\mu = (F_e f_a)/(F_a f_e)$, and the general expressions for the amplitude~\eqref{eq:A2_uncorrelated}, contrast~\eqref{eq:C2_uncorrelated} and phase~\eqref{eq:theta2_uncorrelated} simplify to
\begin{subequations}
\label{eq:I4_classicalXUV}
\begin{equation}
\label{eq:A2_classicalXUV}
    A \approx \dfrac{1}{4} f_0^2 \braket{\aIR^{\dagger}\aIR}  \left\lbrace   |M_e^{(2)} (E)|^2 \: F_e^2 + |M_a^{(2)} (E)|^2 \: F_a^2  \right\rbrace , 
\end{equation}
\begin{equation}
\label{eq:C2_classicalXUV}
    C \approx  \dfrac{2 \big| \xi^{(2)} (E) \big| \mu}{ \left\lbrace \big| \xi^{(2)} (E) \big| \mu \right\rbrace^2 + 1} \times \dfrac{| \braket{\aIR^2} |}{\braket{\aIR^{\dagger} \aIR}} ,
\end{equation}
and 
\begin{equation}
\label{eq:theta2_classicalXUV}
    \theta =  - \arg \left[ \xi^{(2)} (E) \right]  + \arg \braket{\aIR^2} .
\end{equation}
\end{subequations}
Note that they no longer depend on $f_e$ and $f_a$ since they simplify in the product $| \xi^{(2)} (E) | \mu$, and do not contribute in $\arg [ \xi^{(2)} (E) ]$.
However, they still depend on auto-correlations of infrared light through $\braket{\aIR^{\dagger} \aIR}$ and $\braket{\aIR^2}$.

\subsubsection{Expressions for squeezed coherent states}
Interestingly, for a squeezed coherent state characterized by the squeezing parameter $\zeta = r \rme^{\rmi \varphi}$ and the displacement $\alpha_0$, these two-point correlation functions do not factorize as for coherent states.
The initial state can be formally written as
\begin{subequations}
\begin{equation}
\label{eq:squeezed_coherent_state}
    \ket{\psi_0  (-\infty)}  = \opD (\alphaIR) \; \opS (\zeta) \ket{0} ,
\end{equation}
in the convention that displacement
\begin{equation}
    \opD (\alpha) = \rme^{\alpha \aIR^{\dagger} - \alpha^{\star} \aIR} ,
\end{equation}
acts after squeezing 
\begin{equation}
    \opS (\zeta) = \rme^{\frac{1}{2} \left[ \zeta^{\star} \aIR^2 - \zeta (\aIR^{\dagger})^2 \right]} .
\end{equation}
\end{subequations}
We recall that the expectation values of the correlation amplitudes for such fields are
\begin{subequations}
\label{eq:BSV_properties}
\begin{eqnarray}
    && \braket{\aIR} = \alphaIR , \\
    && \braket{\aIR^2} = \alphaIR^2 - \dfrac{\rme^{\rmi \varphi}}{2} \sinh (2 r )  , \label{eq:a2_BSV} \\ 
    && \braket{\aIR^{\dagger} \aIR} = |\alphaIR |^2 + \sinh ( r )^2  , \label{eq:adaggera_SCS}
\end{eqnarray}
\end{subequations}
with $\nIR = \braket{\aIR^{\dagger} \aIR}$ the total number of infrared photons. 
We also introduce the squeezing ratio $\ratioIR \in [0,1]$ defined from the squeezing strength as 
\begin{equation}
\label{eq:ratioIR}
    \ratioIR = \dfrac{\sinh (r)^2}{\nIR} ,
\end{equation}
that fully determines the modulus of displacement through $|\alpha_0|^2 = (1-\ratioIR) \nIR$ and allows us to consider various squeezed states at a constant number of photons.
Note that $\ratioIR = 0$ and  $\ratioIR = 1$ correspond to the coherent state and the BSV limit~\cite{Spasibko2017, Manceau2019, Heimerl2025}, respectively.
In Figs.~\ref{fig:delays_RABBIT}, \ref{fig:energy_RABBIT}, \ref{fig:ratio_RABBIT} and~\ref{fig:sidebands_squeezed_coherent_state}, we fix the number of photons to $\nIR = 10^{16}$ and $\arg \alphaIR = 0$, and as a consequence, the initial Wigner distribution is centered on the positive real axis (see Fig.~\ref{fig:squeezed_light}). 
The infrared field utilizes a vacuum amplitude $f_0 = 10^{-12}$ a.u. corresponding to a quantization volume of $5 \; \rm mm^3$ in~\eqref{eq:gk}.
This photon flux corresponds to an intensity of approximately $10^{9} \; \rm W \ cm^{-2}$, placing the interaction firmly within the regime of second-order perturbation theory. Notably, this remains valid even though the fluctuations of the squeezed coherent state can span several orders of magnitude.
In this regime, even a modest squeezing parameter results in a highly eccentric Wigner distribution, and the phase-space ellipse becomes sufficiently elongated to appear essentially as a line.

\subsubsection{Numerical methods}
We employ two distinct numerical approaches to validate our theoretical model. The first one is based on TDPT, where we evaluate the semiclassical transition matrix elements and incorporate the auto-correlations of infrared light in~\eqref{eq:BSV_properties}. The second approach consists of ab initio simulations of the TDSE.
For our numerical applications, we consider linearly polarized fields and a one-dimensional Hydrogen atom modeled by a soft-Coulomb potential~\cite{Javanainen1988}
\begin{equation}
\label{eq:HA_H_application}
    \opH_{A} = - \dfrac{\Delta}{2} - \dfrac{1}{\sqrt{x^2+2}} ,
\end{equation}
which leads to a ground-state energy $E_{\rm g} = -0.5$ a.u.
The infrared wavelength is set to 800~nm ($\wIR \approx 0.057$ a.u.) and the harmonic intensity is taken as $10^{7} \; \rm W\ cm^{-2}$ (corresponding to field amplitudes $F_{\beta} \approx 1.7 \times 10^{-5}$ a.u. for $\beta=\lbrace a, e \rbrace$).
We include harmonic orders $q =  \lbrace 9,11,\ldots ,25\rbrace$.
The corresponding \textit{semiclassical} two-photon transition matrix elements for the potential in Eq.~\eqref{eq:HA_H_application} are summarized in Table~\ref{tab:M2_RABBIT} as a function of the sideband energy $E = E_{\rm g} + q \, \wIR$.

To obtain the intensity spectra ab initio, we perform an incoherent summation of the spectra generated by numerically solving the TDSE. These simulations utilize an effective classical infrared field weighted by its Wigner distribution as detailed in the Appendices~\ref{sec:Wigner} and~\ref{sec:TDSE}. The validity of computing the total intensity as an incoherent sum of semiclassical simulations stems from the fact that the spectrum represents a direct electron count at specific energies. This approach is further enabled by the strict positivity of the Wigner distribution, a property characteristic of the squeezed coherent states considered here, which allows for a well-defined probabilistic weighting of classical trajectories, as previously noted in~\cite{Heimerl2025}.
The TDSE was solved using high-order split-operator methods~\cite{McLachlan2022}~\footnote{The underlying code is available in the GitHub repository~\url{https://github.com/jonathan-dubois/TDSE_ND_MPI}}. Our simulations were performed on a spatial grid of $2^{16}$ nodes with a spacing of $0.2$ a.u. and a time step of $0.1$ a.u.
As demonstrated in Fig.~\ref{fig:delays_RABBIT} this method reproduces with high fidelity the RABBIT signal in standard form~\eqref{eq:I4}, reminiscent of the one obtained in \textit{semiclassical} simulations and experiments~\cite{Veniard1996, Maquet1998, Klunder2011, Dahlstrom2013, Busto2019, Berkane2024}.
Furthermore, the excellent agreement between the TDSE results (circles) and the TDPT predictions (lines) in Figs.~\ref{fig:delays_RABBIT}, \ref{fig:energy_RABBIT}, and~\ref{fig:ratio_RABBIT} validates our perturbative framework and confirms its reliability for extending RABBIT analysis into the quantum light regime.

\begin{table}
    \newcommand{\complexnum}[2]{#1 \; \exp( #2 \; \rmi )}
    \centering
    \begin{tabular}{|c||c|c|}
        $E$ (in eV) & $M_a^{(2)} (E)$ & $M_e^{(2)} (E)$ \\
        \hline
         1.892 & $\complexnum{86.8374}{0.3692}$ & $\complexnum{56.7437}{-0.3165}$  \\
         4.990 & $\complexnum{79.2683}{0.2293}$ & $\complexnum{55.1106}{-0.2117}$ \\
         8.088 & $\complexnum{67.3156}{0.1585}$ & $\complexnum{48.4818}{-0.1542}$ \\
         11.19 & $\complexnum{55.9270}{0.1192}$ & $\complexnum{41.6357}{-0.1188}$ \\
         14.29 & $\complexnum{46.4232}{0.0944}$ & $\complexnum{35.4227}{- 0.0953}$ \\
         17.38 & $\complexnum{38.9820}{0.769}$ & $\complexnum{30.1262}{- 0.0788}$ \\
         20.48 & $\complexnum{32.7244}{0.0647}$ & $\complexnum{25.7704}{- 0.0666}$ \\
         23.58 & $\complexnum{27.5185}{0.0560}$ & $\complexnum{22.1855}{-0.0571}$ \\
    \end{tabular}
    \caption{Values of the \textit{semiclassical} two-photon transition matrices as a function of the sideband energy $E$ for H in one dimension and a laser wavelength of 800 nm (corresponding to $\wIR\approx 0.057$) computed numerically~\cite{Paul2001, Muller2002}.}
    \label{tab:M2_RABBIT}
\end{table}

\subsubsection{Analysis of RABBIT spectra}

\begin{figure}[htb!]
    \centering
    \includegraphics[width=.45\textwidth]{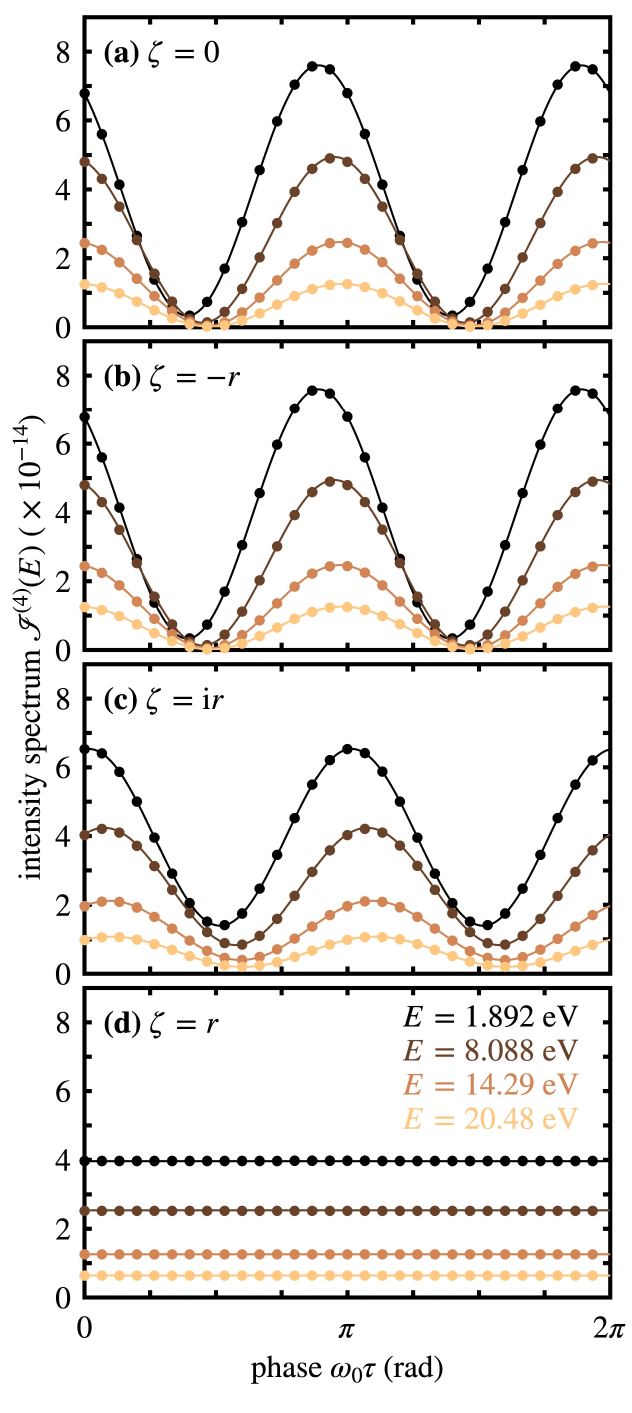}
    \caption{Comparison of the intensity spectrum obtained from the TDPT (solid lines) [Eq.~\eqref{eq:I4}] and the TDSE (colored circles). Note that all the solid curves are scaled by $0.7729$ in order to take into account the finite-pulse duration. The number of photons is $\nIR = 10^{16}$ and $\arg \alphaIR = 0$. The different subplots correspond to different squeezing amplitudes: $\ratioIR = 0$ in (a), corresponding to a coherent state with $\zeta = 0$, and $\ratioIR = 1/2$ in subsequent cases (corresponding to $r \sim 19$). (b) $\varphi = \pi$ corresponding to a phase-squeezed state ($\zeta = -r$), (c) $\varphi = \pi/2$ corresponding to an intermediate phase ($\zeta = \rmi \; r$), and (d) $\varphi=0$ corresponding to an amplitude-squeezed state ($\zeta = r$).}
    \label{fig:delays_RABBIT}
\end{figure}

\begin{figure}
    \centering
    \includegraphics[width=.45\textwidth]{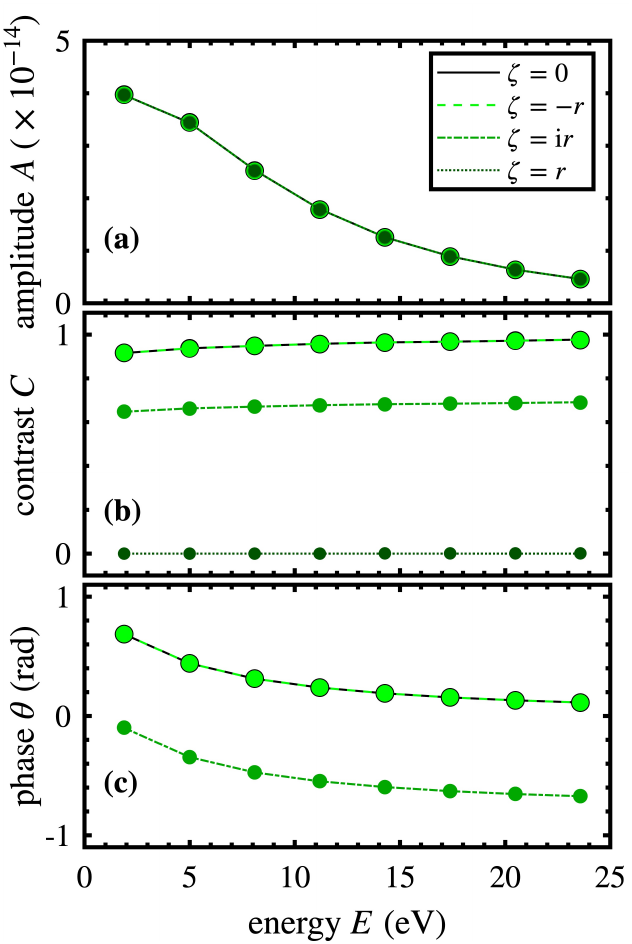}
    \caption{Comparison of the (a) amplitude, (b) contrast and (c) phase of the RABBIT signal as a function of the sideband energy. The solid lines are obtained from the TDPT [Eq.~\eqref{eq:I4}] and the colored circles from the TDSE.
    The parameters are the same as in Fig.~\ref{fig:delays_RABBIT}, i.e. $r \sim 19$, the number of photons is $\nIR = 10^{16}$ and $\arg \alphaIR = 0$. 
    Note that in (a) the solid curves were scaled by $0.7729$, as in Fig.~\ref{fig:delays_RABBIT}.
    In (c) the phase is not displayed for $\zeta=r$ since the contrast is zero.}
    \label{fig:energy_RABBIT}
\end{figure}

We begin by examining four representative cases in Figs.~\ref{fig:delays_RABBIT}, which show RABBIT spectra as a function of the phase $\wIR \tau$ for four values of sideband energies. 
Case (a) corresponds to a standard coherent state ($\ratioIR = 0$), i.e. the regime of classical light usually studied in attosecond science~\cite{Veniard1996, Maquet1998, Klunder2011, Dahlstrom2013, Busto2019, Berkane2024}. 
In subsequent cases, the squeezing ratio is fixed at $\ratioIR = 1/2$ (corresponding to $r \sim 19$) to explore the impact of the squeezing phase: $\varphi = \pi$ in (b) represents a \textit{phase-squeezed state} ($\zeta = -r$), $\varphi = \pi /2$ in (c) represents a state with intermediate squeezing ($\zeta = \rmi r$), and $\varphi=0$ in (d) corresponds to an \textit{amplitude-squeezed state} ($\zeta = r$).
For these parameters of light,
Fig.~\ref{fig:energy_RABBIT} shows the amplitude, contrast, and phase of the RABBIT spectrum as a function of all the sideband energies that have been considered.

We first observe in Fig.~\ref{fig:delays_RABBIT}a that we recover the standard patterns of classical light, namely that the contrasts increase for increasing energies (see also Fig.~\ref{fig:energy_RABBIT}b) and the phase decreases to zero (see also Fig.~\ref{fig:energy_RABBIT}c) as a result of $|\xi^{(2)} (E)|$ tending to the value 1 (see Tab.~\ref{tab:M2_RABBIT}). 
Similarly, the amplitudes of the RABBIT oscillations decrease with increasing energies, since both $|M^{(2)}_{\beta} (E)|$ decrease for $\beta=e$ and $a$ (see Tab.~\ref{tab:M2_RABBIT}) and the intensity of the harmonic fields is the same for all $q$ in Eq.~\eqref{eq:Fht}.
Moreover, Fig.~\ref{fig:delays_RABBIT}a and Fig.~\ref{fig:delays_RABBIT}b are indistinguishable, reflecting the fact that phase-squeezed states do not affect the RABBIT spectrum. 
This can be understood by considering that the squeezed coherent state encompasses a distribution of ``classical'' field components with equal phases, each contributing equally to the RABBIT phase.
As a consequence, the global contrast and phase are unchanged, as can also be observed in Fig.~\ref{fig:energy_RABBIT}.
This can also be easily understood from the TDPT perspective [see Eqs.~\eqref{eq:I4_classicalXUV}] using the following simplification
\begin{equation}
\label{eq:a02_simple}
    \braket{\aIR^2} \approx \braket{\aIR^{\dagger} \aIR} \left[ (1- \ratioIR ) \rme^{2\rmi \arg \alphaIR} - \ratioIR \rme^{\rmi \varphi} \right] ,
\end{equation}
which is valid in the limit of large squeezing amplitude ($\ratioIR\nIR \gg 1$) for which $\sinh r \approx \rme^r/2$. 
We observe in Eq.~\eqref{eq:a02_simple} that $\braket{\aIR^2} \approx \braket{\aIR^{\dagger} \aIR}$ for $\varphi = \pi + 2 \arg \alphaIR$ (modulo $2\pi$), which reduces to the case of coherent states.

In the intermediate regime where $\varphi = \pi/2$, Fig.~\ref{fig:delays_RABBIT}c show a decrease in contrast (see also Fig.~\ref{fig:energy_RABBIT}b) and a decrease in phase (see also Fig.~\ref{fig:energy_RABBIT}c).
This can be understood by considering that the squeezed coherent state encompasses a distribution of ``classical'' field components with varying phases, each contributing differently to the RABBIT phase. The interference of these varying phase contributions leads to an effective averaging, which results in the observed reduction of the signal contrast and modification of the phase.
This decrease can be such that the oscillations disappear, as observed in Fig.~\ref{fig:delays_RABBIT}d and Fig.~\ref{fig:energy_RABBIT}b for the amplitude-squeezed state.
This is the case for $\braket{\aIR^2} = 0$ [see Eq.~\eqref{eq:C2_classicalXUV}], which occurs in general for $\ratioIR = 1/2$ and $\varphi = 2 \arg \alphaIR$ (modulo $2\pi$) [see Eq.~\eqref{eq:a02_simple}].

\begin{figure}
    \centering
    \includegraphics[width=.45\textwidth]{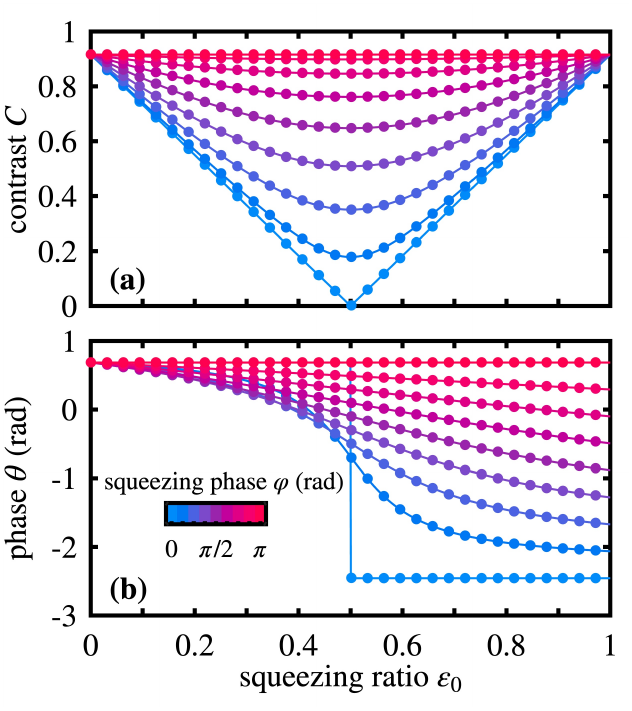}
    \caption{RABBIT (a) contrast and (b) phase as a function of the squeezing ratio $\ratioIR$ [see Eq.~\eqref{eq:ratioIR}] and for different squeezing phases $\varphi$ (in rad), for the sideband energy $E= 1.892$ eV. The number of photons is $\nIR = 10^{16}$ and $\arg \alphaIR = 0$.
    The results are for the analytical results (solid lines) [see Eqs.~\eqref{eq:C2} and~\eqref{eq:theta2}] and from the one-dimensional TDSE simulations (colored circles). Note that all curves from the perturbation theory are scaled by $0.7729$ in order to take into account the finite-pulse duration.}
    \label{fig:ratio_RABBIT}
\end{figure}

We now turn to a more systematic study of the influence of the squeezing amplitude.
Figure~\ref{fig:ratio_RABBIT} shows the contrast and phase of the sideband 10 at energy $E = 1.892$ eV as a function of the squeezing ratio $\ratioIR$ for different values of the squeezing phase $\varphi$. 
As observed earlier, both the contrast (see Fig.~\ref{fig:ratio_RABBIT}a) and the phase (see Fig.~\ref{fig:ratio_RABBIT}b) are unchanged for $\varphi = \pi$, and thus, regardless of the value of $\ratioIR$.
For $\varphi \neq \pi$, the contrast exhibits a non-monotonic dependence on the intensity ratio $\ratioIR$. Specifically, the contrast decreases for $\ratioIR \in [0,1/2]$, reaching a minimum at $\ratioIR=1/2$, before recovering as $\ratioIR$ increases toward 1.
The disappearance of the RABBIT oscillations occurs only for the case represented in Fig.~\ref{fig:delays_RABBIT}d.
For amplitude-squeezed states ($\varphi = 2\arg \alpha_0$), 
the transition from $\ratioIR < 1/2$ to $\ratioIR > 1/2$ is accompanied by a $\pi$-jump in the phase of the RABBIT signal (see Fig.~\ref{fig:ratio_RABBIT}b). 
It is clear that this comes from a sign change from the term on the right-hand side of Eq.~\eqref{eq:a02_simple}.

In the BSV limit~\cite{Spasibko2017, Manceau2019, Heimerl2025} ($\ratioIR = 1$), we observe that RABBIT contrast is the same as for the classical limit, but the RABBIT phase follows the value of $\varphi$. Hence, the role played by the squeezing phase is equivalent to that played by $\tau$ in traditional RABBIT.

\begin{figure*}
    \centering
    \includegraphics[width=.9\textwidth]{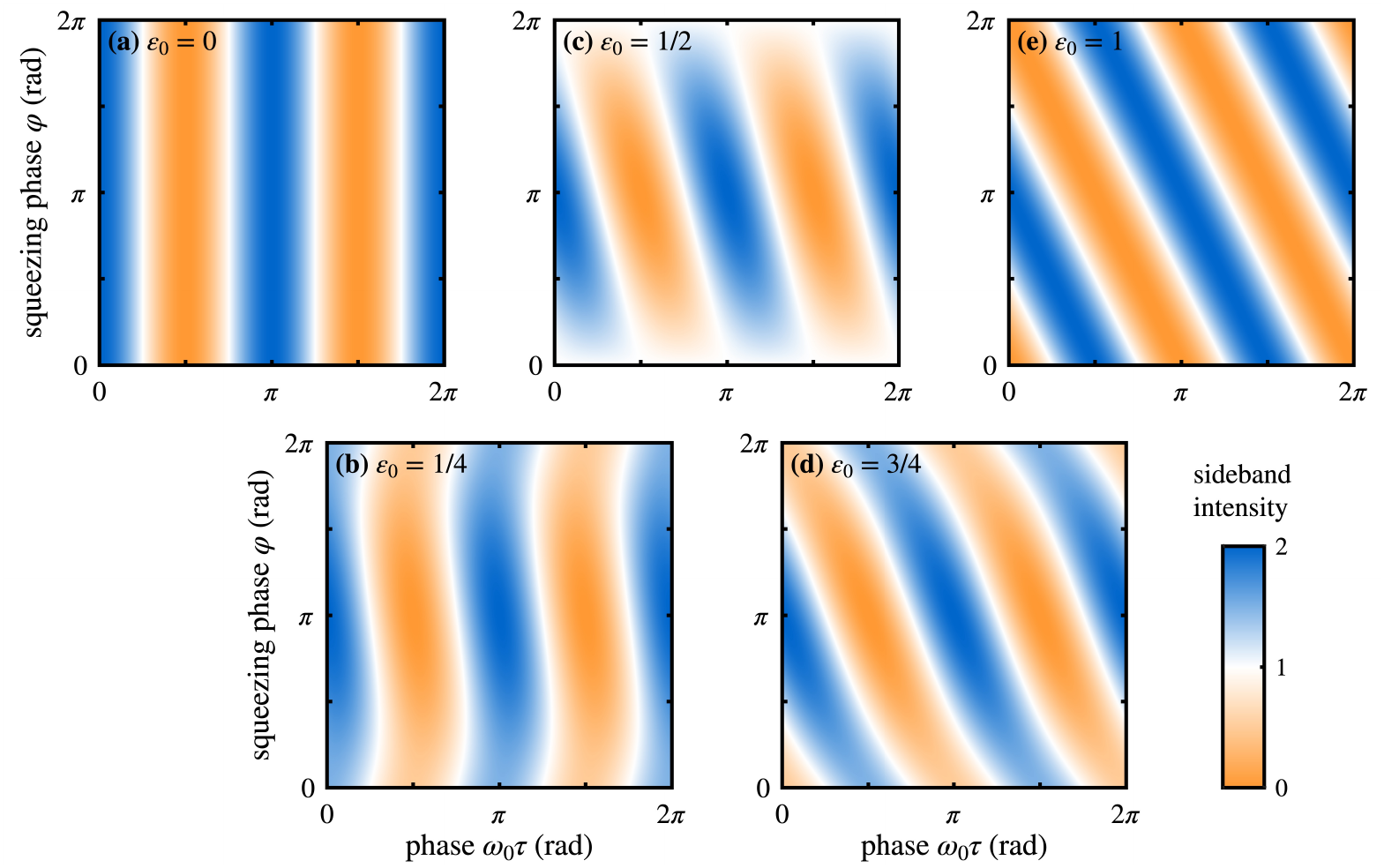}
    \caption{Intensity of the sidebands scaled by the amplitude $\ISpectrum^{(4)} / A$ for a linearly polarized squeezed coherent state [see Eq.~\eqref{eq:squeezed_coherent_state}] as a function of the phase delay $\wIR \tau$ and the squeezing phase $\varphi$. The second-order coefficient is taken as $\xi^{(2)} = 1$. 
    The number of photons is $\nIR = 10^{16}$ and $\arg \alphaIR = 0$. 
    The squeezing ratio is (a) $\ratioIR = 0$, (b) $\ratioIR = 1/4$, (c) $\ratioIR = 1/2$, (d) $\ratioIR = 3/4$, and (e) $\ratioIR = 1$.} \label{fig:sidebands_squeezed_coherent_state}
\end{figure*}

For a more visual summary of the above conclusions, Fig.~\ref{fig:sidebands_squeezed_coherent_state} shows the intensity of the sidebands scaled by the amplitude $\ISpectrum^{(4)} / A$, where we have set $\xi^{(2)} = 1$, as a function of the phase delay $\omega \tau$ and the squeezing phase $\varphi$.
The different subplots are for different values of $\ratioIR$ for $\nIR = 10^{16}$ and $\arg \alpha_0 = 0$.

\subsubsection{Beyond the second-order perturbative regime \label{sec:classicalXUV_nonperturbative}}

\begin{figure}[htb!]
    \centering
    \includegraphics[width=.45\textwidth]{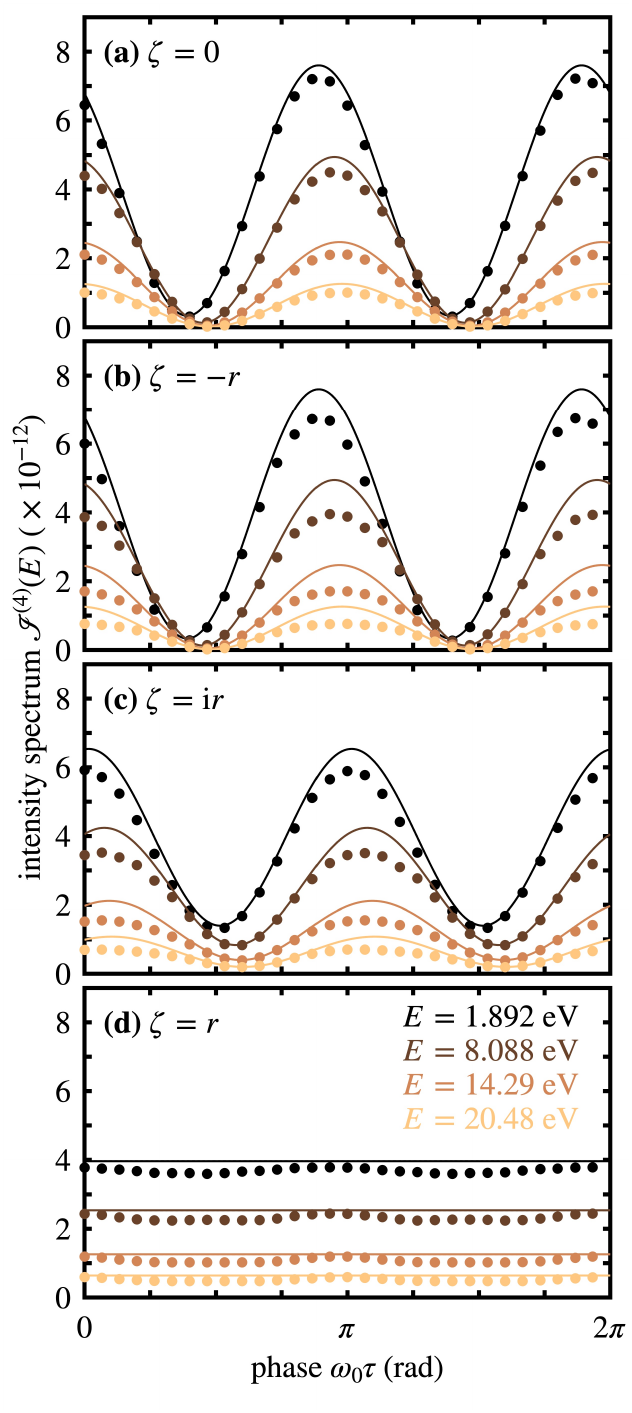}
    \caption{Same as Fig.~\ref{fig:delays_RABBIT} but for an infrared photon number of $\nIR = 10^{18}$.}
    \label{fig:nonperturbative_delays_RABBIT}
\end{figure}

Figure~\ref{fig:nonperturbative_delays_RABBIT} presents the RABBIT sidebands obtained using TDPT (solid lines) and TDSE (colored circles), analogous to Fig.~\ref{fig:delays_RABBIT}, but for a higher infrared photon occupancy of $\nIR = 10^{18}$.
At this photon number, the TDPT and TDSE results no longer exhibit perfect overlap. In particular, the discrepancy is more pronounced for squeezed states (panels b–c) than for the coherent state (panel a). This deviation arises because the Wigner distributions for squeezed states span several orders of magnitude in the effective field intensity. For example, while the coherent state is localized around an intensity of $10^{11} \; \rm W \ cm^{-2}$, the squeezed states extend into the $10^{12} \; \rm W \ cm^{-2}$ regime.
Consequently, the second-order perturbative regime, and thus the TDPT, becomes less valued at lower mean photon numbers for squeezed light than for coherent light.

\section{Conclusions \label{sec:conclusions}}

In this work, we have theoretically investigated the imprints of both classical and quantum light statistics on interferometric photoemission measurements. By employing a TDPT framework, we have established a unified description that accounts for the quantum nature of the electromagnetic field within the photoelectron dynamics. 
In the semiclassical limit of a product of coherent states~\eqref{eq:psi_classical}, this approach converges to the standard semiclassical results, as it involves the same conceptual quantities such as the one- and two-photon transition matrix elements~\cite{Veniard1996, Maquet1998, Klunder2011, Dahlstrom2013, Busto2019, Berkane2024}, $M_{\beta}^{(1)} (E)$ and $M_{\beta}^{(2)} (E)$. Beyond this limit, the formalism extends to non-classical and quantum states of light, providing the necessary flexibility to treat few-photon regimes, entangled states, and non-classical field statistics such as squeezing.

Applying this formalism to the RABBIT scheme, we derived a generalized expression for the main harmonic peak intensity spectrum in terms of photon number operators. Furthermore, we recovered the conventional sideband modulation as a function of the phase $\wIR \tau$. Crucially, our results reveal that the sideband parameters, amplitude $A$, contrast $C$, and phase $\theta$, are intrinsically linked to the high-order correlations between harmonic and infrared fields [see Eqs.~\eqref{eq:I4_all}]. We have demonstrated that ``cross-mode bunching'' between the infrared and harmonic modes can significantly amplify the RABBIT oscillation amplitude, suggesting a viable pathway for spectroscopic probing of the joint properties of the driving field and the emitted radiation. In contrast, we have shown that a lack of coherence, either between the harmonics and the infrared field or between the harmonics themselves, leads to a marked reduction in signal contrast. Specifically, a non-zero interferometric signal requires a stable phase relationship between the harmonics. In the absence of this inter-harmonic coherence, such as in the case of independent Fock states, the RABBIT oscillations are entirely suppressed.

The application of our theory to infrared light in squeezed coherent states further demonstrates the sensitivity of the RABBIT signal to underlying photon statistics. Our analysis shows that the squeezing phase $\varphi$ and the ratio $\ratioIR$ play a decisive role in the resulting interferogram. Specifically, we identified a regime of diminished contrast at intermediate squeezing phases, which we attribute to the reduction in the magnitude of the field's second-order moment. The excellent agreement between our TDPT results and the ab-initio numerical solutions (see Section~\ref{sec:TDSE}) validates the robustness of our perturbative approach.

In conclusion, this study demonstrates that interferometric photoemission is not only a tool for temporal metrology but also a promising avenue for probing the quantum nature of light. 
The general formalism for polychromatic quantum light enables the mapping of higher-order correlation terms directly onto the photoelectron spectrum. 
While our current implementation using the two-photon RABBIT scheme accesses correlations up to fourth-order, the framework is scalable, i.e. the inclusion of transitions involving a higher number of photons would allow for the extraction of even higher-order statistical moments.
Furthermore, this approach provides a robust framework for reconstructing the photoelectron density matrix~\cite{Bourassin2020, Laurell2022} and resolving angular distributions.
The ability to map photon statistics onto photoelectron observables opens new perspectives for quantum-enhanced spectroscopy and the characterization of complex quantum states of light at attosecond timescales.


\begin{thebibliography}{67}%
\makeatletter
\providecommand \@ifxundefined [1]{%
 \@ifx{#1\undefined}
}%
\providecommand \@ifnum [1]{%
 \ifnum #1\expandafter \@firstoftwo
 \else \expandafter \@secondoftwo
 \fi
}%
\providecommand \@ifx [1]{%
 \ifx #1\expandafter \@firstoftwo
 \else \expandafter \@secondoftwo
 \fi
}%
\providecommand \natexlab [1]{#1}%
\providecommand \enquote  [1]{``#1''}%
\providecommand \bibnamefont  [1]{#1}%
\providecommand \bibfnamefont [1]{#1}%
\providecommand \citenamefont [1]{#1}%
\providecommand \href@noop [0]{\@secondoftwo}%
\providecommand \href [0]{\begingroup \@sanitize@url \@href}%
\providecommand \@href[1]{\@@startlink{#1}\@@href}%
\providecommand \@@href[1]{\endgroup#1\@@endlink}%
\providecommand \@sanitize@url [0]{\catcode `\\12\catcode `\$12\catcode
  `\&12\catcode `\#12\catcode `\^12\catcode `\_12\catcode `\%12\relax}%
\providecommand \@@startlink[1]{}%
\providecommand \@@endlink[0]{}%
\providecommand \url  [0]{\begingroup\@sanitize@url \@url }%
\providecommand \@url [1]{\endgroup\@href {#1}{\urlprefix }}%
\providecommand \urlprefix  [0]{URL }%
\providecommand \Eprint [0]{\href }%
\providecommand \doibase [0]{http://dx.doi.org/}%
\providecommand \selectlanguage [0]{\@gobble}%
\providecommand \bibinfo  [0]{\@secondoftwo}%
\providecommand \bibfield  [0]{\@secondoftwo}%
\providecommand \translation [1]{[#1]}%
\providecommand \BibitemOpen [0]{}%
\providecommand \bibitemStop [0]{}%
\providecommand \bibitemNoStop [0]{.\EOS\space}%
\providecommand \EOS [0]{\spacefactor3000\relax}%
\providecommand \BibitemShut  [1]{\csname bibitem#1\endcsname}%
\let\auto@bib@innerbib\@empty
\bibitem [{\citenamefont {Ferray}\ \emph {et~al.}(1988)\citenamefont {Ferray},
  \citenamefont {L'Huillier}, \citenamefont {Li}, \citenamefont {Lompre},
  \citenamefont {Mainfray},\ and\ \citenamefont {Manus}}]{Ferray1988}%
  \BibitemOpen
  \bibfield  {author} {\bibinfo {author} {\bibfnamefont {M.}~\bibnamefont
  {Ferray}}, \bibinfo {author} {\bibfnamefont {A.}~\bibnamefont {L'Huillier}},
  \bibinfo {author} {\bibfnamefont {X.~F.}\ \bibnamefont {Li}}, \bibinfo
  {author} {\bibfnamefont {L.~A.}\ \bibnamefont {Lompre}}, \bibinfo {author}
  {\bibfnamefont {G.}~\bibnamefont {Mainfray}}, \ and\ \bibinfo {author}
  {\bibfnamefont {C.}~\bibnamefont {Manus}},\ }\href {\doibase
  10.1088/0953-4075/21/3/001} {\bibfield  {journal} {\bibinfo  {journal}
  {Journal of Physics B: Atomic, Molecular and Optical Physics}\ }\textbf
  {\bibinfo {volume} {21}},\ \bibinfo {pages} {L31} (\bibinfo {year}
  {1988})}\BibitemShut {NoStop}%
\bibitem [{\citenamefont {Paul}\ \emph {et~al.}(2001)\citenamefont {Paul},
  \citenamefont {Toma}, \citenamefont {Breger}, \citenamefont {Mullot},
  \citenamefont {Augé}, \citenamefont {Balcou}, \citenamefont {Muller},\ and\
  \citenamefont {Agostini}}]{Paul2001}%
  \BibitemOpen
  \bibfield  {author} {\bibinfo {author} {\bibfnamefont {P.~M.}\ \bibnamefont
  {Paul}}, \bibinfo {author} {\bibfnamefont {E.~S.}\ \bibnamefont {Toma}},
  \bibinfo {author} {\bibfnamefont {P.}~\bibnamefont {Breger}}, \bibinfo
  {author} {\bibfnamefont {G.}~\bibnamefont {Mullot}}, \bibinfo {author}
  {\bibfnamefont {F.}~\bibnamefont {Augé}}, \bibinfo {author} {\bibfnamefont
  {P.}~\bibnamefont {Balcou}}, \bibinfo {author} {\bibfnamefont {H.~G.}\
  \bibnamefont {Muller}}, \ and\ \bibinfo {author} {\bibfnamefont
  {P.}~\bibnamefont {Agostini}},\ }\href {\doibase 10.1126/science.1059413}
  {\bibfield  {journal} {\bibinfo  {journal} {Science}\ }\textbf {\bibinfo
  {volume} {292}},\ \bibinfo {pages} {1689} (\bibinfo {year}
  {2001})}\BibitemShut {NoStop}%
\bibitem [{\citenamefont {Hentschel}\ \emph {et~al.}(2001)\citenamefont
  {Hentschel}, \citenamefont {Kienberger}, \citenamefont {Spielmann},
  \citenamefont {Reider}, \citenamefont {Milosevic}, \citenamefont {Brabec},
  \citenamefont {Corkum}, \citenamefont {Heinzmann}, \citenamefont {Drescher},\
  and\ \citenamefont {Krausz}}]{Hentschel2001}%
  \BibitemOpen
  \bibfield  {author} {\bibinfo {author} {\bibfnamefont {M.}~\bibnamefont
  {Hentschel}}, \bibinfo {author} {\bibfnamefont {R.}~\bibnamefont
  {Kienberger}}, \bibinfo {author} {\bibfnamefont {C.}~\bibnamefont
  {Spielmann}}, \bibinfo {author} {\bibfnamefont {G.~A.}\ \bibnamefont
  {Reider}}, \bibinfo {author} {\bibfnamefont {N.}~\bibnamefont {Milosevic}},
  \bibinfo {author} {\bibfnamefont {T.}~\bibnamefont {Brabec}}, \bibinfo
  {author} {\bibfnamefont {P.}~\bibnamefont {Corkum}}, \bibinfo {author}
  {\bibfnamefont {U.}~\bibnamefont {Heinzmann}}, \bibinfo {author}
  {\bibfnamefont {M.}~\bibnamefont {Drescher}}, \ and\ \bibinfo {author}
  {\bibfnamefont {F.}~\bibnamefont {Krausz}},\ }\href {\doibase
  10.1038/35107000} {\bibfield  {journal} {\bibinfo  {journal} {Nature}\
  }\textbf {\bibinfo {volume} {414}},\ \bibinfo {pages} {509} (\bibinfo {year}
  {2001})}\BibitemShut {NoStop}%
\bibitem [{\citenamefont {Sansone}\ \emph {et~al.}(2006)\citenamefont
  {Sansone}, \citenamefont {Benedetti}, \citenamefont {Calegari}, \citenamefont
  {Vozzi}, \citenamefont {Avaldi}, \citenamefont {Flammini}, \citenamefont
  {Poletto}, \citenamefont {Villoresi}, \citenamefont {Altucci}, \citenamefont
  {Velotta}, \citenamefont {Stagira}, \citenamefont {Silvestri},\ and\
  \citenamefont {Nisoli}}]{Sansone2006}%
  \BibitemOpen
  \bibfield  {author} {\bibinfo {author} {\bibfnamefont {G.}~\bibnamefont
  {Sansone}}, \bibinfo {author} {\bibfnamefont {E.}~\bibnamefont {Benedetti}},
  \bibinfo {author} {\bibfnamefont {F.}~\bibnamefont {Calegari}}, \bibinfo
  {author} {\bibfnamefont {C.}~\bibnamefont {Vozzi}}, \bibinfo {author}
  {\bibfnamefont {L.}~\bibnamefont {Avaldi}}, \bibinfo {author} {\bibfnamefont
  {R.}~\bibnamefont {Flammini}}, \bibinfo {author} {\bibfnamefont
  {L.}~\bibnamefont {Poletto}}, \bibinfo {author} {\bibfnamefont
  {P.}~\bibnamefont {Villoresi}}, \bibinfo {author} {\bibfnamefont
  {C.}~\bibnamefont {Altucci}}, \bibinfo {author} {\bibfnamefont
  {R.}~\bibnamefont {Velotta}}, \bibinfo {author} {\bibfnamefont
  {S.}~\bibnamefont {Stagira}}, \bibinfo {author} {\bibfnamefont {S.~D.}\
  \bibnamefont {Silvestri}}, \ and\ \bibinfo {author} {\bibfnamefont
  {M.}~\bibnamefont {Nisoli}},\ }\href {\doibase 10.1126/science.1132838}
  {\bibfield  {journal} {\bibinfo  {journal} {Science}\ }\textbf {\bibinfo
  {volume} {314}},\ \bibinfo {pages} {443} (\bibinfo {year}
  {2006})}\BibitemShut {NoStop}%
\bibitem [{\citenamefont {Calegari}\ \emph {et~al.}(2014)\citenamefont
  {Calegari}, \citenamefont {Ayuso}, \citenamefont {Trabattoni}, \citenamefont
  {Belshaw}, \citenamefont {Camillis}, \citenamefont {Anumula}, \citenamefont
  {Frassetto}, \citenamefont {Poletto}, \citenamefont {Palacios}, \citenamefont
  {Decleva}, \citenamefont {Greenwood}, \citenamefont {Martín},\ and\
  \citenamefont {Nisoli}}]{Calegari2014}%
  \BibitemOpen
  \bibfield  {author} {\bibinfo {author} {\bibfnamefont {F.}~\bibnamefont
  {Calegari}}, \bibinfo {author} {\bibfnamefont {D.}~\bibnamefont {Ayuso}},
  \bibinfo {author} {\bibfnamefont {A.}~\bibnamefont {Trabattoni}}, \bibinfo
  {author} {\bibfnamefont {L.}~\bibnamefont {Belshaw}}, \bibinfo {author}
  {\bibfnamefont {S.~D.}\ \bibnamefont {Camillis}}, \bibinfo {author}
  {\bibfnamefont {S.}~\bibnamefont {Anumula}}, \bibinfo {author} {\bibfnamefont
  {F.}~\bibnamefont {Frassetto}}, \bibinfo {author} {\bibfnamefont
  {L.}~\bibnamefont {Poletto}}, \bibinfo {author} {\bibfnamefont
  {A.}~\bibnamefont {Palacios}}, \bibinfo {author} {\bibfnamefont
  {P.}~\bibnamefont {Decleva}}, \bibinfo {author} {\bibfnamefont {J.~B.}\
  \bibnamefont {Greenwood}}, \bibinfo {author} {\bibfnamefont {F.}~\bibnamefont
  {Martín}}, \ and\ \bibinfo {author} {\bibfnamefont {M.}~\bibnamefont
  {Nisoli}},\ }\href {\doibase 10.1126/science.1254061} {\bibfield  {journal}
  {\bibinfo  {journal} {Science}\ }\textbf {\bibinfo {volume} {346}},\ \bibinfo
  {pages} {336} (\bibinfo {year} {2014})}\BibitemShut {NoStop}%
\bibitem [{\citenamefont {Nisoli}\ \emph {et~al.}(2017)\citenamefont {Nisoli},
  \citenamefont {Decleva}, \citenamefont {Calegari}, \citenamefont {Palacios},\
  and\ \citenamefont {Mart{\'\i}n}}]{Nisoli2017}%
  \BibitemOpen
  \bibfield  {author} {\bibinfo {author} {\bibfnamefont {M.}~\bibnamefont
  {Nisoli}}, \bibinfo {author} {\bibfnamefont {P.}~\bibnamefont {Decleva}},
  \bibinfo {author} {\bibfnamefont {F.}~\bibnamefont {Calegari}}, \bibinfo
  {author} {\bibfnamefont {A.}~\bibnamefont {Palacios}}, \ and\ \bibinfo
  {author} {\bibfnamefont {F.}~\bibnamefont {Mart{\'\i}n}},\ }\href {\doibase
  10.1021/acs.chemrev.6b00453} {\bibfield  {journal} {\bibinfo  {journal}
  {Chemical Reviews}\ }\textbf {\bibinfo {volume} {117}},\ \bibinfo {pages}
  {10760} (\bibinfo {year} {2017})}\BibitemShut {NoStop}%
\bibitem [{\citenamefont {Merritt}\ \emph {et~al.}(2021)\citenamefont
  {Merritt}, \citenamefont {Jacquemin},\ and\ \citenamefont
  {Vacher}}]{Merritt2021}%
  \BibitemOpen
  \bibfield  {author} {\bibinfo {author} {\bibfnamefont {I.~C.~D.}\
  \bibnamefont {Merritt}}, \bibinfo {author} {\bibfnamefont {D.}~\bibnamefont
  {Jacquemin}}, \ and\ \bibinfo {author} {\bibfnamefont {M.}~\bibnamefont
  {Vacher}},\ }\href {\doibase 10.1021/acs.jpclett.1c02016} {\bibfield
  {journal} {\bibinfo  {journal} {The Journal of Physical Chemistry Letters}\
  }\textbf {\bibinfo {volume} {12}},\ \bibinfo {pages} {8404} (\bibinfo {year}
  {2021})}\BibitemShut {NoStop}%
\bibitem [{\citenamefont {Alexander}\ \emph {et~al.}(2023)\citenamefont
  {Alexander}, \citenamefont {Marangos}, \citenamefont {Ruberti},\ and\
  \citenamefont {Vacher}}]{Alexander2023}%
  \BibitemOpen
  \bibfield  {author} {\bibinfo {author} {\bibfnamefont {O.~G.}\ \bibnamefont
  {Alexander}}, \bibinfo {author} {\bibfnamefont {J.~P.}\ \bibnamefont
  {Marangos}}, \bibinfo {author} {\bibfnamefont {M.}~\bibnamefont {Ruberti}}, \
  and\ \bibinfo {author} {\bibfnamefont {M.}~\bibnamefont {Vacher}},\ }in\
  \href {\doibase https://doi.org/10.1016/bs.aamop.2023.05.001} {\emph
  {\bibinfo {booktitle} {Advances in Atomic, Molecular, and Optical
  Physics}}},\ \bibinfo {series} {Advances In Atomic, Molecular, and Optical
  Physics}, Vol.~\bibinfo {volume} {72},\ \bibinfo {editor} {edited by\
  \bibinfo {editor} {\bibfnamefont {L.~F.}\ \bibnamefont {DiMauro}}, \bibinfo
  {editor} {\bibfnamefont {H.}~\bibnamefont {Perrin}}, \ and\ \bibinfo {editor}
  {\bibfnamefont {S.~F.}\ \bibnamefont {Yelin}}}\ (\bibinfo  {publisher}
  {Academic Press},\ \bibinfo {year} {2023})\ pp.\ \bibinfo {pages}
  {183--251}\BibitemShut {NoStop}%
\bibitem [{\citenamefont {Goulielmakis}\ \emph {et~al.}(2004)\citenamefont
  {Goulielmakis}, \citenamefont {Uiberacker}, \citenamefont {Kienberger},
  \citenamefont {Baltuska}, \citenamefont {Yakovlev}, \citenamefont {Scrinzi},
  \citenamefont {Westerwalbesloh}, \citenamefont {Kleineberg}, \citenamefont
  {Heinzmann}, \citenamefont {Drescher},\ and\ \citenamefont
  {Krausz}}]{Goulielmakis2004}%
  \BibitemOpen
  \bibfield  {author} {\bibinfo {author} {\bibfnamefont {E.}~\bibnamefont
  {Goulielmakis}}, \bibinfo {author} {\bibfnamefont {M.}~\bibnamefont
  {Uiberacker}}, \bibinfo {author} {\bibfnamefont {R.}~\bibnamefont
  {Kienberger}}, \bibinfo {author} {\bibfnamefont {A.}~\bibnamefont
  {Baltuska}}, \bibinfo {author} {\bibfnamefont {V.}~\bibnamefont {Yakovlev}},
  \bibinfo {author} {\bibfnamefont {A.}~\bibnamefont {Scrinzi}}, \bibinfo
  {author} {\bibfnamefont {T.}~\bibnamefont {Westerwalbesloh}}, \bibinfo
  {author} {\bibfnamefont {U.}~\bibnamefont {Kleineberg}}, \bibinfo {author}
  {\bibfnamefont {U.}~\bibnamefont {Heinzmann}}, \bibinfo {author}
  {\bibfnamefont {M.}~\bibnamefont {Drescher}}, \ and\ \bibinfo {author}
  {\bibfnamefont {F.}~\bibnamefont {Krausz}},\ }\href {\doibase
  10.1126/science.1100866} {\bibfield  {journal} {\bibinfo  {journal}
  {Science}\ }\textbf {\bibinfo {volume} {305}},\ \bibinfo {pages} {1267}
  (\bibinfo {year} {2004})}\BibitemShut {NoStop}%
\bibitem [{\citenamefont {Mairesse}\ and\ \citenamefont
  {Qu\'er\'e}(2005)}]{Mairesse2005}%
  \BibitemOpen
  \bibfield  {author} {\bibinfo {author} {\bibfnamefont {Y.}~\bibnamefont
  {Mairesse}}\ and\ \bibinfo {author} {\bibfnamefont {F.}~\bibnamefont
  {Qu\'er\'e}},\ }\href {\doibase 10.1103/PhysRevA.71.011401} {\bibfield
  {journal} {\bibinfo  {journal} {Phys. Rev. A}\ }\textbf {\bibinfo {volume}
  {71}},\ \bibinfo {pages} {011401} (\bibinfo {year} {2005})}\BibitemShut
  {NoStop}%
\bibitem [{\citenamefont {V\'eniard}\ \emph {et~al.}(1996)\citenamefont
  {V\'eniard}, \citenamefont {Ta\"{\i}eb},\ and\ \citenamefont
  {Maquet}}]{Veniard1996}%
  \BibitemOpen
  \bibfield  {author} {\bibinfo {author} {\bibfnamefont {V.}~\bibnamefont
  {V\'eniard}}, \bibinfo {author} {\bibfnamefont {R.}~\bibnamefont
  {Ta\"{\i}eb}}, \ and\ \bibinfo {author} {\bibfnamefont {A.}~\bibnamefont
  {Maquet}},\ }\href {\doibase 10.1103/PhysRevA.54.721} {\bibfield  {journal}
  {\bibinfo  {journal} {Phys. Rev. A}\ }\textbf {\bibinfo {volume} {54}},\
  \bibinfo {pages} {721} (\bibinfo {year} {1996})}\BibitemShut {NoStop}%
\bibitem [{\citenamefont {Maquet}\ \emph {et~al.}(1998)\citenamefont {Maquet},
  \citenamefont {Véniard},\ and\ \citenamefont {Marian}}]{Maquet1998}%
  \BibitemOpen
  \bibfield  {author} {\bibinfo {author} {\bibfnamefont {A.}~\bibnamefont
  {Maquet}}, \bibinfo {author} {\bibfnamefont {V.}~\bibnamefont {Véniard}}, \
  and\ \bibinfo {author} {\bibfnamefont {T.~A.}\ \bibnamefont {Marian}},\
  }\href {\doibase 10.1088/0953-4075/31/17/004} {\bibfield  {journal} {\bibinfo
   {journal} {Journal of Physics B: Atomic, Molecular and Optical Physics}\
  }\textbf {\bibinfo {volume} {31}},\ \bibinfo {pages} {3743} (\bibinfo {year}
  {1998})}\BibitemShut {NoStop}%
\bibitem [{\citenamefont {Dahlström}\ \emph {et~al.}(2013)\citenamefont
  {Dahlström}, \citenamefont {Guénot}, \citenamefont {Klünder},
  \citenamefont {Gisselbrecht}, \citenamefont {Mauritsson}, \citenamefont
  {L’Huillier}, \citenamefont {Maquet},\ and\ \citenamefont
  {Taïeb}}]{Dahlstrom2013}%
  \BibitemOpen
  \bibfield  {author} {\bibinfo {author} {\bibfnamefont {J.}~\bibnamefont
  {Dahlström}}, \bibinfo {author} {\bibfnamefont {D.}~\bibnamefont {Guénot}},
  \bibinfo {author} {\bibfnamefont {K.}~\bibnamefont {Klünder}}, \bibinfo
  {author} {\bibfnamefont {M.}~\bibnamefont {Gisselbrecht}}, \bibinfo {author}
  {\bibfnamefont {J.}~\bibnamefont {Mauritsson}}, \bibinfo {author}
  {\bibfnamefont {A.}~\bibnamefont {L’Huillier}}, \bibinfo {author}
  {\bibfnamefont {A.}~\bibnamefont {Maquet}}, \ and\ \bibinfo {author}
  {\bibfnamefont {R.}~\bibnamefont {Taïeb}},\ }\href {\doibase
  https://doi.org/10.1016/j.chemphys.2012.01.017} {\bibfield  {journal}
  {\bibinfo  {journal} {Chemical Physics}\ }\textbf {\bibinfo {volume} {414}},\
  \bibinfo {pages} {53} (\bibinfo {year} {2013})},\ \bibinfo {note} {attosecond
  spectroscopy}\BibitemShut {NoStop}%
\bibitem [{\citenamefont {Busto}\ \emph {et~al.}(2019)\citenamefont {Busto},
  \citenamefont {Vinbladh}, \citenamefont {Zhong}, \citenamefont {Isinger},
  \citenamefont {Nandi}, \citenamefont {Maclot}, \citenamefont {Johnsson},
  \citenamefont {Gisselbrecht}, \citenamefont {L'Huillier}, \citenamefont
  {Lindroth},\ and\ \citenamefont {Dahlstr\"om}}]{Busto2019}%
  \BibitemOpen
  \bibfield  {author} {\bibinfo {author} {\bibfnamefont {D.}~\bibnamefont
  {Busto}}, \bibinfo {author} {\bibfnamefont {J.}~\bibnamefont {Vinbladh}},
  \bibinfo {author} {\bibfnamefont {S.}~\bibnamefont {Zhong}}, \bibinfo
  {author} {\bibfnamefont {M.}~\bibnamefont {Isinger}}, \bibinfo {author}
  {\bibfnamefont {S.}~\bibnamefont {Nandi}}, \bibinfo {author} {\bibfnamefont
  {S.}~\bibnamefont {Maclot}}, \bibinfo {author} {\bibfnamefont
  {P.}~\bibnamefont {Johnsson}}, \bibinfo {author} {\bibfnamefont
  {M.}~\bibnamefont {Gisselbrecht}}, \bibinfo {author} {\bibfnamefont
  {A.}~\bibnamefont {L'Huillier}}, \bibinfo {author} {\bibfnamefont
  {E.}~\bibnamefont {Lindroth}}, \ and\ \bibinfo {author} {\bibfnamefont
  {J.~M.}\ \bibnamefont {Dahlstr\"om}},\ }\href {\doibase
  10.1103/PhysRevLett.123.133201} {\bibfield  {journal} {\bibinfo  {journal}
  {Phys. Rev. Lett.}\ }\textbf {\bibinfo {volume} {123}},\ \bibinfo {pages}
  {133201} (\bibinfo {year} {2019})}\BibitemShut {NoStop}%
\bibitem [{\citenamefont {Berkane}\ \emph {et~al.}(2024)\citenamefont
  {Berkane}, \citenamefont {L\'ev\^eque}, \citenamefont {Ta\"{\i}eb},
  \citenamefont {Caillat},\ and\ \citenamefont {Dubois}}]{Berkane2024}%
  \BibitemOpen
  \bibfield  {author} {\bibinfo {author} {\bibfnamefont {M.}~\bibnamefont
  {Berkane}}, \bibinfo {author} {\bibfnamefont {C.}~\bibnamefont
  {L\'ev\^eque}}, \bibinfo {author} {\bibfnamefont {R.}~\bibnamefont
  {Ta\"{\i}eb}}, \bibinfo {author} {\bibfnamefont {J.}~\bibnamefont {Caillat}},
  \ and\ \bibinfo {author} {\bibfnamefont {J.}~\bibnamefont {Dubois}},\ }\href
  {\doibase 10.1103/PhysRevA.110.013120} {\bibfield  {journal} {\bibinfo
  {journal} {Phys. Rev. A}\ }\textbf {\bibinfo {volume} {110}},\ \bibinfo
  {pages} {013120} (\bibinfo {year} {2024})}\BibitemShut {NoStop}%
\bibitem [{\citenamefont {Gorlach}\ \emph {et~al.}(2020)\citenamefont
  {Gorlach}, \citenamefont {Neufeld}, \citenamefont {Rivera}, \citenamefont
  {Cohen},\ and\ \citenamefont {Kaminer}}]{Gorlach2020}%
  \BibitemOpen
  \bibfield  {author} {\bibinfo {author} {\bibfnamefont {A.}~\bibnamefont
  {Gorlach}}, \bibinfo {author} {\bibfnamefont {O.}~\bibnamefont {Neufeld}},
  \bibinfo {author} {\bibfnamefont {N.}~\bibnamefont {Rivera}}, \bibinfo
  {author} {\bibfnamefont {O.}~\bibnamefont {Cohen}}, \ and\ \bibinfo {author}
  {\bibfnamefont {I.}~\bibnamefont {Kaminer}},\ }\href {\doibase
  10.1038/s41467-020-18218-w} {\bibfield  {journal} {\bibinfo  {journal}
  {Nature Communications}\ }\textbf {\bibinfo {volume} {11}},\ \bibinfo {pages}
  {4598} (\bibinfo {year} {2020})}\BibitemShut {NoStop}%
\bibitem [{\citenamefont {Stammer}\ \emph {et~al.}(2024)\citenamefont
  {Stammer}, \citenamefont {Rivera-Dean}, \citenamefont {Maxwell},
  \citenamefont {Lamprou}, \citenamefont {Arg\"uello-Luengo}, \citenamefont
  {Tzallas}, \citenamefont {Ciappina},\ and\ \citenamefont
  {Lewenstein}}]{Stammer2024}%
  \BibitemOpen
  \bibfield  {author} {\bibinfo {author} {\bibfnamefont {P.}~\bibnamefont
  {Stammer}}, \bibinfo {author} {\bibfnamefont {J.}~\bibnamefont
  {Rivera-Dean}}, \bibinfo {author} {\bibfnamefont {A.~S.}\ \bibnamefont
  {Maxwell}}, \bibinfo {author} {\bibfnamefont {T.}~\bibnamefont {Lamprou}},
  \bibinfo {author} {\bibfnamefont {J.}~\bibnamefont {Arg\"uello-Luengo}},
  \bibinfo {author} {\bibfnamefont {P.}~\bibnamefont {Tzallas}}, \bibinfo
  {author} {\bibfnamefont {M.~F.}\ \bibnamefont {Ciappina}}, \ and\ \bibinfo
  {author} {\bibfnamefont {M.}~\bibnamefont {Lewenstein}},\ }\href {\doibase
  10.1103/PhysRevLett.132.143603} {\bibfield  {journal} {\bibinfo  {journal}
  {Phys. Rev. Lett.}\ }\textbf {\bibinfo {volume} {132}},\ \bibinfo {pages}
  {143603} (\bibinfo {year} {2024})}\BibitemShut {NoStop}%
\bibitem [{\citenamefont {Gombk\"ot\ifmmode~\mbox{\H{o}}\else \H{o}\fi{}}\
  \emph {et~al.}(2021)\citenamefont {Gombk\"ot\ifmmode~\mbox{\H{o}}\else
  \H{o}\fi{}}, \citenamefont {F\"oldi},\ and\ \citenamefont
  {Varr\'o}}]{Gombkoto2021}%
  \BibitemOpen
  \bibfield  {author} {\bibinfo {author} {\bibfnamefont {A.}~\bibnamefont
  {Gombk\"ot\ifmmode~\mbox{\H{o}}\else \H{o}\fi{}}}, \bibinfo {author}
  {\bibfnamefont {P.}~\bibnamefont {F\"oldi}}, \ and\ \bibinfo {author}
  {\bibfnamefont {S.}~\bibnamefont {Varr\'o}},\ }\href {\doibase
  10.1103/PhysRevA.104.033703} {\bibfield  {journal} {\bibinfo  {journal}
  {Phys. Rev. A}\ }\textbf {\bibinfo {volume} {104}},\ \bibinfo {pages}
  {033703} (\bibinfo {year} {2021})}\BibitemShut {NoStop}%
\bibitem [{\citenamefont {Stammer}\ \emph {et~al.}(2023)\citenamefont
  {Stammer}, \citenamefont {Rivera-Dean}, \citenamefont {Maxwell},
  \citenamefont {Lamprou}, \citenamefont {Ord\'o\~nez}, \citenamefont
  {Ciappina}, \citenamefont {Tzallas},\ and\ \citenamefont
  {Lewenstein}}]{Stammer2023}%
  \BibitemOpen
  \bibfield  {author} {\bibinfo {author} {\bibfnamefont {P.}~\bibnamefont
  {Stammer}}, \bibinfo {author} {\bibfnamefont {J.}~\bibnamefont
  {Rivera-Dean}}, \bibinfo {author} {\bibfnamefont {A.}~\bibnamefont
  {Maxwell}}, \bibinfo {author} {\bibfnamefont {T.}~\bibnamefont {Lamprou}},
  \bibinfo {author} {\bibfnamefont {A.}~\bibnamefont {Ord\'o\~nez}}, \bibinfo
  {author} {\bibfnamefont {M.~F.}\ \bibnamefont {Ciappina}}, \bibinfo {author}
  {\bibfnamefont {P.}~\bibnamefont {Tzallas}}, \ and\ \bibinfo {author}
  {\bibfnamefont {M.}~\bibnamefont {Lewenstein}},\ }\href {\doibase
  10.1103/PRXQuantum.4.010201} {\bibfield  {journal} {\bibinfo  {journal} {PRX
  Quantum}\ }\textbf {\bibinfo {volume} {4}},\ \bibinfo {pages} {010201}
  (\bibinfo {year} {2023})}\BibitemShut {NoStop}%
\bibitem [{\citenamefont {Gonoskov}\ \emph {et~al.}(2024)\citenamefont
  {Gonoskov}, \citenamefont {Sondenheimer}, \citenamefont {H\"unecke},
  \citenamefont {Kartashov}, \citenamefont {Peschel},\ and\ \citenamefont
  {Gr\"afe}}]{Gonoskov2024}%
  \BibitemOpen
  \bibfield  {author} {\bibinfo {author} {\bibfnamefont {I.}~\bibnamefont
  {Gonoskov}}, \bibinfo {author} {\bibfnamefont {R.}~\bibnamefont
  {Sondenheimer}}, \bibinfo {author} {\bibfnamefont {C.}~\bibnamefont
  {H\"unecke}}, \bibinfo {author} {\bibfnamefont {D.}~\bibnamefont
  {Kartashov}}, \bibinfo {author} {\bibfnamefont {U.}~\bibnamefont {Peschel}},
  \ and\ \bibinfo {author} {\bibfnamefont {S.}~\bibnamefont {Gr\"afe}},\ }\href
  {\doibase 10.1103/PhysRevB.109.125110} {\bibfield  {journal} {\bibinfo
  {journal} {Phys. Rev. B}\ }\textbf {\bibinfo {volume} {109}},\ \bibinfo
  {pages} {125110} (\bibinfo {year} {2024})}\BibitemShut {NoStop}%
\bibitem [{\citenamefont {Rivera-Dean}\ \emph {et~al.}(2024)\citenamefont
  {Rivera-Dean}, \citenamefont {Crispin}, \citenamefont {Stammer},
  \citenamefont {Lamprou}, \citenamefont {Pisanty}, \citenamefont {Kr\"uger},
  \citenamefont {Tzallas}, \citenamefont {Lewenstein},\ and\ \citenamefont
  {Ciappina}}]{Rivera2024}%
  \BibitemOpen
  \bibfield  {author} {\bibinfo {author} {\bibfnamefont {J.}~\bibnamefont
  {Rivera-Dean}}, \bibinfo {author} {\bibfnamefont {H.~B.}\ \bibnamefont
  {Crispin}}, \bibinfo {author} {\bibfnamefont {P.}~\bibnamefont {Stammer}},
  \bibinfo {author} {\bibfnamefont {T.}~\bibnamefont {Lamprou}}, \bibinfo
  {author} {\bibfnamefont {E.}~\bibnamefont {Pisanty}}, \bibinfo {author}
  {\bibfnamefont {M.}~\bibnamefont {Kr\"uger}}, \bibinfo {author}
  {\bibfnamefont {P.}~\bibnamefont {Tzallas}}, \bibinfo {author} {\bibfnamefont
  {M.}~\bibnamefont {Lewenstein}}, \ and\ \bibinfo {author} {\bibfnamefont
  {M.~F.}\ \bibnamefont {Ciappina}},\ }\href {\doibase
  10.1103/PhysRevA.110.063118} {\bibfield  {journal} {\bibinfo  {journal}
  {Phys. Rev. A}\ }\textbf {\bibinfo {volume} {110}},\ \bibinfo {pages}
  {063118} (\bibinfo {year} {2024})}\BibitemShut {NoStop}%
\bibitem [{\citenamefont {Theidel}\ \emph {et~al.}(2024)\citenamefont
  {Theidel}, \citenamefont {Cotte}, \citenamefont {Sondenheimer}, \citenamefont
  {Shiriaeva}, \citenamefont {Froidevaux}, \citenamefont {Severin},
  \citenamefont {Merdji-Larue}, \citenamefont {Mosel}, \citenamefont
  {Fr\"ohlich}, \citenamefont {Weber}, \citenamefont {Morgner}, \citenamefont
  {Kovacev}, \citenamefont {Biegert},\ and\ \citenamefont
  {Merdji}}]{Theidel2024}%
  \BibitemOpen
  \bibfield  {author} {\bibinfo {author} {\bibfnamefont {D.}~\bibnamefont
  {Theidel}}, \bibinfo {author} {\bibfnamefont {V.}~\bibnamefont {Cotte}},
  \bibinfo {author} {\bibfnamefont {R.}~\bibnamefont {Sondenheimer}}, \bibinfo
  {author} {\bibfnamefont {V.}~\bibnamefont {Shiriaeva}}, \bibinfo {author}
  {\bibfnamefont {M.}~\bibnamefont {Froidevaux}}, \bibinfo {author}
  {\bibfnamefont {V.}~\bibnamefont {Severin}}, \bibinfo {author} {\bibfnamefont
  {A.}~\bibnamefont {Merdji-Larue}}, \bibinfo {author} {\bibfnamefont
  {P.}~\bibnamefont {Mosel}}, \bibinfo {author} {\bibfnamefont
  {S.}~\bibnamefont {Fr\"ohlich}}, \bibinfo {author} {\bibfnamefont {K.-A.}\
  \bibnamefont {Weber}}, \bibinfo {author} {\bibfnamefont {U.}~\bibnamefont
  {Morgner}}, \bibinfo {author} {\bibfnamefont {M.}~\bibnamefont {Kovacev}},
  \bibinfo {author} {\bibfnamefont {J.}~\bibnamefont {Biegert}}, \ and\
  \bibinfo {author} {\bibfnamefont {H.}~\bibnamefont {Merdji}},\ }\href
  {\doibase 10.1103/PRXQuantum.5.040319} {\bibfield  {journal} {\bibinfo
  {journal} {PRX Quantum}\ }\textbf {\bibinfo {volume} {5}},\ \bibinfo {pages}
  {040319} (\bibinfo {year} {2024})}\BibitemShut {NoStop}%
\bibitem [{\citenamefont {Theidel}\ \emph {et~al.}(2025)\citenamefont
  {Theidel}, \citenamefont {Cotte}, \citenamefont {Heinzel}, \citenamefont
  {Griguer}, \citenamefont {Weis}, \citenamefont {Sondenheimer},\ and\
  \citenamefont {Merdji}}]{Theidel2025}%
  \BibitemOpen
  \bibfield  {author} {\bibinfo {author} {\bibfnamefont {D.}~\bibnamefont
  {Theidel}}, \bibinfo {author} {\bibfnamefont {V.}~\bibnamefont {Cotte}},
  \bibinfo {author} {\bibfnamefont {P.}~\bibnamefont {Heinzel}}, \bibinfo
  {author} {\bibfnamefont {H.}~\bibnamefont {Griguer}}, \bibinfo {author}
  {\bibfnamefont {M.}~\bibnamefont {Weis}}, \bibinfo {author} {\bibfnamefont
  {R.}~\bibnamefont {Sondenheimer}}, \ and\ \bibinfo {author} {\bibfnamefont
  {H.}~\bibnamefont {Merdji}},\ }\href {\doibase 10.1103/6r6n-pxfp} {\bibfield
  {journal} {\bibinfo  {journal} {Phys. Rev. Res.}\ }\textbf {\bibinfo {volume}
  {7}},\ \bibinfo {pages} {033223} (\bibinfo {year} {2025})}\BibitemShut
  {NoStop}%
\bibitem [{\citenamefont {Gorlach}\ \emph {et~al.}(2023)\citenamefont
  {Gorlach}, \citenamefont {Tzur}, \citenamefont {Birk}, \citenamefont
  {Kr{\"u}ger}, \citenamefont {Rivera}, \citenamefont {Cohen},\ and\
  \citenamefont {Kaminer}}]{Gorlach2023}%
  \BibitemOpen
  \bibfield  {author} {\bibinfo {author} {\bibfnamefont {A.}~\bibnamefont
  {Gorlach}}, \bibinfo {author} {\bibfnamefont {M.~E.}\ \bibnamefont {Tzur}},
  \bibinfo {author} {\bibfnamefont {M.}~\bibnamefont {Birk}}, \bibinfo {author}
  {\bibfnamefont {M.}~\bibnamefont {Kr{\"u}ger}}, \bibinfo {author}
  {\bibfnamefont {N.}~\bibnamefont {Rivera}}, \bibinfo {author} {\bibfnamefont
  {O.}~\bibnamefont {Cohen}}, \ and\ \bibinfo {author} {\bibfnamefont
  {I.}~\bibnamefont {Kaminer}},\ }\href {\doibase 10.1038/s41567-023-02127-y}
  {\bibfield  {journal} {\bibinfo  {journal} {Nature Physics}\ }\textbf
  {\bibinfo {volume} {19}},\ \bibinfo {pages} {1689} (\bibinfo {year}
  {2023})}\BibitemShut {NoStop}%
\bibitem [{\citenamefont {Even~Tzur}\ \emph {et~al.}(2023)\citenamefont
  {Even~Tzur}, \citenamefont {Birk}, \citenamefont {Gorlach}, \citenamefont
  {Kr{\"u}ger}, \citenamefont {Kaminer},\ and\ \citenamefont
  {Cohen}}]{Tzur2023}%
  \BibitemOpen
  \bibfield  {author} {\bibinfo {author} {\bibfnamefont {M.}~\bibnamefont
  {Even~Tzur}}, \bibinfo {author} {\bibfnamefont {M.}~\bibnamefont {Birk}},
  \bibinfo {author} {\bibfnamefont {A.}~\bibnamefont {Gorlach}}, \bibinfo
  {author} {\bibfnamefont {M.}~\bibnamefont {Kr{\"u}ger}}, \bibinfo {author}
  {\bibfnamefont {I.}~\bibnamefont {Kaminer}}, \ and\ \bibinfo {author}
  {\bibfnamefont {O.}~\bibnamefont {Cohen}},\ }\href {\doibase
  10.1038/s41566-023-01209-w} {\bibfield  {journal} {\bibinfo  {journal}
  {Nature Photonics}\ }\textbf {\bibinfo {volume} {17}},\ \bibinfo {pages}
  {501} (\bibinfo {year} {2023})}\BibitemShut {NoStop}%
\bibitem [{\citenamefont {Rasputnyi}\ \emph {et~al.}(2024)\citenamefont
  {Rasputnyi}, \citenamefont {Chen}, \citenamefont {Birk}, \citenamefont
  {Cohen}, \citenamefont {Kaminer}, \citenamefont {Kr{\"u}ger}, \citenamefont
  {Seletskiy}, \citenamefont {Chekhova},\ and\ \citenamefont
  {Tani}}]{Rasputnyi2024}%
  \BibitemOpen
  \bibfield  {author} {\bibinfo {author} {\bibfnamefont {A.}~\bibnamefont
  {Rasputnyi}}, \bibinfo {author} {\bibfnamefont {Z.}~\bibnamefont {Chen}},
  \bibinfo {author} {\bibfnamefont {M.}~\bibnamefont {Birk}}, \bibinfo {author}
  {\bibfnamefont {O.}~\bibnamefont {Cohen}}, \bibinfo {author} {\bibfnamefont
  {I.}~\bibnamefont {Kaminer}}, \bibinfo {author} {\bibfnamefont
  {M.}~\bibnamefont {Kr{\"u}ger}}, \bibinfo {author} {\bibfnamefont
  {D.}~\bibnamefont {Seletskiy}}, \bibinfo {author} {\bibfnamefont
  {M.}~\bibnamefont {Chekhova}}, \ and\ \bibinfo {author} {\bibfnamefont
  {F.}~\bibnamefont {Tani}},\ }\href {\doibase 10.1038/s41567-024-02659-x}
  {\bibfield  {journal} {\bibinfo  {journal} {Nature Physics}\ }\textbf
  {\bibinfo {volume} {20}},\ \bibinfo {pages} {1960} (\bibinfo {year}
  {2024})}\BibitemShut {NoStop}%
\bibitem [{\citenamefont {Tzur}\ \emph {et~al.}(2024)\citenamefont {Tzur},
  \citenamefont {Birk}, \citenamefont {Gorlach}, \citenamefont {Kaminer},
  \citenamefont {Kr\"uger},\ and\ \citenamefont {Cohen}}]{Tzur2024}%
  \BibitemOpen
  \bibfield  {author} {\bibinfo {author} {\bibfnamefont {M.~E.}\ \bibnamefont
  {Tzur}}, \bibinfo {author} {\bibfnamefont {M.}~\bibnamefont {Birk}}, \bibinfo
  {author} {\bibfnamefont {A.}~\bibnamefont {Gorlach}}, \bibinfo {author}
  {\bibfnamefont {I.}~\bibnamefont {Kaminer}}, \bibinfo {author} {\bibfnamefont
  {M.}~\bibnamefont {Kr\"uger}}, \ and\ \bibinfo {author} {\bibfnamefont
  {O.}~\bibnamefont {Cohen}},\ }\href {\doibase
  10.1103/PhysRevResearch.6.033079} {\bibfield  {journal} {\bibinfo  {journal}
  {Phys. Rev. Res.}\ }\textbf {\bibinfo {volume} {6}},\ \bibinfo {pages}
  {033079} (\bibinfo {year} {2024})}\BibitemShut {NoStop}%
\bibitem [{\citenamefont {Wang}\ \emph {et~al.}(2025)\citenamefont {Wang},
  \citenamefont {Yang}, \citenamefont {Cui}, \citenamefont {He}, \citenamefont
  {Ou}, \citenamefont {Jin}, \citenamefont {Liao}, \citenamefont {Lan},\ and\
  \citenamefont {Lu}}]{Wang2025}%
  \BibitemOpen
  \bibfield  {author} {\bibinfo {author} {\bibfnamefont {F.}~\bibnamefont
  {Wang}}, \bibinfo {author} {\bibfnamefont {C.}~\bibnamefont {Yang}}, \bibinfo
  {author} {\bibfnamefont {X.}~\bibnamefont {Cui}}, \bibinfo {author}
  {\bibfnamefont {L.}~\bibnamefont {He}}, \bibinfo {author} {\bibfnamefont
  {T.}~\bibnamefont {Ou}}, \bibinfo {author} {\bibfnamefont {R.-B.}\
  \bibnamefont {Jin}}, \bibinfo {author} {\bibfnamefont {Q.}~\bibnamefont
  {Liao}}, \bibinfo {author} {\bibfnamefont {P.}~\bibnamefont {Lan}}, \ and\
  \bibinfo {author} {\bibfnamefont {P.}~\bibnamefont {Lu}},\ }\href {\doibase
  10.1103/hph3-kzb2} {\bibfield  {journal} {\bibinfo  {journal} {Phys. Rev. A}\
  }\textbf {\bibinfo {volume} {112}},\ \bibinfo {pages} {043119} (\bibinfo
  {year} {2025})}\BibitemShut {NoStop}%
\bibitem [{\citenamefont {Li}\ \emph {et~al.}(2025)\citenamefont {Li},
  \citenamefont {Lyu}, \citenamefont {Liu},\ and\ \citenamefont
  {Liu}}]{Li2025}%
  \BibitemOpen
  \bibfield  {author} {\bibinfo {author} {\bibfnamefont {J.}~\bibnamefont
  {Li}}, \bibinfo {author} {\bibfnamefont {Z.}~\bibnamefont {Lyu}}, \bibinfo
  {author} {\bibfnamefont {H.}~\bibnamefont {Liu}}, \ and\ \bibinfo {author}
  {\bibfnamefont {Y.}~\bibnamefont {Liu}},\ }\href {\doibase 10.1103/vmk1-qjpj}
  {\bibfield  {journal} {\bibinfo  {journal} {Phys. Rev. A}\ }\textbf {\bibinfo
  {volume} {112}},\ \bibinfo {pages} {033507} (\bibinfo {year}
  {2025})}\BibitemShut {NoStop}%
\bibitem [{\citenamefont {Rivera-Dean}\ \emph
  {et~al.}(2025{\natexlab{a}})\citenamefont {Rivera-Dean}, \citenamefont
  {Stammer}, \citenamefont {Ciappina},\ and\ \citenamefont
  {Lewenstein}}]{Rivera2025str}%
  \BibitemOpen
  \bibfield  {author} {\bibinfo {author} {\bibfnamefont {J.}~\bibnamefont
  {Rivera-Dean}}, \bibinfo {author} {\bibfnamefont {P.}~\bibnamefont
  {Stammer}}, \bibinfo {author} {\bibfnamefont {M.~F.}\ \bibnamefont
  {Ciappina}}, \ and\ \bibinfo {author} {\bibfnamefont {M.}~\bibnamefont
  {Lewenstein}},\ }\href {\doibase 10.1103/4hdl-bdwj} {\bibfield  {journal}
  {\bibinfo  {journal} {Phys. Rev. Lett.}\ }\textbf {\bibinfo {volume} {135}},\
  \bibinfo {pages} {013801} (\bibinfo {year} {2025}{\natexlab{a}})}\BibitemShut
  {NoStop}%
\bibitem [{\citenamefont {Lemieux}\ \emph {et~al.}(2025)\citenamefont
  {Lemieux}, \citenamefont {Jalil}, \citenamefont {Purschke}, \citenamefont
  {Boroumand}, \citenamefont {Hammond}, \citenamefont {Villeneuve},
  \citenamefont {Naumov}, \citenamefont {Brabec},\ and\ \citenamefont
  {Vampa}}]{Lemieux2025}%
  \BibitemOpen
  \bibfield  {author} {\bibinfo {author} {\bibfnamefont {S.}~\bibnamefont
  {Lemieux}}, \bibinfo {author} {\bibfnamefont {S.~A.}\ \bibnamefont {Jalil}},
  \bibinfo {author} {\bibfnamefont {D.~N.}\ \bibnamefont {Purschke}}, \bibinfo
  {author} {\bibfnamefont {N.}~\bibnamefont {Boroumand}}, \bibinfo {author}
  {\bibfnamefont {T.~J.}\ \bibnamefont {Hammond}}, \bibinfo {author}
  {\bibfnamefont {D.}~\bibnamefont {Villeneuve}}, \bibinfo {author}
  {\bibfnamefont {A.}~\bibnamefont {Naumov}}, \bibinfo {author} {\bibfnamefont
  {T.}~\bibnamefont {Brabec}}, \ and\ \bibinfo {author} {\bibfnamefont
  {G.}~\bibnamefont {Vampa}},\ }\href {\doibase 10.1038/s41566-025-01673-6}
  {\bibfield  {journal} {\bibinfo  {journal} {Nature Photonics}\ }\textbf
  {\bibinfo {volume} {19}},\ \bibinfo {pages} {767} (\bibinfo {year}
  {2025})}\BibitemShut {NoStop}%
\bibitem [{\citenamefont {Rivera-Dean}\ \emph
  {et~al.}(2025{\natexlab{b}})\citenamefont {Rivera-Dean}, \citenamefont
  {Stammer}, \citenamefont {Faria},\ and\ \citenamefont
  {Lewenstein}}]{Rivera2025mic}%
  \BibitemOpen
  \bibfield  {author} {\bibinfo {author} {\bibfnamefont {J.}~\bibnamefont
  {Rivera-Dean}}, \bibinfo {author} {\bibfnamefont {P.}~\bibnamefont
  {Stammer}}, \bibinfo {author} {\bibfnamefont {C.~F. d.~M.}\ \bibnamefont
  {Faria}}, \ and\ \bibinfo {author} {\bibfnamefont {M.}~\bibnamefont
  {Lewenstein}},\ }\href {\doibase 10.1103/hb3n-h2zy} {\bibfield  {journal}
  {\bibinfo  {journal} {Phys. Rev. A}\ }\textbf {\bibinfo {volume} {112}},\
  \bibinfo {pages} {063101} (\bibinfo {year} {2025}{\natexlab{b}})}\BibitemShut
  {NoStop}%
\bibitem [{\citenamefont {Glauber}(1963)}]{Glauber1963}%
  \BibitemOpen
  \bibfield  {author} {\bibinfo {author} {\bibfnamefont {R.~J.}\ \bibnamefont
  {Glauber}},\ }\href {\doibase 10.1103/PhysRev.131.2766} {\bibfield  {journal}
  {\bibinfo  {journal} {Phys. Rev.}\ }\textbf {\bibinfo {volume} {131}},\
  \bibinfo {pages} {2766} (\bibinfo {year} {1963})}\BibitemShut {NoStop}%
\bibitem [{\citenamefont {Schnabel}(2017)}]{Schnabel2017}%
  \BibitemOpen
  \bibfield  {author} {\bibinfo {author} {\bibfnamefont {R.}~\bibnamefont
  {Schnabel}},\ }\href {\doibase https://doi.org/10.1016/j.physrep.2017.04.001}
  {\bibfield  {journal} {\bibinfo  {journal} {Physics Reports}\ }\textbf
  {\bibinfo {volume} {684}},\ \bibinfo {pages} {1} (\bibinfo {year} {2017})},\
  \bibinfo {note} {squeezed states of light and their applications in laser
  interferometers}\BibitemShut {NoStop}%
\bibitem [{\citenamefont {Iskhakov}\ \emph {et~al.}(2012)\citenamefont
  {Iskhakov}, \citenamefont {P\'{e}rez}, \citenamefont {Spasibko},
  \citenamefont {Chekhova},\ and\ \citenamefont {Leuchs}}]{Iskhakov2012}%
  \BibitemOpen
  \bibfield  {author} {\bibinfo {author} {\bibfnamefont {T.~S.}\ \bibnamefont
  {Iskhakov}}, \bibinfo {author} {\bibfnamefont {A.~M.}\ \bibnamefont
  {P\'{e}rez}}, \bibinfo {author} {\bibfnamefont {K.~Y.}\ \bibnamefont
  {Spasibko}}, \bibinfo {author} {\bibfnamefont {M.~V.}\ \bibnamefont
  {Chekhova}}, \ and\ \bibinfo {author} {\bibfnamefont {G.}~\bibnamefont
  {Leuchs}},\ }\href {\doibase 10.1364/OL.37.001919} {\bibfield  {journal}
  {\bibinfo  {journal} {Opt. Lett.}\ }\textbf {\bibinfo {volume} {37}},\
  \bibinfo {pages} {1919} (\bibinfo {year} {2012})}\BibitemShut {NoStop}%
\bibitem [{\citenamefont {Manceau}\ \emph {et~al.}(2019)\citenamefont
  {Manceau}, \citenamefont {Spasibko}, \citenamefont {Leuchs}, \citenamefont
  {Filip},\ and\ \citenamefont {Chekhova}}]{Manceau2019}%
  \BibitemOpen
  \bibfield  {author} {\bibinfo {author} {\bibfnamefont {M.}~\bibnamefont
  {Manceau}}, \bibinfo {author} {\bibfnamefont {K.~Y.}\ \bibnamefont
  {Spasibko}}, \bibinfo {author} {\bibfnamefont {G.}~\bibnamefont {Leuchs}},
  \bibinfo {author} {\bibfnamefont {R.}~\bibnamefont {Filip}}, \ and\ \bibinfo
  {author} {\bibfnamefont {M.~V.}\ \bibnamefont {Chekhova}},\ }\href {\doibase
  10.1103/PhysRevLett.123.123606} {\bibfield  {journal} {\bibinfo  {journal}
  {Phys. Rev. Lett.}\ }\textbf {\bibinfo {volume} {123}},\ \bibinfo {pages}
  {123606} (\bibinfo {year} {2019})}\BibitemShut {NoStop}%
\bibitem [{\citenamefont {Heimerl}\ \emph {et~al.}(2024)\citenamefont
  {Heimerl}, \citenamefont {Mikhaylov}, \citenamefont {Meier}, \citenamefont
  {H{\"o}llerer}, \citenamefont {Kaminer}, \citenamefont {Chekhova},\ and\
  \citenamefont {Hommelhoff}}]{Heimerl2024}%
  \BibitemOpen
  \bibfield  {author} {\bibinfo {author} {\bibfnamefont {J.}~\bibnamefont
  {Heimerl}}, \bibinfo {author} {\bibfnamefont {A.}~\bibnamefont {Mikhaylov}},
  \bibinfo {author} {\bibfnamefont {S.}~\bibnamefont {Meier}}, \bibinfo
  {author} {\bibfnamefont {H.}~\bibnamefont {H{\"o}llerer}}, \bibinfo {author}
  {\bibfnamefont {I.}~\bibnamefont {Kaminer}}, \bibinfo {author} {\bibfnamefont
  {M.}~\bibnamefont {Chekhova}}, \ and\ \bibinfo {author} {\bibfnamefont
  {P.}~\bibnamefont {Hommelhoff}},\ }\href {\doibase
  10.1038/s41567-024-02472-6} {\bibfield  {journal} {\bibinfo  {journal}
  {Nature Physics}\ }\textbf {\bibinfo {volume} {20}},\ \bibinfo {pages} {945}
  (\bibinfo {year} {2024})}\BibitemShut {NoStop}%
\bibitem [{\citenamefont {Heimerl}\ \emph {et~al.}(2025)\citenamefont
  {Heimerl}, \citenamefont {Rasputnyi}, \citenamefont {P{\"o}lloth},
  \citenamefont {Meier}, \citenamefont {Chekhova},\ and\ \citenamefont
  {Hommelhoff}}]{Heimerl2025}%
  \BibitemOpen
  \bibfield  {author} {\bibinfo {author} {\bibfnamefont {J.}~\bibnamefont
  {Heimerl}}, \bibinfo {author} {\bibfnamefont {A.}~\bibnamefont {Rasputnyi}},
  \bibinfo {author} {\bibfnamefont {J.}~\bibnamefont {P{\"o}lloth}}, \bibinfo
  {author} {\bibfnamefont {S.}~\bibnamefont {Meier}}, \bibinfo {author}
  {\bibfnamefont {M.}~\bibnamefont {Chekhova}}, \ and\ \bibinfo {author}
  {\bibfnamefont {P.}~\bibnamefont {Hommelhoff}},\ }\href {\doibase
  10.1038/s41567-025-03087-1} {\bibfield  {journal} {\bibinfo  {journal}
  {Nature Physics}\ }\textbf {\bibinfo {volume} {21}},\ \bibinfo {pages} {1899}
  (\bibinfo {year} {2025})}\BibitemShut {NoStop}%
\bibitem [{\citenamefont {Aasi}\ and\ \citenamefont {\textit{et
  al.}}(2013)}]{Aasi2013}%
  \BibitemOpen
  \bibfield  {author} {\bibinfo {author} {\bibfnamefont {J.}~\bibnamefont
  {Aasi}}\ and\ \bibinfo {author} {\bibnamefont {\textit{et al.}}},\ }\href
  {\doibase 10.1038/nphoton.2013.177} {\bibfield  {journal} {\bibinfo
  {journal} {Nature Photonics}\ }\textbf {\bibinfo {volume} {7}},\ \bibinfo
  {pages} {613} (\bibinfo {year} {2013})}\BibitemShut {NoStop}%
\bibitem [{\citenamefont {Mandel}\ and\ \citenamefont
  {Wolf}(1995)}]{Mandel1995}%
  \BibitemOpen
  \bibfield  {author} {\bibinfo {author} {\bibfnamefont {L.}~\bibnamefont
  {Mandel}}\ and\ \bibinfo {author} {\bibfnamefont {E.}~\bibnamefont {Wolf}},\
  }\href@noop {} {\emph {\bibinfo {title} {Optical Coherence and Quantum
  Optics}}}\ (\bibinfo  {publisher} {Cambridge University Press},\ \bibinfo
  {address} {Cambridge},\ \bibinfo {year} {1995})\BibitemShut {NoStop}%
\bibitem [{\citenamefont {Dahan}\ \emph {et~al.}(2021)\citenamefont {Dahan},
  \citenamefont {Gorlach}, \citenamefont {Haeusler}, \citenamefont {Karnieli},
  \citenamefont {Eyal}, \citenamefont {Yousefi}, \citenamefont {Segev},
  \citenamefont {Arie}, \citenamefont {Eisenstein}, \citenamefont
  {Hommelhoff},\ and\ \citenamefont {Kaminer}}]{Dahan2021}%
  \BibitemOpen
  \bibfield  {author} {\bibinfo {author} {\bibfnamefont {R.}~\bibnamefont
  {Dahan}}, \bibinfo {author} {\bibfnamefont {A.}~\bibnamefont {Gorlach}},
  \bibinfo {author} {\bibfnamefont {U.}~\bibnamefont {Haeusler}}, \bibinfo
  {author} {\bibfnamefont {A.}~\bibnamefont {Karnieli}}, \bibinfo {author}
  {\bibfnamefont {O.}~\bibnamefont {Eyal}}, \bibinfo {author} {\bibfnamefont
  {P.}~\bibnamefont {Yousefi}}, \bibinfo {author} {\bibfnamefont
  {M.}~\bibnamefont {Segev}}, \bibinfo {author} {\bibfnamefont
  {A.}~\bibnamefont {Arie}}, \bibinfo {author} {\bibfnamefont {G.}~\bibnamefont
  {Eisenstein}}, \bibinfo {author} {\bibfnamefont {P.}~\bibnamefont
  {Hommelhoff}}, \ and\ \bibinfo {author} {\bibfnamefont {I.}~\bibnamefont
  {Kaminer}},\ }\href {\doibase 10.1126/science.abj7128} {\bibfield  {journal}
  {\bibinfo  {journal} {Science}\ }\textbf {\bibinfo {volume} {373}},\ \bibinfo
  {pages} {eabj7128} (\bibinfo {year} {2021})}\BibitemShut {NoStop}%
\bibitem [{\citenamefont {Kl\"under}\ \emph {et~al.}(2011)\citenamefont
  {Kl\"under}, \citenamefont {Dahlstr\"om}, \citenamefont {Gisselbrecht},
  \citenamefont {Fordell}, \citenamefont {Swoboda}, \citenamefont {Gu\'enot},
  \citenamefont {Johnsson}, \citenamefont {Caillat}, \citenamefont
  {Mauritsson}, \citenamefont {Maquet}, \citenamefont {Ta\"{\i}eb},\ and\
  \citenamefont {L'Huillier}}]{Klunder2011}%
  \BibitemOpen
  \bibfield  {author} {\bibinfo {author} {\bibfnamefont {K.}~\bibnamefont
  {Kl\"under}}, \bibinfo {author} {\bibfnamefont {J.~M.}\ \bibnamefont
  {Dahlstr\"om}}, \bibinfo {author} {\bibfnamefont {M.}~\bibnamefont
  {Gisselbrecht}}, \bibinfo {author} {\bibfnamefont {T.}~\bibnamefont
  {Fordell}}, \bibinfo {author} {\bibfnamefont {M.}~\bibnamefont {Swoboda}},
  \bibinfo {author} {\bibfnamefont {D.}~\bibnamefont {Gu\'enot}}, \bibinfo
  {author} {\bibfnamefont {P.}~\bibnamefont {Johnsson}}, \bibinfo {author}
  {\bibfnamefont {J.}~\bibnamefont {Caillat}}, \bibinfo {author} {\bibfnamefont
  {J.}~\bibnamefont {Mauritsson}}, \bibinfo {author} {\bibfnamefont
  {A.}~\bibnamefont {Maquet}}, \bibinfo {author} {\bibfnamefont
  {R.}~\bibnamefont {Ta\"{\i}eb}}, \ and\ \bibinfo {author} {\bibfnamefont
  {A.}~\bibnamefont {L'Huillier}},\ }\href {\doibase
  10.1103/PhysRevLett.106.143002} {\bibfield  {journal} {\bibinfo  {journal}
  {Phys. Rev. Lett.}\ }\textbf {\bibinfo {volume} {106}},\ \bibinfo {pages}
  {143002} (\bibinfo {year} {2011})}\BibitemShut {NoStop}%
\bibitem [{\citenamefont {Bourassin-Bouchet}\ \emph {et~al.}(2020)\citenamefont
  {Bourassin-Bouchet}, \citenamefont {Barreau}, \citenamefont {Gruson},
  \citenamefont {Hergott}, \citenamefont {Qu\'er\'e}, \citenamefont
  {Sali\`eres},\ and\ \citenamefont {Ruchon}}]{Bourassin2020}%
  \BibitemOpen
  \bibfield  {author} {\bibinfo {author} {\bibfnamefont {C.}~\bibnamefont
  {Bourassin-Bouchet}}, \bibinfo {author} {\bibfnamefont {L.}~\bibnamefont
  {Barreau}}, \bibinfo {author} {\bibfnamefont {V.}~\bibnamefont {Gruson}},
  \bibinfo {author} {\bibfnamefont {J.-F.}\ \bibnamefont {Hergott}}, \bibinfo
  {author} {\bibfnamefont {F.}~\bibnamefont {Qu\'er\'e}}, \bibinfo {author}
  {\bibfnamefont {P.}~\bibnamefont {Sali\`eres}}, \ and\ \bibinfo {author}
  {\bibfnamefont {T.}~\bibnamefont {Ruchon}},\ }\href {\doibase
  10.1103/PhysRevX.10.031048} {\bibfield  {journal} {\bibinfo  {journal} {Phys.
  Rev. X}\ }\textbf {\bibinfo {volume} {10}},\ \bibinfo {pages} {031048}
  (\bibinfo {year} {2020})}\BibitemShut {NoStop}%
\bibitem [{\citenamefont {Berkane}\ \emph {et~al.}(2025)\citenamefont
  {Berkane}, \citenamefont {Ta\"{\i}eb}, \citenamefont {Granveau},
  \citenamefont {Sali\`eres}, \citenamefont {Bourassin-Bouchet}, \citenamefont
  {L\'ev\^eque},\ and\ \citenamefont {Caillat}}]{Berkane2025}%
  \BibitemOpen
  \bibfield  {author} {\bibinfo {author} {\bibfnamefont {M.}~\bibnamefont
  {Berkane}}, \bibinfo {author} {\bibfnamefont {R.}~\bibnamefont {Ta\"{\i}eb}},
  \bibinfo {author} {\bibfnamefont {G.}~\bibnamefont {Granveau}}, \bibinfo
  {author} {\bibfnamefont {P.}~\bibnamefont {Sali\`eres}}, \bibinfo {author}
  {\bibfnamefont {C.}~\bibnamefont {Bourassin-Bouchet}}, \bibinfo {author}
  {\bibfnamefont {C.}~\bibnamefont {L\'ev\^eque}}, \ and\ \bibinfo {author}
  {\bibfnamefont {J.}~\bibnamefont {Caillat}},\ }\href {\doibase
  10.1103/PhysRevA.111.L041101} {\bibfield  {journal} {\bibinfo  {journal}
  {Phys. Rev. A}\ }\textbf {\bibinfo {volume} {111}},\ \bibinfo {pages}
  {L041101} (\bibinfo {year} {2025})}\BibitemShut {NoStop}%
\bibitem [{\citenamefont {Lange}\ \emph {et~al.}(2024)\citenamefont {Lange},
  \citenamefont {Hansen},\ and\ \citenamefont {Madsen}}]{Lange2024}%
  \BibitemOpen
  \bibfield  {author} {\bibinfo {author} {\bibfnamefont {C.~S.}\ \bibnamefont
  {Lange}}, \bibinfo {author} {\bibfnamefont {T.}~\bibnamefont {Hansen}}, \
  and\ \bibinfo {author} {\bibfnamefont {L.~B.}\ \bibnamefont {Madsen}},\
  }\href {\doibase 10.1103/PhysRevA.109.033110} {\bibfield  {journal} {\bibinfo
   {journal} {Phys. Rev. A}\ }\textbf {\bibinfo {volume} {109}},\ \bibinfo
  {pages} {033110} (\bibinfo {year} {2024})}\BibitemShut {NoStop}%
\bibitem [{\citenamefont {Messiah}(1962)}]{Messiah1962}%
  \BibitemOpen
  \bibfield  {author} {\bibinfo {author} {\bibfnamefont {A.}~\bibnamefont
  {Messiah}},\ }\href@noop {} {\emph {\bibinfo {title} {Quantum Mechanics}}},\
  Vol.~\bibinfo {volume} {2}\ (\bibinfo  {publisher} {North-Holland Publishing
  Company},\ \bibinfo {address} {Amsterdam},\ \bibinfo {year} {1962})\ \bibinfo
  {note} {translated from the French by J. Anderson}\BibitemShut {NoStop}%
\bibitem [{\citenamefont {Scully}\ and\ \citenamefont
  {Zubairy}(1997)}]{Scully_Zubairy_1997}%
  \BibitemOpen
  \bibfield  {author} {\bibinfo {author} {\bibfnamefont {M.~O.}\ \bibnamefont
  {Scully}}\ and\ \bibinfo {author} {\bibfnamefont {M.~S.}\ \bibnamefont
  {Zubairy}},\ }\href@noop {} {\emph {\bibinfo {title} {Quantum Optics}}}\
  (\bibinfo  {publisher} {Cambridge University Press},\ \bibinfo {year}
  {1997})\BibitemShut {NoStop}%
\bibitem [{\citenamefont {Laurell}\ \emph {et~al.}(2022)\citenamefont
  {Laurell}, \citenamefont {Finkelstein-Shapiro}, \citenamefont {Dittel},
  \citenamefont {Guo}, \citenamefont {Demjaha}, \citenamefont {Ammitzb\"oll},
  \citenamefont {Weissenbilder}, \citenamefont {Neori\ifmmode \check{c}\else
  \v{c}\fi{}i\ifmmode~\acute{c}\else \'{c}\fi{}}, \citenamefont {Luo},
  \citenamefont {Gisselbrecht}, \citenamefont {Arnold}, \citenamefont
  {Buchleitner}, \citenamefont {Pullerits}, \citenamefont {L'Huillier},\ and\
  \citenamefont {Busto}}]{Laurell2022}%
  \BibitemOpen
  \bibfield  {author} {\bibinfo {author} {\bibfnamefont {H.}~\bibnamefont
  {Laurell}}, \bibinfo {author} {\bibfnamefont {D.}~\bibnamefont
  {Finkelstein-Shapiro}}, \bibinfo {author} {\bibfnamefont {C.}~\bibnamefont
  {Dittel}}, \bibinfo {author} {\bibfnamefont {C.}~\bibnamefont {Guo}},
  \bibinfo {author} {\bibfnamefont {R.}~\bibnamefont {Demjaha}}, \bibinfo
  {author} {\bibfnamefont {M.}~\bibnamefont {Ammitzb\"oll}}, \bibinfo {author}
  {\bibfnamefont {R.}~\bibnamefont {Weissenbilder}}, \bibinfo {author}
  {\bibfnamefont {L.}~\bibnamefont {Neori\ifmmode \check{c}\else
  \v{c}\fi{}i\ifmmode~\acute{c}\else \'{c}\fi{}}}, \bibinfo {author}
  {\bibfnamefont {S.}~\bibnamefont {Luo}}, \bibinfo {author} {\bibfnamefont
  {M.}~\bibnamefont {Gisselbrecht}}, \bibinfo {author} {\bibfnamefont {C.~L.}\
  \bibnamefont {Arnold}}, \bibinfo {author} {\bibfnamefont {A.}~\bibnamefont
  {Buchleitner}}, \bibinfo {author} {\bibfnamefont {T.}~\bibnamefont
  {Pullerits}}, \bibinfo {author} {\bibfnamefont {A.}~\bibnamefont
  {L'Huillier}}, \ and\ \bibinfo {author} {\bibfnamefont {D.}~\bibnamefont
  {Busto}},\ }\href {\doibase 10.1103/PhysRevResearch.4.033220} {\bibfield
  {journal} {\bibinfo  {journal} {Phys. Rev. Res.}\ }\textbf {\bibinfo {volume}
  {4}},\ \bibinfo {pages} {033220} (\bibinfo {year} {2022})}\BibitemShut
  {NoStop}%
\bibitem [{\citenamefont {Laurell}\ \emph {et~al.}(2025)\citenamefont
  {Laurell}, \citenamefont {Luo}, \citenamefont {Weissenbilder}, \citenamefont
  {Ammitzb{\"o}ll}, \citenamefont {Ahmed}, \citenamefont {S{\"o}derberg},
  \citenamefont {Petersson}, \citenamefont {Poulain}, \citenamefont {Guo},
  \citenamefont {Dittel}, \citenamefont {Finkelstein-Shapiro}, \citenamefont
  {Squibb}, \citenamefont {Feifel}, \citenamefont {Gisselbrecht}, \citenamefont
  {Arnold}, \citenamefont {Buchleitner}, \citenamefont {Lindroth},
  \citenamefont {Frisk~Kockum}, \citenamefont {L'Huillier},\ and\ \citenamefont
  {Busto}}]{Laurell2025}%
  \BibitemOpen
  \bibfield  {author} {\bibinfo {author} {\bibfnamefont {H.}~\bibnamefont
  {Laurell}}, \bibinfo {author} {\bibfnamefont {S.}~\bibnamefont {Luo}},
  \bibinfo {author} {\bibfnamefont {R.}~\bibnamefont {Weissenbilder}}, \bibinfo
  {author} {\bibfnamefont {M.}~\bibnamefont {Ammitzb{\"o}ll}}, \bibinfo
  {author} {\bibfnamefont {S.}~\bibnamefont {Ahmed}}, \bibinfo {author}
  {\bibfnamefont {H.}~\bibnamefont {S{\"o}derberg}}, \bibinfo {author}
  {\bibfnamefont {C.~L.~M.}\ \bibnamefont {Petersson}}, \bibinfo {author}
  {\bibfnamefont {V.}~\bibnamefont {Poulain}}, \bibinfo {author} {\bibfnamefont
  {C.}~\bibnamefont {Guo}}, \bibinfo {author} {\bibfnamefont {C.}~\bibnamefont
  {Dittel}}, \bibinfo {author} {\bibfnamefont {D.}~\bibnamefont
  {Finkelstein-Shapiro}}, \bibinfo {author} {\bibfnamefont {R.~J.}\
  \bibnamefont {Squibb}}, \bibinfo {author} {\bibfnamefont {R.}~\bibnamefont
  {Feifel}}, \bibinfo {author} {\bibfnamefont {M.}~\bibnamefont
  {Gisselbrecht}}, \bibinfo {author} {\bibfnamefont {C.~L.}\ \bibnamefont
  {Arnold}}, \bibinfo {author} {\bibfnamefont {A.}~\bibnamefont {Buchleitner}},
  \bibinfo {author} {\bibfnamefont {E.}~\bibnamefont {Lindroth}}, \bibinfo
  {author} {\bibfnamefont {A.}~\bibnamefont {Frisk~Kockum}}, \bibinfo {author}
  {\bibfnamefont {A.}~\bibnamefont {L'Huillier}}, \ and\ \bibinfo {author}
  {\bibfnamefont {D.}~\bibnamefont {Busto}},\ }\href {\doibase
  10.1038/s41566-024-01607-8} {\bibfield  {journal} {\bibinfo  {journal}
  {Nature Photonics}\ }\textbf {\bibinfo {volume} {19}},\ \bibinfo {pages}
  {352} (\bibinfo {year} {2025})}\BibitemShut {NoStop}%
\bibitem [{\citenamefont {Caillat}\ \emph {et~al.}(2011)\citenamefont
  {Caillat}, \citenamefont {Maquet}, \citenamefont {Haessler}, \citenamefont
  {Fabre}, \citenamefont {Ruchon}, \citenamefont {Sali\`eres}, \citenamefont
  {Mairesse},\ and\ \citenamefont {Ta\"{\i}eb}}]{Caillat2011}%
  \BibitemOpen
  \bibfield  {author} {\bibinfo {author} {\bibfnamefont {J.}~\bibnamefont
  {Caillat}}, \bibinfo {author} {\bibfnamefont {A.}~\bibnamefont {Maquet}},
  \bibinfo {author} {\bibfnamefont {S.}~\bibnamefont {Haessler}}, \bibinfo
  {author} {\bibfnamefont {B.}~\bibnamefont {Fabre}}, \bibinfo {author}
  {\bibfnamefont {T.}~\bibnamefont {Ruchon}}, \bibinfo {author} {\bibfnamefont
  {P.}~\bibnamefont {Sali\`eres}}, \bibinfo {author} {\bibfnamefont
  {Y.}~\bibnamefont {Mairesse}}, \ and\ \bibinfo {author} {\bibfnamefont
  {R.}~\bibnamefont {Ta\"{\i}eb}},\ }\href {\doibase
  10.1103/PhysRevLett.106.093002} {\bibfield  {journal} {\bibinfo  {journal}
  {Phys. Rev. Lett.}\ }\textbf {\bibinfo {volume} {106}},\ \bibinfo {pages}
  {093002} (\bibinfo {year} {2011})}\BibitemShut {NoStop}%
\bibitem [{\citenamefont {Beaulieu}\ \emph {et~al.}(2017)\citenamefont
  {Beaulieu}, \citenamefont {Comby}, \citenamefont {Clergerie}, \citenamefont
  {Caillat}, \citenamefont {Descamps}, \citenamefont {Dudovich}, \citenamefont
  {Fabre}, \citenamefont {Géneaux}, \citenamefont {Légaré}, \citenamefont
  {Petit}, \citenamefont {Pons}, \citenamefont {Porat}, \citenamefont {Ruchon},
  \citenamefont {Taïeb}, \citenamefont {Blanchet},\ and\ \citenamefont
  {Mairesse}}]{Beaulieu2017}%
  \BibitemOpen
  \bibfield  {author} {\bibinfo {author} {\bibfnamefont {S.}~\bibnamefont
  {Beaulieu}}, \bibinfo {author} {\bibfnamefont {A.}~\bibnamefont {Comby}},
  \bibinfo {author} {\bibfnamefont {A.}~\bibnamefont {Clergerie}}, \bibinfo
  {author} {\bibfnamefont {J.}~\bibnamefont {Caillat}}, \bibinfo {author}
  {\bibfnamefont {D.}~\bibnamefont {Descamps}}, \bibinfo {author}
  {\bibfnamefont {N.}~\bibnamefont {Dudovich}}, \bibinfo {author}
  {\bibfnamefont {B.}~\bibnamefont {Fabre}}, \bibinfo {author} {\bibfnamefont
  {R.}~\bibnamefont {Géneaux}}, \bibinfo {author} {\bibfnamefont
  {F.}~\bibnamefont {Légaré}}, \bibinfo {author} {\bibfnamefont
  {S.}~\bibnamefont {Petit}}, \bibinfo {author} {\bibfnamefont
  {B.}~\bibnamefont {Pons}}, \bibinfo {author} {\bibfnamefont {G.}~\bibnamefont
  {Porat}}, \bibinfo {author} {\bibfnamefont {T.}~\bibnamefont {Ruchon}},
  \bibinfo {author} {\bibfnamefont {R.}~\bibnamefont {Taïeb}}, \bibinfo
  {author} {\bibfnamefont {V.}~\bibnamefont {Blanchet}}, \ and\ \bibinfo
  {author} {\bibfnamefont {Y.}~\bibnamefont {Mairesse}},\ }\href {\doibase
  10.1126/science.aao5624} {\bibfield  {journal} {\bibinfo  {journal}
  {Science}\ }\textbf {\bibinfo {volume} {358}},\ \bibinfo {pages} {1288}
  (\bibinfo {year} {2017})}\BibitemShut {NoStop}%
\bibitem [{\citenamefont {Autuori}\ \emph {et~al.}(2022)\citenamefont
  {Autuori}, \citenamefont {Platzer}, \citenamefont {Lejman}, \citenamefont
  {Gallician}, \citenamefont {Maëder}, \citenamefont {Covolo}, \citenamefont
  {Bosse}, \citenamefont {Dalui}, \citenamefont {Bresteau}, \citenamefont
  {Hergott}, \citenamefont {Tcherbakoff}, \citenamefont {Marroux},
  \citenamefont {Loriot}, \citenamefont {Lépine}, \citenamefont {Poisson},
  \citenamefont {Taïeb}, \citenamefont {Caillat},\ and\ \citenamefont
  {Salières}}]{Autuori2022}%
  \BibitemOpen
  \bibfield  {author} {\bibinfo {author} {\bibfnamefont {A.}~\bibnamefont
  {Autuori}}, \bibinfo {author} {\bibfnamefont {D.}~\bibnamefont {Platzer}},
  \bibinfo {author} {\bibfnamefont {M.}~\bibnamefont {Lejman}}, \bibinfo
  {author} {\bibfnamefont {G.}~\bibnamefont {Gallician}}, \bibinfo {author}
  {\bibfnamefont {L.}~\bibnamefont {Maëder}}, \bibinfo {author} {\bibfnamefont
  {A.}~\bibnamefont {Covolo}}, \bibinfo {author} {\bibfnamefont
  {L.}~\bibnamefont {Bosse}}, \bibinfo {author} {\bibfnamefont
  {M.}~\bibnamefont {Dalui}}, \bibinfo {author} {\bibfnamefont
  {D.}~\bibnamefont {Bresteau}}, \bibinfo {author} {\bibfnamefont {J.-F.}\
  \bibnamefont {Hergott}}, \bibinfo {author} {\bibfnamefont {O.}~\bibnamefont
  {Tcherbakoff}}, \bibinfo {author} {\bibfnamefont {H.~J.~B.}\ \bibnamefont
  {Marroux}}, \bibinfo {author} {\bibfnamefont {V.}~\bibnamefont {Loriot}},
  \bibinfo {author} {\bibfnamefont {F.}~\bibnamefont {Lépine}}, \bibinfo
  {author} {\bibfnamefont {L.}~\bibnamefont {Poisson}}, \bibinfo {author}
  {\bibfnamefont {R.}~\bibnamefont {Taïeb}}, \bibinfo {author} {\bibfnamefont
  {J.}~\bibnamefont {Caillat}}, \ and\ \bibinfo {author} {\bibfnamefont
  {P.}~\bibnamefont {Salières}},\ }\href {\doibase 10.1126/sciadv.abl7594}
  {\bibfield  {journal} {\bibinfo  {journal} {Science Advances}\ }\textbf
  {\bibinfo {volume} {8}},\ \bibinfo {pages} {eabl7594} (\bibinfo {year}
  {2022})}\BibitemShut {NoStop}%
\bibitem [{\citenamefont {Muller}(2002)}]{Muller2002}%
  \BibitemOpen
  \bibfield  {author} {\bibinfo {author} {\bibfnamefont {H.~G.}\ \bibnamefont
  {Muller}},\ }\href {\doibase 10.1007/s00340-002-0894-8} {\bibfield  {journal}
  {\bibinfo  {journal} {Applied Physics B}\ }\textbf {\bibinfo {volume} {74}},\
  \bibinfo {pages} {s17} (\bibinfo {year} {2002})}\BibitemShut {NoStop}%
\bibitem [{\citenamefont {Zurek}(2003)}]{Zurek2003}%
  \BibitemOpen
  \bibfield  {author} {\bibinfo {author} {\bibfnamefont {W.~H.}\ \bibnamefont
  {Zurek}},\ }\href {\doibase 10.1103/RevModPhys.75.715} {\bibfield  {journal}
  {\bibinfo  {journal} {Rev. Mod. Phys.}\ }\textbf {\bibinfo {volume} {75}},\
  \bibinfo {pages} {715} (\bibinfo {year} {2003})}\BibitemShut {NoStop}%
\bibitem [{\citenamefont {Breuer}\ and\ \citenamefont
  {Petruccione}(2007)}]{Breuer2007}%
  \BibitemOpen
  \bibfield  {author} {\bibinfo {author} {\bibfnamefont {H.-P.}\ \bibnamefont
  {Breuer}}\ and\ \bibinfo {author} {\bibfnamefont {F.}~\bibnamefont
  {Petruccione}},\ }\href {\doibase 10.1093/acprof:oso/9780199213900.001.0001}
  {\emph {\bibinfo {title} {The Theory of Open Quantum Systems}}}\ (\bibinfo
  {publisher} {Oxford University Press},\ \bibinfo {year} {2007})\BibitemShut
  {NoStop}%
\bibitem [{\citenamefont {Fabre}\ and\ \citenamefont
  {Treps}(2020)}]{Fabre2020}%
  \BibitemOpen
  \bibfield  {author} {\bibinfo {author} {\bibfnamefont {C.}~\bibnamefont
  {Fabre}}\ and\ \bibinfo {author} {\bibfnamefont {N.}~\bibnamefont {Treps}},\
  }\href {\doibase 10.1103/RevModPhys.92.035005} {\bibfield  {journal}
  {\bibinfo  {journal} {Rev. Mod. Phys.}\ }\textbf {\bibinfo {volume} {92}},\
  \bibinfo {pages} {035005} (\bibinfo {year} {2020})}\BibitemShut {NoStop}%
\bibitem [{\citenamefont {Isinger}\ \emph {et~al.}(2017)\citenamefont
  {Isinger}, \citenamefont {Squibb}, \citenamefont {Busto}, \citenamefont
  {Zhong}, \citenamefont {Harth}, \citenamefont {Kroon}, \citenamefont {Nandi},
  \citenamefont {Arnold}, \citenamefont {Miranda}, \citenamefont {Dahlström},
  \citenamefont {Lindroth}, \citenamefont {Feifel}, \citenamefont
  {Gisselbrecht},\ and\ \citenamefont {L’Huillier}}]{Isinger2017}%
  \BibitemOpen
  \bibfield  {author} {\bibinfo {author} {\bibfnamefont {M.}~\bibnamefont
  {Isinger}}, \bibinfo {author} {\bibfnamefont {R.~J.}\ \bibnamefont {Squibb}},
  \bibinfo {author} {\bibfnamefont {D.}~\bibnamefont {Busto}}, \bibinfo
  {author} {\bibfnamefont {S.}~\bibnamefont {Zhong}}, \bibinfo {author}
  {\bibfnamefont {A.}~\bibnamefont {Harth}}, \bibinfo {author} {\bibfnamefont
  {D.}~\bibnamefont {Kroon}}, \bibinfo {author} {\bibfnamefont
  {S.}~\bibnamefont {Nandi}}, \bibinfo {author} {\bibfnamefont {C.~L.}\
  \bibnamefont {Arnold}}, \bibinfo {author} {\bibfnamefont {M.}~\bibnamefont
  {Miranda}}, \bibinfo {author} {\bibfnamefont {J.~M.}\ \bibnamefont
  {Dahlström}}, \bibinfo {author} {\bibfnamefont {E.}~\bibnamefont
  {Lindroth}}, \bibinfo {author} {\bibfnamefont {R.}~\bibnamefont {Feifel}},
  \bibinfo {author} {\bibfnamefont {M.}~\bibnamefont {Gisselbrecht}}, \ and\
  \bibinfo {author} {\bibfnamefont {A.}~\bibnamefont {L’Huillier}},\ }\href
  {\doibase 10.1126/science.aao7043} {\bibfield  {journal} {\bibinfo  {journal}
  {Science}\ }\textbf {\bibinfo {volume} {358}},\ \bibinfo {pages} {893}
  (\bibinfo {year} {2017})}\BibitemShut {NoStop}%
\bibitem [{\citenamefont {Wahlstr\"{o}m}\ \emph {et~al.}(1993)\citenamefont
  {Wahlstr\"{o}m}, \citenamefont {Larsson}, \citenamefont {Persson},
  \citenamefont {Starczewski}, \citenamefont {Svanberg}, \citenamefont
  {Sali\`{e}res}, \citenamefont {Balcou},\ and\ \citenamefont
  {L'Huillier}}]{Wahlstrom1993}%
  \BibitemOpen
  \bibfield  {author} {\bibinfo {author} {\bibfnamefont {C.~G.}\ \bibnamefont
  {Wahlstr\"{o}m}}, \bibinfo {author} {\bibfnamefont {J.}~\bibnamefont
  {Larsson}}, \bibinfo {author} {\bibfnamefont {A.}~\bibnamefont {Persson}},
  \bibinfo {author} {\bibfnamefont {T.}~\bibnamefont {Starczewski}}, \bibinfo
  {author} {\bibnamefont {Svanberg}}, \bibinfo {author} {\bibfnamefont
  {P.}~\bibnamefont {Sali\`{e}res}}, \bibinfo {author} {\bibfnamefont
  {P.}~\bibnamefont {Balcou}}, \ and\ \bibinfo {author} {\bibfnamefont
  {A.}~\bibnamefont {L'Huillier}},\ }\href@noop {} {\bibfield  {journal}
  {\bibinfo  {journal} {Phys. Rev. A}\ }\textbf {\bibinfo {volume} {48}}
  (\bibinfo {year} {1993})}\BibitemShut {NoStop}%
\bibitem [{\citenamefont {Weissenbilder}\ \emph {et~al.}(2022)\citenamefont
  {Weissenbilder}, \citenamefont {Carlstr{\"o}m}, \citenamefont {Rego},
  \citenamefont {Guo}, \citenamefont {Heyl}, \citenamefont {Smorenburg},
  \citenamefont {Constant}, \citenamefont {Arnold},\ and\ \citenamefont
  {L'Huillier}}]{Weissenbilder2022}%
  \BibitemOpen
  \bibfield  {author} {\bibinfo {author} {\bibfnamefont {R.}~\bibnamefont
  {Weissenbilder}}, \bibinfo {author} {\bibfnamefont {S.}~\bibnamefont
  {Carlstr{\"o}m}}, \bibinfo {author} {\bibfnamefont {L.}~\bibnamefont {Rego}},
  \bibinfo {author} {\bibfnamefont {C.}~\bibnamefont {Guo}}, \bibinfo {author}
  {\bibfnamefont {C.~M.}\ \bibnamefont {Heyl}}, \bibinfo {author}
  {\bibfnamefont {P.}~\bibnamefont {Smorenburg}}, \bibinfo {author}
  {\bibfnamefont {E.}~\bibnamefont {Constant}}, \bibinfo {author}
  {\bibfnamefont {C.~L.}\ \bibnamefont {Arnold}}, \ and\ \bibinfo {author}
  {\bibfnamefont {A.}~\bibnamefont {L'Huillier}},\ }\href {\doibase
  10.1038/s42254-022-00522-7} {\bibfield  {journal} {\bibinfo  {journal}
  {Nature Reviews Physics}\ }\textbf {\bibinfo {volume} {4}},\ \bibinfo {pages}
  {713} (\bibinfo {year} {2022})}\BibitemShut {NoStop}%
\bibitem [{\citenamefont {Ruo-Berchera}\ \emph {et~al.}(2015)\citenamefont
  {Ruo-Berchera}, \citenamefont {Degiovanni}, \citenamefont {Olivares},
  \citenamefont {Samantaray}, \citenamefont {Traina},\ and\ \citenamefont
  {Genovese}}]{Ruo2015}%
  \BibitemOpen
  \bibfield  {author} {\bibinfo {author} {\bibfnamefont {I.}~\bibnamefont
  {Ruo-Berchera}}, \bibinfo {author} {\bibfnamefont {I.~P.}\ \bibnamefont
  {Degiovanni}}, \bibinfo {author} {\bibfnamefont {S.}~\bibnamefont
  {Olivares}}, \bibinfo {author} {\bibfnamefont {N.}~\bibnamefont
  {Samantaray}}, \bibinfo {author} {\bibfnamefont {P.}~\bibnamefont {Traina}},
  \ and\ \bibinfo {author} {\bibfnamefont {M.}~\bibnamefont {Genovese}},\
  }\href {\doibase 10.1103/PhysRevA.92.053821} {\bibfield  {journal} {\bibinfo
  {journal} {Phys. Rev. A}\ }\textbf {\bibinfo {volume} {92}},\ \bibinfo
  {pages} {053821} (\bibinfo {year} {2015})}\BibitemShut {NoStop}%
\bibitem [{\citenamefont {Heinrich}\ \emph {et~al.}(2021)\citenamefont
  {Heinrich}, \citenamefont {Saule}, \citenamefont {H{\"o}gner}, \citenamefont
  {Cui}, \citenamefont {Yakovlev}, \citenamefont {Pupeza},\ and\ \citenamefont
  {Kleineberg}}]{Heinrich2021}%
  \BibitemOpen
  \bibfield  {author} {\bibinfo {author} {\bibfnamefont {S.}~\bibnamefont
  {Heinrich}}, \bibinfo {author} {\bibfnamefont {T.}~\bibnamefont {Saule}},
  \bibinfo {author} {\bibfnamefont {M.}~\bibnamefont {H{\"o}gner}}, \bibinfo
  {author} {\bibfnamefont {Y.}~\bibnamefont {Cui}}, \bibinfo {author}
  {\bibfnamefont {V.~S.}\ \bibnamefont {Yakovlev}}, \bibinfo {author}
  {\bibfnamefont {I.}~\bibnamefont {Pupeza}}, \ and\ \bibinfo {author}
  {\bibfnamefont {U.}~\bibnamefont {Kleineberg}},\ }\href {\doibase
  10.1038/s41467-021-23650-7} {\bibfield  {journal} {\bibinfo  {journal}
  {Nature Communications}\ }\textbf {\bibinfo {volume} {12}},\ \bibinfo {pages}
  {3404} (\bibinfo {year} {2021})}\BibitemShut {NoStop}%
\bibitem [{\citenamefont {Berrah}\ \emph {et~al.}(2025)\citenamefont {Berrah},
  \citenamefont {Cryan}, \citenamefont {Robles}, \citenamefont {Driver},
  \citenamefont {Marinelli},\ and\ \citenamefont {Bucksbaum}}]{Berrah2025}%
  \BibitemOpen
  \bibfield  {author} {\bibinfo {author} {\bibfnamefont {N.}~\bibnamefont
  {Berrah}}, \bibinfo {author} {\bibfnamefont {J.}~\bibnamefont {Cryan}},
  \bibinfo {author} {\bibfnamefont {R.}~\bibnamefont {Robles}}, \bibinfo
  {author} {\bibfnamefont {T.}~\bibnamefont {Driver}}, \bibinfo {author}
  {\bibfnamefont {A.}~\bibnamefont {Marinelli}}, \ and\ \bibinfo {author}
  {\bibfnamefont {P.}~\bibnamefont {Bucksbaum}},\ }\href {\doibase
  10.1364/AOP.540527} {\bibfield  {journal} {\bibinfo  {journal} {Adv. Opt.
  Photon.}\ }\textbf {\bibinfo {volume} {17}},\ \bibinfo {pages} {623}
  (\bibinfo {year} {2025})}\BibitemShut {NoStop}%
\bibitem [{\citenamefont {Spasibko}\ \emph {et~al.}(2017)\citenamefont
  {Spasibko}, \citenamefont {Kopylov}, \citenamefont {Krutyanskiy},
  \citenamefont {Murzina}, \citenamefont {Leuchs},\ and\ \citenamefont
  {Chekhova}}]{Spasibko2017}%
  \BibitemOpen
  \bibfield  {author} {\bibinfo {author} {\bibfnamefont {K.~Y.}\ \bibnamefont
  {Spasibko}}, \bibinfo {author} {\bibfnamefont {D.~A.}\ \bibnamefont
  {Kopylov}}, \bibinfo {author} {\bibfnamefont {V.~L.}\ \bibnamefont
  {Krutyanskiy}}, \bibinfo {author} {\bibfnamefont {T.~V.}\ \bibnamefont
  {Murzina}}, \bibinfo {author} {\bibfnamefont {G.}~\bibnamefont {Leuchs}}, \
  and\ \bibinfo {author} {\bibfnamefont {M.~V.}\ \bibnamefont {Chekhova}},\
  }\href {\doibase 10.1103/PhysRevLett.119.223603} {\bibfield  {journal}
  {\bibinfo  {journal} {Phys. Rev. Lett.}\ }\textbf {\bibinfo {volume} {119}},\
  \bibinfo {pages} {223603} (\bibinfo {year} {2017})}\BibitemShut {NoStop}%
\bibitem [{\citenamefont {Javanainen}\ \emph {et~al.}(1988)\citenamefont
  {Javanainen}, \citenamefont {Eberly},\ and\ \citenamefont
  {Su}}]{Javanainen1988}%
  \BibitemOpen
  \bibfield  {author} {\bibinfo {author} {\bibfnamefont {J.}~\bibnamefont
  {Javanainen}}, \bibinfo {author} {\bibfnamefont {J.~H.}\ \bibnamefont
  {Eberly}}, \ and\ \bibinfo {author} {\bibfnamefont {Q.}~\bibnamefont {Su}},\
  }\href {\doibase 10.1103/PhysRevA.38.3430} {\bibfield  {journal} {\bibinfo
  {journal} {Phys. Rev. A}\ }\textbf {\bibinfo {volume} {38}},\ \bibinfo
  {pages} {3430} (\bibinfo {year} {1988})}\BibitemShut {NoStop}%
\bibitem [{\citenamefont {McLachlan}(2022)}]{McLachlan2022}%
  \BibitemOpen
  \bibfield  {author} {\bibinfo {author} {\bibfnamefont {R.~I.}\ \bibnamefont
  {McLachlan}},\ }\href {https://api.semanticscholar.org/CorpusID:233324517}
  {\bibfield  {journal} {\bibinfo  {journal} {Comm. Comp. Phys}\ }\textbf
  {\bibinfo {volume} {31}},\ \bibinfo {pages} {987} (\bibinfo {year}
  {2022})}\BibitemShut {NoStop}%
\bibitem [{Note1()}]{Note1}%
  \BibitemOpen
  \bibinfo {note} {The underlying code is available in the GitHub
  repository~\protect \url
  {https://github.com/jonathan-dubois/TDSE_ND_MPI}}\BibitemShut {NoStop}%
\bibitem [{\citenamefont {Bishop}\ and\ \citenamefont
  {Vourdas}(1994)}]{Bishop1994}%
  \BibitemOpen
  \bibfield  {author} {\bibinfo {author} {\bibfnamefont {R.~F.}\ \bibnamefont
  {Bishop}}\ and\ \bibinfo {author} {\bibfnamefont {A.}~\bibnamefont
  {Vourdas}},\ }\href {\doibase 10.1103/PhysRevA.50.4488} {\bibfield  {journal}
  {\bibinfo  {journal} {Phys. Rev. A}\ }\textbf {\bibinfo {volume} {50}},\
  \bibinfo {pages} {4488} (\bibinfo {year} {1994})}\BibitemShut {NoStop}%
\end{thebibliography}

%

\appendix
\section{Light-matter Hamilton operators in different representations \label{sec:derivation_H}}
\subsection{Generic representation}
We start from the standard light-matter Hamilton operator~\cite{Gorlach2020, Lange2024}
\begin{equation}
    \tilde{\opH} = \tilde{\opH}_A + \tilde{\opH}_{\rm int} + \tilde{\opH}_F ,
\end{equation}
with the Hamilton operators for matter $\tilde{\opH}_A$, for light $\tilde{\opH}_F$, and for light-matter interaction $\tilde{\opH}_{\rm int}$.
We consider a general quantized light described by the vector potential
\begin{equation}
    \tilde{\mathbf{A}} (\mathbf{r}) = \sum_{\mathbf{k} \sigma} \dfrac{f_{\bf k}}{\omega_{\mathbf{k}}} \left( \mathbf{e}_{\sigma} 
    \opa_{\mathbf{k}\sigma} \rme^{\rmi \mathbf{r} \cdot \mathbf{k}} + \mathbf{e}_{\sigma}^{\star} \opa_{\mathbf{k}\sigma}^{\dagger} \rme^{- \rmi \mathbf{r} \cdot \mathbf{k}} \right) ,
\end{equation}
with $f_{\bf k}$ the vacuum electric field amplitude given by Eq.~\eqref{eq:gk}.
The Hamilton operators are 
\begin{equation}
    \tilde{\opH}_F = \sum_{\mathbf{k}\sigma} \omega_{\mathbf{k}} \: \opa_{\mathbf{k}\sigma}^{\dagger} \opa_{\mathbf{k}\sigma} .
\end{equation}
for the field and
\begin{equation}
    \tilde{\opH}_{\rm int} = \dfrac{1}{2} \left[ \mathbf{P} \cdot \tilde{\mathbf{A}} (\mathbf{r}) + \tilde{\mathbf{A}} (\mathbf{r}) \cdot \mathbf{P} + \tilde{\mathbf{A}}(\mathbf{r})^2 \right] ,
\end{equation}
for the light-matter interaction, with $\mathbf{P}$ the dipole momentum conjugated to the dipole moment $\mathbf{D}$ of the atom or the molecule.
The time evolution of the density matrix of the total system is governed by the von Neumann equation of motion
\begin{equation}
\label{eq:vonNeumann}
    \dfrac{\rmd \tilde{\rho}}{\rmd t} = - \rmi [ \tilde{\opH} , \tilde{\rho} ] .
\end{equation}

\subsection{Interaction picture of the field}
We perform the transformation to the interaction picture of the field
\begin{equation}
    \bar{\rho} = \opU_F^{\dagger} (t) \: \tilde{\rho} \: \opU_F (t) , \qquad \opU_F (t) = \rme^{-\rmi \tilde{\opH}_F t} .
\end{equation}
Clearly, $[\tilde{\opH}_A , \opU_F(t)] = 0$.
The equation of motion~\eqref{eq:vonNeumann} becomes 
\begin{equation}
    \dfrac{\rmd \bar{\rho}}{\rmd t} = - \rmi [ \bar{\opH} (t) , \bar{\rho} ] ,
\end{equation}
with $\bar{\opH} (t) = \opU_F^{\dagger} \tilde{\opH} \opU_F + \rmi \dot{\opU}_F^{\dagger} \opU_F$, i.e.,
\begin{equation}
    \bar{\opH} (t) = \bar{\opH}_A + \bar{\opH}_{\rm int} (t) .
\end{equation}
The Hamilton operator for matter $\bar{\opH}_A =  \tilde{\opH}_A$ is unchanged and that for the light-matter interaction becomes 
\begin{equation}
    \bar{\opH}_{\rm int} = \dfrac{1}{2} \left[ \mathbf{P} \cdot \bar{\mathbf{A}} (\mathbf{r},t) + \bar{\mathbf{A}}(\mathbf{r},t)^2 + \bar{\mathbf{A}} (\mathbf{r},t) \cdot \mathbf{P} \right]  ,
\end{equation}
with a quantum vector potential 
\begin{equation}
    \bar{\mathbf{A}} (\mathbf{r},t) = \opU_F^{\dagger} (t) \tilde{\mathbf{A}} (\mathbf{r}) \opU_F (t) ,
\end{equation}
that reads explicitly as
\begin{equation}
\label{eq:AF_interaction}
    \bar{\mathbf{A}} (\mathbf{r},t) = \sum_{\mathbf{k} \sigma} \dfrac{f_{\bf k}}{\omega_{\mathbf{k}}} \left( \mathbf{e}_{\sigma} 
    \opa_{\mathbf{k}\sigma} \rme^{\rmi ( \mathbf{r} \cdot \mathbf{k} - \omega_{\mathbf{k}} t)} + \mathbf{e}_{\sigma}^{\star} \opa_{\mathbf{k}\sigma}^{\dagger} \rme^{- \rmi ( \mathbf{r} \cdot \mathbf{k} - \omega_{\mathbf{k}} t )} \right) .
\end{equation}

\subsection{Dipole approximation and length gauge}
Within the dipole approximation, the spatial variations of the vector potential are neglected and the vector potential~\eqref{eq:AF_interaction} becomes 
\begin{equation}
    \bar{\mathbf{A}} (\mathbf{r},t) \approx \mathbf{A}(t) ,
\end{equation}
with vector potential
\begin{equation}
    \mathbf{A} (t) = \sum_{\mathbf{k} \sigma} \dfrac{f_{\bf k}}{\omega_{\mathbf{k}}} \left( \mathbf{e}_{\sigma} \opa_{\mathbf{k}\sigma} \rme^{- \rmi \omega_{\mathbf{k}}t} + \mathbf{e}_{\sigma}^{\star} \opa_{\mathbf{k}\sigma}^{\dagger}  \rme^{\rmi \omega_{\mathbf{k}}t} \right) ,
\end{equation}
and electric field $\mathbf{F}(t) = - \partial_t \mathbf{A}(t)$ given by Eq.~\eqref{eq:main_Ft}.
Performing the gauge transformation 
\begin{equation}
    \rho (t) = \opU_{L}^{\dagger} (t) \bar{\rho} (t) \opU_L (t) , \qquad \opU_L(t) = \rme^{-\rmi \mathbf{r} \cdot \mathbf{A} (t)} ,
\end{equation}
leads to the von Neumann equation given by Eq.~\eqref{eq:main_vonNeumann} with the Hamilton operator~\eqref{eq:main_Hamiltonian}. 

\section{Mathematical complements \label{sec:formal_expressions}}
\subsection{Semiclassical one- and two-photon transition matrix elements \label{sec:SemiclassicalM1M2}}
In this appendix, we recall the semiclassical one- and two-photon transition matrix elements in the infinitely long pulse limit~\cite{Veniard1996, Maquet1998, Klunder2011, Dahlstrom2013, Busto2019, Berkane2024}.
We first introduce the eigenbasis of the field-free atom $\lbrace \ket{n} \rbrace$ with $\opH_A \ket{n} = E_n \ket{n}$ which includes the continuum states.
In Eqs.~\eqref{eq:U1_Eg} and~\eqref{eq:U2_Eg}, there are the $n$-photon transition matrix elements from the ground state to a continuum state of energy $E$ such that
\begin{equation}
    M^{(n)}_{\beta} (E) = \lim_{t \to \infty} \bra{E} \opM^{(n)}_{\beta} (t) \ket{\rm g} ,
\end{equation}
for $\beta = \lbrace a,e \rbrace$,
with $\opM^{(n)}_{\beta} (t)$ the $n$-photon time-dependent transition matrix element.
They read explicitly as
\begin{widetext}
\begin{equation}
    \opM^{(1)}_{\beta} (t) = \lim_{\epsilon\to0^+}  \sumint_{nm} \dfrac{\ket{n}\bra{n} \mathbf{D} \cdot \mathbf{e}_{\beta}  \ket{m}\bra{m}}{E_m + \omega_{\beta} - E_n + \rmi \epsilon} \exp \left[  \rmi (E_n - E_m - \omega_{\beta} - \rmi \epsilon) t \right] ,
\end{equation}
for $n=1$ and
\begin{eqnarray}
    && \opM^{(2)}_{\beta} (t) = \lim_{\epsilon\to 0^+}  \sumint_{knm}  \dfrac{\exp \left[  \rmi (E_k - E_m - \omega_{\beta} + \epsilon_{\beta} \wIR - 2 \rmi \epsilon) t \right]}{E_m + \omega_{\beta} - \epsilon_{\beta} \wIR - E_k + 2 \rmi \epsilon}  \nonumber \\
    && \qquad \qquad \qquad \qquad \times  \left\lbrace \dfrac{\ket{k}\bra{k} \mathbf{D} \cdot \mathbf{e}_{0} \ket{n}\bra{n} \mathbf{D} \cdot \mathbf{e}_{\beta}  \ket{m}\bra{m}}{E_m + \omega_{\beta} - E_n + \rmi \epsilon} + \dfrac{\ket{k}\bra{k} \mathbf{D} \cdot \mathbf{e}_{\beta} \ket{n}\bra{n} \mathbf{D} \cdot \mathbf{e}_{0}  \ket{m}\bra{m}}{E_m + \wIR - E_n + \rmi \epsilon} \right\rbrace , 
\end{eqnarray}
\end{widetext}
for $n=2$, with $\epsilon_{a}=-1$ and $\epsilon_e=+1$.

\subsection{Cauchy-Schwarz bounds on the RABBIT contrast \label{sec:Cauchy_Schwartz_derivation}}
Substituting~\eqref{eq:Cauchy_Contrast} into~\eqref{eq:C2} yields
\begin{equation}
    C \leq \dfrac{2 \: \big| \xi^{(2)} (E) \big| \times \sqrt{\dfrac{\braket{\opa_e^{\dagger} \aIR \aIR^{\dagger} \opa_e}}{\braket{\opa_a^{\dagger} \aIR^{\dagger}\aIR \opa_a}}}}{\big| \xi^{(2)} (E) \big|^2 \dfrac{\braket{\opa_e^{\dagger} \aIR \aIR^{\dagger} \opa_e}}{\braket{\opa_a^{\dagger} \aIR^{\dagger}\aIR \opa_a}} + 1} . 
\end{equation}
Introducing the dimensionless quantity
\begin{equation}
    z = \big| \xi^{(2)} (E) \big| \times \sqrt{\dfrac{\braket{\opa_e^{\dagger} \aIR \aIR^{\dagger} \opa_e}}{\braket{\opa_a^{\dagger} \aIR^{\dagger}\aIR \opa_a}}} ,
\end{equation}
the contrast satisfies the compact bound
\begin{equation}
    C \leq \dfrac{2 z}{z^2 + 1} . 
\end{equation}
Since $2 z / (z^2+1) \leq 1$ for all $z\geq 0$, with equality at $z=1$,
the contrast is bounded above by unity. This maximum is achieved when the Cauchy-Schwarz inequality~\eqref{eq:Cauchy_Contrast} is saturated and $z=1$ simultaneously.

\subsection{Wigner distribution for a squeezed coherent state \label{sec:Wigner}}
We consider a squeezed coherent state with squeezing amplitude $\zeta = r \rme^{\rmi \varphi}$ and displacement $\alpha_0$ as written in~\eqref{eq:squeezed_coherent_state}.
The Wigner distribution in the complex phase space is given by~\cite{Bishop1994}
\begin{equation}
    W (\alpha) = \dfrac{2}{\pi} {\rm tr} \left[ \rho \: \opD (\alpha) \: \Pi \: \opD^{\dagger} (\alpha) \right],
\end{equation}
with $\rho$ the density matrix for~\eqref{eq:squeezed_coherent_state} and $\Pi = (-1)^{\opa^{\dagger}\opa}$ the parity operator.
Its Wigner distribution is explicitly given by
\begin{equation}
\label{eq:WD_BSV}
    W (\alpha) = \dfrac{2}{\pi} \exp \left[ - \dfrac{1}{2} (\mathbf{A} - \mathbf{A}_0)^{\dagger} \Sigma^{-1} (\mathbf{A}-\mathbf{A}_0) \right] ,
\end{equation}
with $\mathbf{A} = (\alpha,\alpha^{\star})^{\top}$ and $\mathbf{A}_0 = (\alpha_0,\alpha_0^{\star})^{\top}$.
The covariance matrix is
\begin{equation}
    \Sigma^{-1} = 2 \; \opQ (\varphi/2) 
    \begin{pmatrix}
        \rme^{2r} & 0 \\
        0 & \rme^{-2r} 
    \end{pmatrix} 
    \opQ (\varphi/2)^{\dagger} ,
\end{equation}
with the unitary matrix
\begin{equation}
    \opQ (\theta) = \dfrac{1}{\sqrt{2}} 
    \begin{pmatrix}
        \rme^{\rmi \theta} & - \rmi \rme^{\rmi \theta} \\
        \rme^{-\rmi \theta} & \rmi \rme^{-\rmi \theta}
    \end{pmatrix} .
\end{equation}
Given that $\det \Sigma^{-1} = 4$, we recover the normalization condition 
\begin{equation}
    \int W(\alpha) \; \rmd^2 \alpha = 1 .
\end{equation}
We have verified that, for $\theta=0$, this Wigner distribution leads to the variances $(\Delta x)^2 = \rme^{-2r}/2$ and $(\Delta p)^2 = \rme^{2r}/2$ in the convention $\alpha = (x+\rmi p)/\sqrt{2}$.

In order to retrieve the properties~\eqref{eq:BSV_properties}, one can use the usual rules for the Wigner distribution~\eqref{eq:WD_BSV} which are given by the following 
\begin{equation}
    \braket{(\hat{\mathbf{A}} - \mathbf{A}_0) (\hat{\mathbf{A}} - \mathbf{A}_0)^{\dagger}}_{\rm sym} = \Sigma ,
\end{equation}
where $\hat{\mathbf{A}} = (\opa , \opa^{\dagger})^{\top}$ and where $\braket{\cdot}_{\rm sym}$ denotes the symmetrized bracket such that $\braket{\opa \opa^{\dagger}}_{\rm sym} = \braket{\opa^{\dagger} \opa}_{\rm sym} = \braket{\opa^{\dagger} \opa} + 1/2$.

\section{Ab-initio numerical simulations for an infrared squeezed coherent state \label{sec:TDSE}}
For the simulations with an infrared squeezed coherent state of light in Sections~\ref{sec:classicalXUV} and~\ref{sec:classicalXUV_nonperturbative}, the RABBIT spectrum is obtained from an incoherent sum of spectra weighted by $W(\alpha)$ obtained with the TDSE
\begin{equation}
\label{eq:TDSE}
    \dot{\psi}_{\alpha} (x,t) = - \rmi \Big( \opH_A + \bra{\alpha} \opH_{\rm int}(t) \ket{\alpha} \Big) \psi_{\alpha} (x,t) .
\end{equation}
This method is particularly well adapted for definite-positive Wigner distributions that have the property of a density probability, and when the net exchange of photons during the simulation is negligible compared to the number of photons in the incident field.
Formally speaking, this comes from the calculation of the intensity spectrum in Eq.~\eqref{eq:ISpectrum} using the reconstruction of the density matrix in Eq.~\eqref{eq:main_vonNeumann} in terms of 
\begin{equation}
\label{eq:rho_W}
    \rho (t) = \int \ket{\psi_{\alpha}(t)} W(\alpha) \bra{\psi_{\alpha}(t)} \; \rmd \alpha .
\end{equation}
In this way, the numerical solution of the von Neumann equation~\eqref{eq:main_vonNeumann} reduces to solving the standard TDSE with \textit{semiclassical} light-matter interaction 
\begin{subequations}
\begin{equation}
    \bra{\alpha} \opH_{\rm int}(t) \ket{\alpha} = x \left[ F_{\alpha} (t-\tau) + F_{h} (t) \right] ,    
\end{equation}
where the harmonic field is
\begin{equation}
\label{eq:Fht}
    F_{h} (t) = \sum_{q} F_q \, f (t)  \cos (\omega_q t) ,
\end{equation}
and the effective infrared field is
\begin{equation}
    F_{\alpha} (t) = 2 f_0 \: f (t) \Re \left( \alpha \rme^{-\rmi \wIR t} \right) .
\end{equation}
\end{subequations}
Here, $f(t)$ is a $\sin^2$ laser envelope with a total duration of 3000 a.u. 
To evaluate the $W (\alpha)$-weighted sum, we generated a comprehensive library of $\ket{\alpha}$-dependent spectra. These were obtained by interpolating from a data set sampled in 100 carrier-envelope phases and a log-scaled range of IR intensities spanning $0$ to $10^{13} \; \rm W \ cm^{-2}$.
Note that the intensity of the squeezed coherent states considered here is negligible outside this range of intensities. 

\end{document}